\documentclass[aps,twocolumn,showpacs,preprintnumbers,amsmath,amssymb,nofootinbib,superscriptaddress,showkeys]{revtex4}

\usepackage{epsfig}
\usepackage{graphicx}

\begin{document}

\title{Serber symmetry, Large $N_c$ and Yukawa-like One Boson
  Exchange Potentials}

\author{A. Calle Cord\'on}\email{alvarocalle@ugr.es}
  \affiliation{Departamento de F\'isica At\'omica, Molecular y
  Nuclear, Universidad de Granada, E-18071 Granada, Spain.}
\author{E. Ruiz
  Arriola}\email{earriola@ugr.es}
  \affiliation{Departamento de F\'isica At\'omica, Molecular y Nuclear,
  Universidad de Granada,
  E-18071 Granada, Spain.}
\date{\today}

\begin{abstract} 
\rule{0ex}{3ex} The Serber force has relative orbital parity symmetry
and requires vanishing NN interactions in partial waves with odd
angular momentum. We illustrate how this property is well fulfilled
for spin triplet states with odd angular momentum and violated for odd
singlet states for realistic potentials but fails for chiral
potentials.  We analyze how Serber symmetry can be accommodated within
a large $N_c$ perspective when interpreted as a long distance
symmetry. A prerequisite for this is the numerical similarity of the
scalar and vector meson resonance masses. The conditions under which
the resonance exchange potential can be approximated by a Yukawa form
are also discussed.  While these masses arise as poles on the Second
Riemann in $\pi\pi$ scattering, we find that within the large $N_c$
expansion the corresponding Yukawa masses correspond instead to a well
defined large $N_c$ approximation to the pole which cannot be
distinguished from their location as Breit-Wigner resonances.  
\end{abstract}

\maketitle



\section{Introduction}

The modern theory of nuclear forces~\cite{Epelbaum:2008ga} aims at a
systematic and model-independent derivation of the forces between
nucleons in harmony with the symmetries of Quantum Chromodynamics.
Actually, an outstanding feature of nuclear forces is their exchange
character. Many years ago, Serber postulated~\cite{Serber} an
interesting symmetry for the nucleon-nucleon system based on the
observation that at low energies the proton-proton and neutron-proton
differential cross section are symmetric functions in the Center of
Mass (CM) scattering angle around $90^0$. This orbital parity symmetry
corresponds to the transformation $\theta \to \pi -\theta $ in the
scattering amplitude and was naturally explained by assuming that the
potential was vanishing for partial waves with odd angular
momentum. Specific attempts were directed towards the verification of
such a property~\cite{PhysRev.73.973} (See
Refs.~\cite{1952RPPh...15...68C} and \cite{TheoLib:NUC05} for early
and comprehensive reviews.). This symmetry was shown to hold for the
NN system, up to relatively high energies~\cite{PhysRev.77.441}.
However, such a force was also found to be incompatible with the
requirement of nuclear matter saturation~\cite{PhysRev.77.568} as well
as with the underlying meson forces mediated by one and two pion
exchange~\cite{PhysRev.88.144}. These puzzling inconsistencies were
cleared up when it was understood that only singular Serber forces
could provide saturation~\cite{PhysRev.81.165}. Old phase shift
analyses~\cite{PhysRev.122.1606} confirm the rough Serber exchange
character of nuclear forces. Many nuclear
structure~\cite{delaRipelle:2004ms}, nuclear matter~\cite{Fetter},
nuclear
reactions~\cite{PhysRev.77.647,2007PAN....70.1440L,RevModPhys.57.923},
use Serber forces both for their simplicity as well as their
phenomenological success in the low and medium energy region.  The
possibility of implementing Serber forces in the nuclear potential was
envisaged in Skyrme's seminal paper~\cite{Skyrme:1959zz}. Modern
versions (SLy4) of the Skyrme effective
interactions~\cite{Chabanat:1997qh} implement the symmetry
explicitly. In a recent paper~\cite{Zalewski:2008is} a novel fitting
strategy has been proposed for the coupling constants of the nuclear
energy density functional, which focus on single-particle energies
rather than ground-state bulk properties, yielding naturally an almost
perfect fulfillment of Serber symmetry.

\begin{figure}[bbb]
\includegraphics[height=6cm,width=5cm,angle=270]{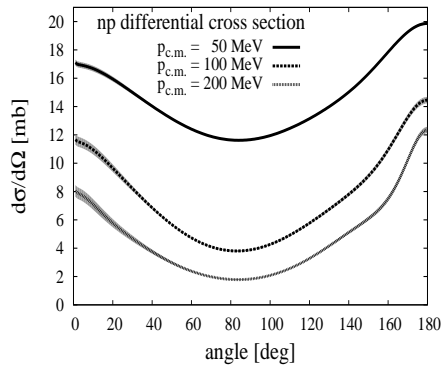}
\caption{Differential cross section for np scattering at several LAB
  energies. The error band reflects the PWA and high quality potentials
  of the Nijmegen group~\cite{Stoks:1993tb,Stoks:1994wp}. Serber
  symmetry implies that the functions should be symmetric in the CM
  scattering angle around $90^o$.}
\label{fig:dsg}
\end{figure}

A vivid demonstration of the Serber symmetry is shown in
Fig.~\ref{fig:dsg} where the np differential cross section is plotted
for several LAB energies using the Partial Wave Analysis and the high
quality potentials~\cite{Stoks:1993tb,Stoks:1994wp} carried out by the
Nijmegen group. While discrepancies regarding the comparison between
forward and backward directions show that this symmetry breaks down at
short distances, the intermediate region does exhibit Serber
symmetry. In any case it is interesting to see that even in the
intermediate energy region departures from the symmetry can be seen,
the most important one is the fact that the symmetry point is shifted
a few degrees towards lower values than $90^0$ for increasing
energies.  While these are well established features of the NN
interaction, it is amazing that such a time honoured force and gross
feature of the NN interaction, even if it does not hold in the entire
range, has no obvious explanation from the more fundamental and QCD
motivated side. To our knowledge this topic has not been explicitly
treated in any detail in the literature and no attempts have been made
to justify this evident but, so far, accidental
symmetry.  The present paper tries to fill this gap unveiling Serber
symmetry at the relevant scales from current theoretical approaches to
the NN problem, looking for its consequences in nuclear physics and
analyzing its possible origin. Of course, a definite explanation
might finally be given by lattice QCD calculations for which incipient
results exist already in the case of S-wave
interactions~\cite{Ishii:2006ec,Beane:2006mx}.

The motivation for the present study arises from our recent
analysis~\cite{CalleCordon:2008cz} of an equally old symmetry, the
SU(4)-spin-isospin symmetry proposed independently by Wigner and
Hund~\cite{Wigner:1936dx, 1937ZPhy..105..202H} by introducing the
concept of long distance symmetry. Specifically we showed how a
symmetry of the potential at any non vanishing but arbitrarily small
distance does not necessarily imply a symmetry of the S-matrix which
may be directly observed at all energies. This provided an
interpretation of the role played by the Wigner symmetry in the
S-waves; the potentials for the two nucleon $^1S_0$ and $^3S_1$ states
are {\it identical} while the corresponding phase shifts are {\it very
  different} at all measurable energies.  Furthermore, we showed how a
sum rule for SU(4) super-multiplet phase shifts splitting due to
spin-orbit and tensor interactions is well fulfilled for non-central
L-even waves, and strongly violated in L-odd waves where a Serber-like
symmetry holds instead.  In Section~\ref{sec:mean-pot} we review the
sum rules obtained from our previous work for the partial wave phase
shifts and show how their potential counterpart is also well verified
by high quality NN potentials, i.e. potentials which have $\chi^2/DOF
\sim 1$.  Obviously, {\it any} NN potential explaining the data will
necessarily comply to odd-L Serber symmetry as a whole; a less trivial
matter is to check whether this is displayed {\it explicitly} by the
potential and what are the relevant ranges where the symmetry is
located.  At long distances the interaction is given by One Pion
Exchange (OPE) which is Wigner symmetric for even-L waves and provides
a $1/9$ violation of Serber symmetry for odd-L waves at long
distances. Phenomenological potentials seem to provide different
ranges for each symmetry. In Section~\ref{sec:searching} we analyze
the signatures of the symmetry and its range from several viewpoints
including the PWA of the Nijmegen group, the $V_{\rm low \, k}$
approach and chiral two pion exchange. In Section~\ref{sec:ct} we
digress on the meaning of counterterms as a diagnostics tool to
characterize a long distance symmetry both from a perturbative as well
as from a Wilsonian renormalization point of view, the implications
for Skyrme forces and the resonance saturation of chiral forces.

In any case, the evidence for both even-L Wigner and odd-L Serber
symmetries is so overwhelming that we feel a pressing need for an
explanation more closely based on our current knowledge of strong
interactions and QCD. Actually, central to our analysis will be the
use of the large $N_c$ expansion~\cite{'tHooft:1973jz,Witten:1979kh}
(for comprehensive reviews see
e.g. \cite{Manohar:1998xv,Jenkins:1998wy,Lebed:1998st}). Here $N_c$ is
the number of colours and in this limit the strong coupling constant
scales as $\alpha_s \sim 1/N_c$. For color singlet states the picture
is that of infinitely many stable mesons and glue-balls, which masses
behave as $m \sim N_c^0$ and widths as $\Gamma \sim 1/N_c$, and heavy
baryons, with their masses scaling as $M \sim N_c$. This limit also
fixes the interactions among hadrons. Meson-meson interactions are
weak since they scale as $1/N_c$, meson-baryon interactions $\sim
N_c^0$ and baryon-baryon interactions are strong as they scale as
$\sim N_c$.  While the pattern of $SU(4)$-symmetry breaking complies
to the large $N_c$ expectations~\cite{Kaplan:1995yg,Kaplan:1996rk}, a
somewhat unexpected conclusion, we also pointed out that Serber
symmetry, while not excluded for odd-L waves, {\it was not} a
necessary consequence of large $N_c$. The search for an explanation of
the Serber force requires more detailed information than in the case
of Wigner symmetry.

In Section~\ref{sec:serber-largeN} we approach the Serber symmetry
from a large $N_c$ perspective, and make explicit use of the fact that
the meson exchange picture seems
justified~\cite{Banerjee:2001js}. Actually, within such a realization
a necessary prerequisite for the validity of the symmetry would be a
numerical similarity of the scalar and vector meson masses. This poses
a puzzle since, as is well known, these mesons arise as resonances in
$\pi\pi$ scattering as poles in the second Riemann sheet yielding the
values $\sqrt{s_\sigma}= m_\sigma - i \Gamma_\sigma /2 =
441^{+16}_{-8} - i 272^{+9}_{-12} {\rm MeV}$~\cite{Caprini:2005zr} and
$\sqrt{s_\rho}= m_\rho - i \Gamma_\rho /2 = 775.49\pm 0.34 - i
149.4\pm 1.0 {\rm MeV}$~\cite{Amsler:2008zz}. The scalar and vector
masses and widths are sufficiently different as to make one question
if one is close to the Serber limit. In Section~\ref{sec:yukawa} we
review the role of two pion exchange and analyze the resonances as
well as the generation of Yukawa potentials from the exchange of
$\pi\pi$ resonances from a large $N_c$ viewpoint. An important result
of the present paper is to show that the Yukawa masses are determined
as large $N_c$ approximations to the pole position which cannot be
distinguished from the Breit-Wigner resonance. On the light of this
result it is possible indeed from the large $N_c$ side to envisage a
rationale for the Serber symmetry. Finally, in
Section~\ref{sec:conclusions} we come to the conclusions and summarize
our main points.

\section{Long distance symmetry and weighted average potentials}
\label{sec:mean-pot}

When discussing and analyzing symmetries in Nuclear Physics
quantitatively we find it convenient to delineate the scale where they
supposedly operate.  As we see from Fig.~\ref{fig:dsg}, Serber
symmetry does not work all over the range and equally well for all
energies. Thus, we expect to see the symmetry in the medium and long
distance region, precisely where a reliable theoretical description in
terms of potentials becomes possible. While this is easily understood,
it is less trivial to implement these features in a model independent
formulation. Furthermore, as we see from Fig.~\ref{fig:dsg} there are
some small deviations and it may be advisable to find out not only the
origin of the symmetry but also the sources of this symmetry breaking.

For the purpose of the present discussion we will separate the NN
potential as the sum of central components and non-central ones which
will be assumed to be small,
\begin{eqnarray} 
{\cal V}_{NN} = V_0 + V_1 \, , 
\end{eqnarray}
where $[ \vec L, V_0] =0 $ whereas $[ \vec J, V_1] = 0 $ and $[ \vec
  L, V_1] \neq 0 $. Specifically, for the central part we take 
\begin{eqnarray}
V_0= V_C + \tau W_C  + \sigma V_S + \tau \sigma W_S  \, , 
\end{eqnarray} 
while the non-central part is 
\begin{eqnarray}
V_1= (V_T  + \tau W_T ) S_{12}  +  (V_{LS}  + \tau W_{LS}) 
L \cdot S   \, . 
\end{eqnarray} 
Here $\tau= \tau_1 \cdot \tau_2 $ and $\sigma= \sigma_1 \cdot \sigma_2
$ with $\sigma_i$ and $\tau_i$ are the Pauli matrices representing the
spin and isospin respectively of the nucleon $i$. The tensor operator
is $S_{12} = 3 \sigma_1 \cdot \hat x \sigma_2 \cdot \hat x - \sigma_1
\cdot \sigma_2 $ while $L \cdot S $ corresponds to the spin-orbit
term. The total potential commutes with the total angular momentum $J
= L+S$. However, we will start assuming that the potential is central,
and that the breaking in orbital angular momentum is small.

\begin{figure}
\includegraphics[height=4cm,width=4cm,angle=270]{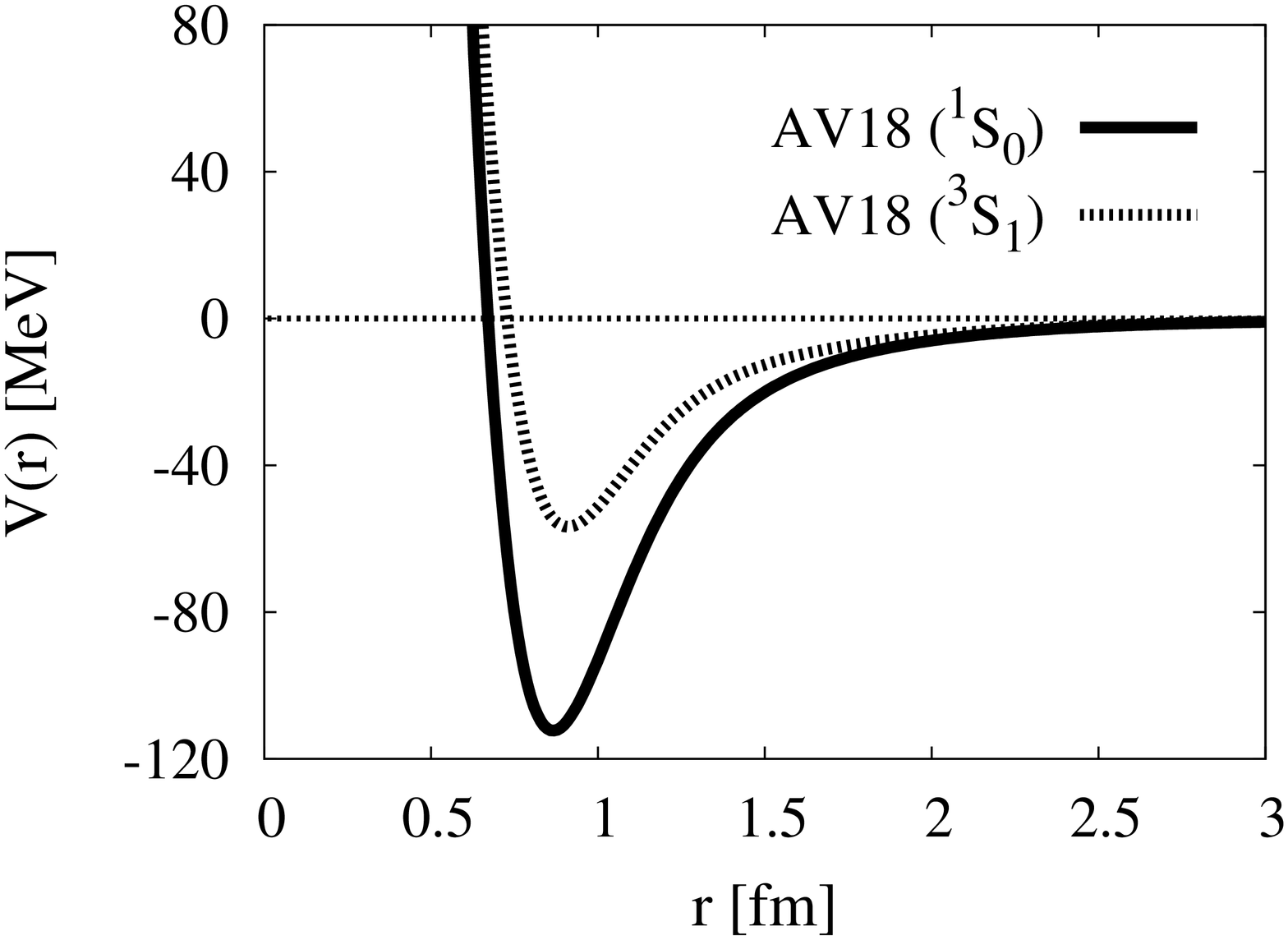}
\includegraphics[height=4cm,width=4cm,angle=270]{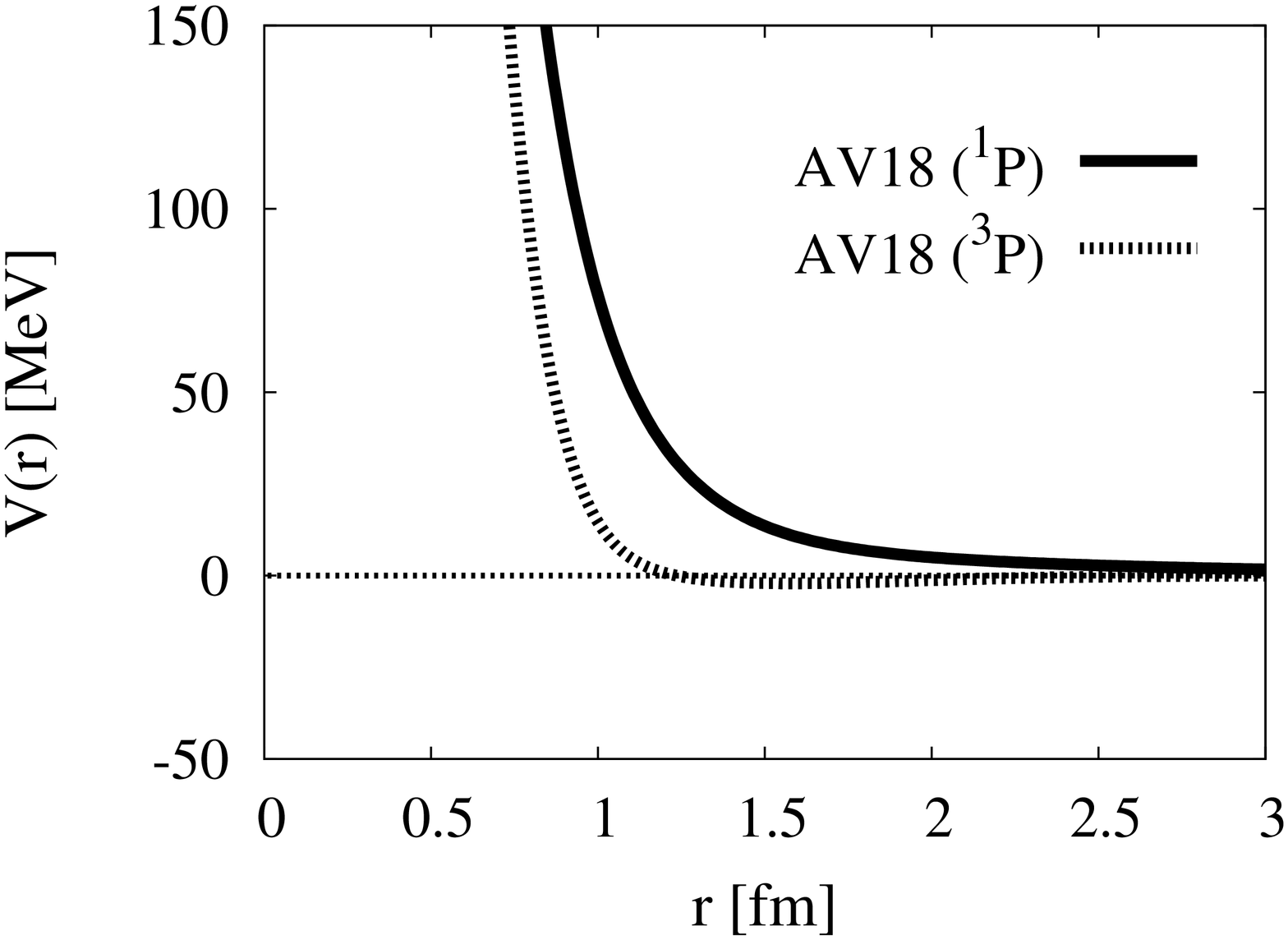}\\ \includegraphics[height=4cm,width=4cm,angle=270]{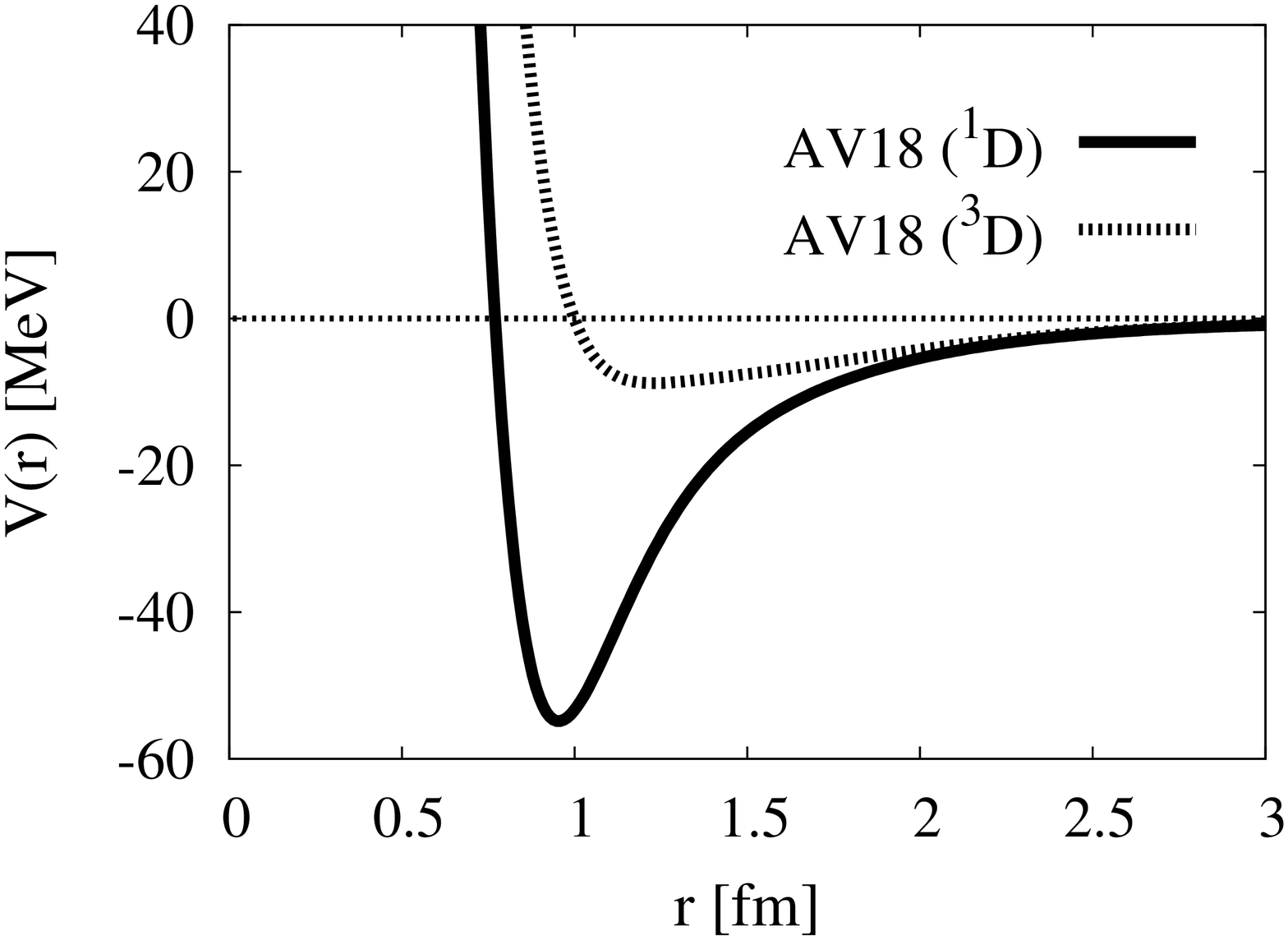}
\includegraphics[height=4cm,width=4cm,angle=270]{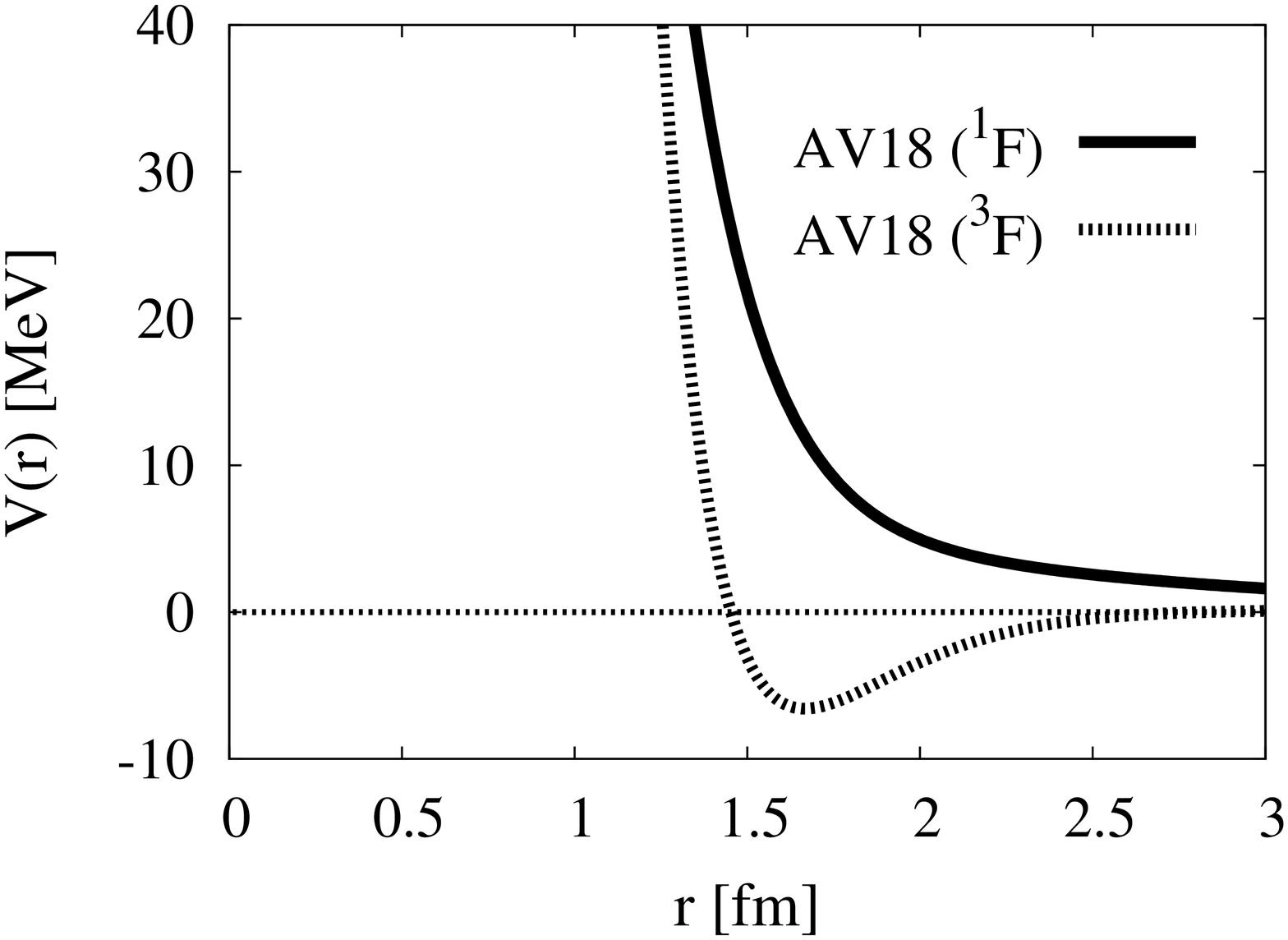}\\ \includegraphics[height=4cm,width=4cm,angle=270]{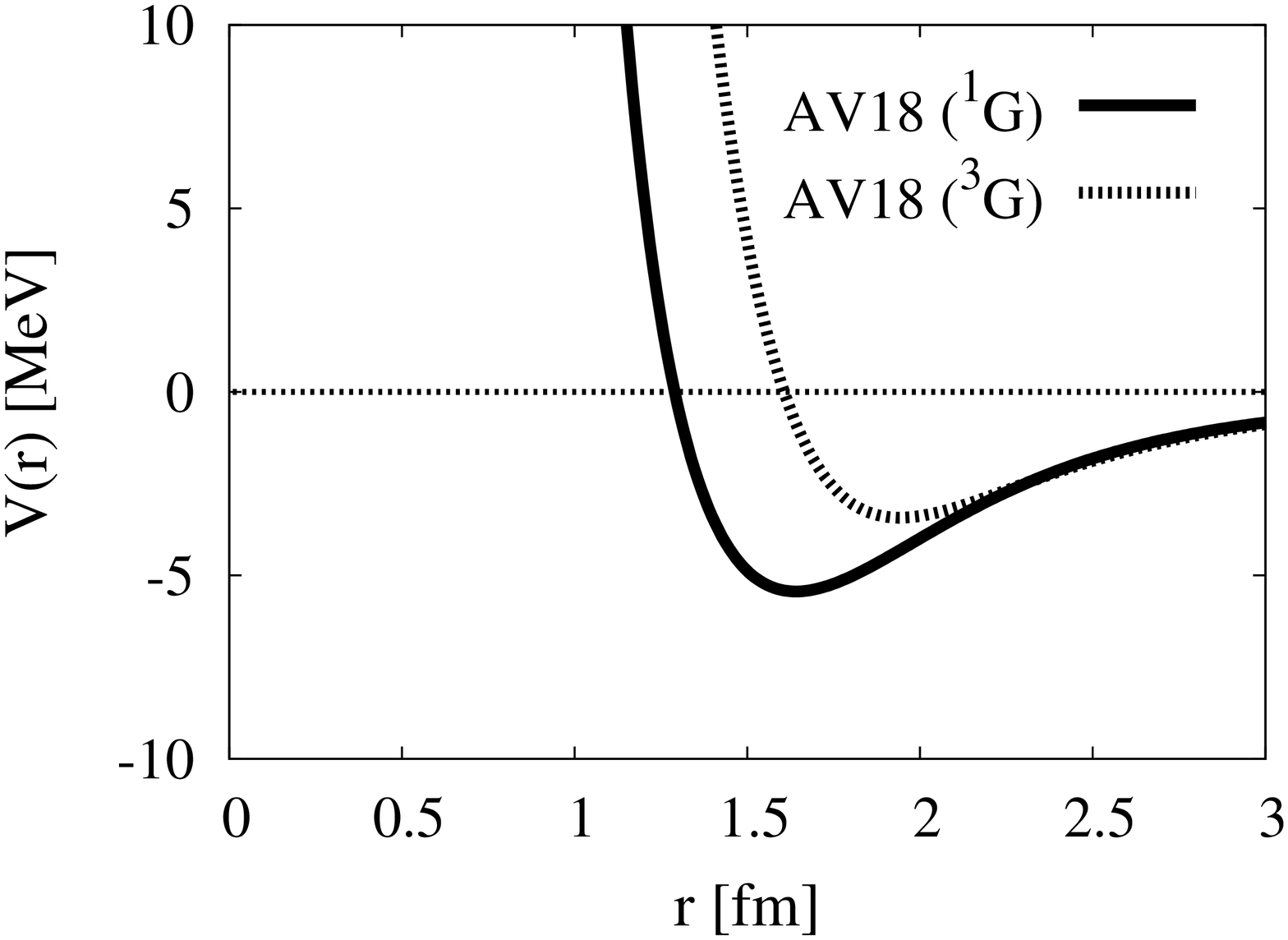}
\includegraphics[height=4cm,width=4cm,angle=270]{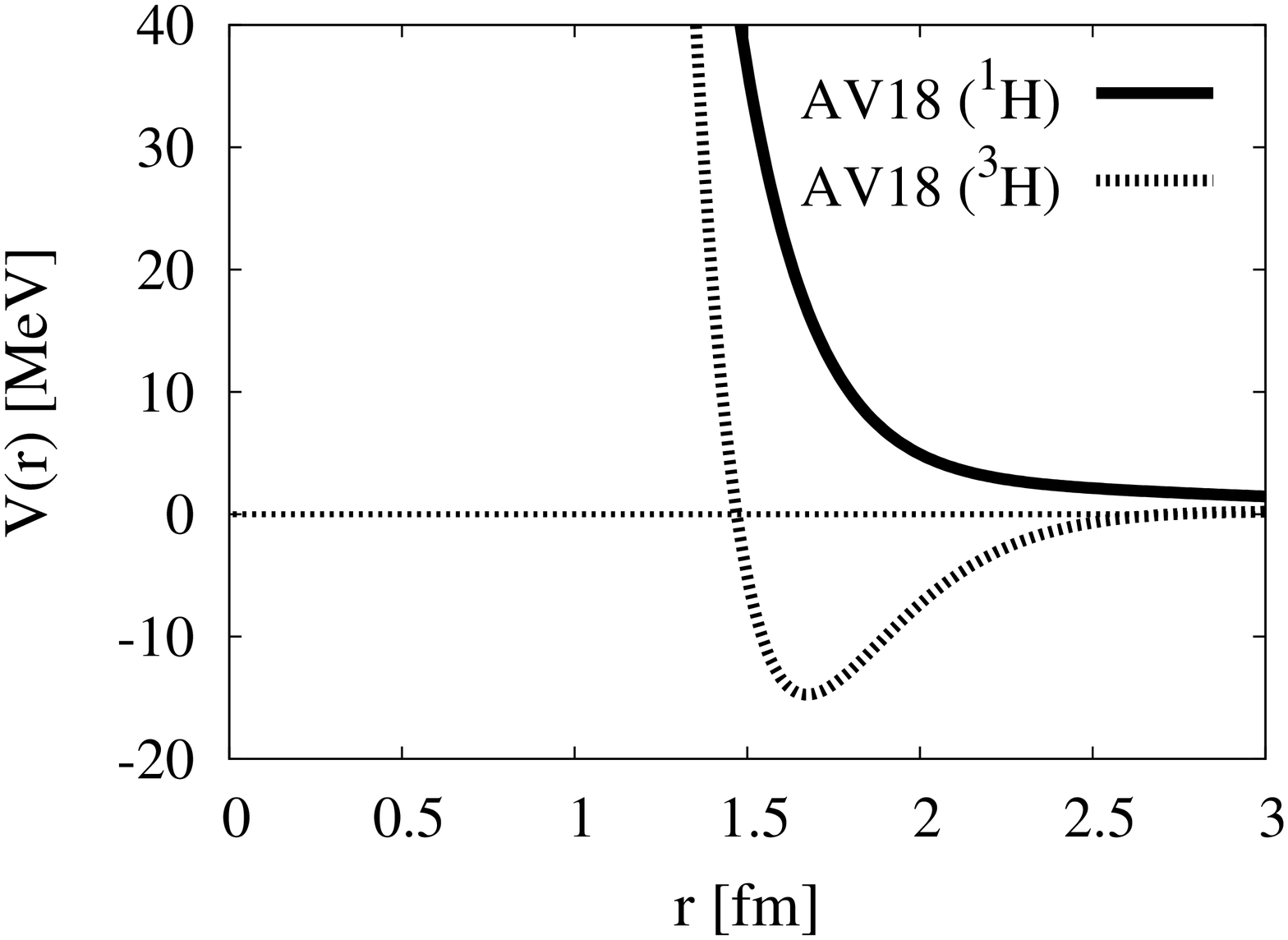}
\caption{ Argonne V-18 potentials~\cite{Wiringa:1994wb} for the center
  of the Serber-Wigner multiplets. Wigner symmetry requires singlet
  and triplet potentials to coincide. Serber symmetry implies
  vanishing odd-L partial waves.  Even-L waves posses Wigner symmetry
  while odd-L triplet waves exhibit Serber symmetry.}
\label{fig:wigner-serber}
\end{figure}

\begin{figure}[tbc]
\includegraphics[height=4cm,width=4cm,angle=270]{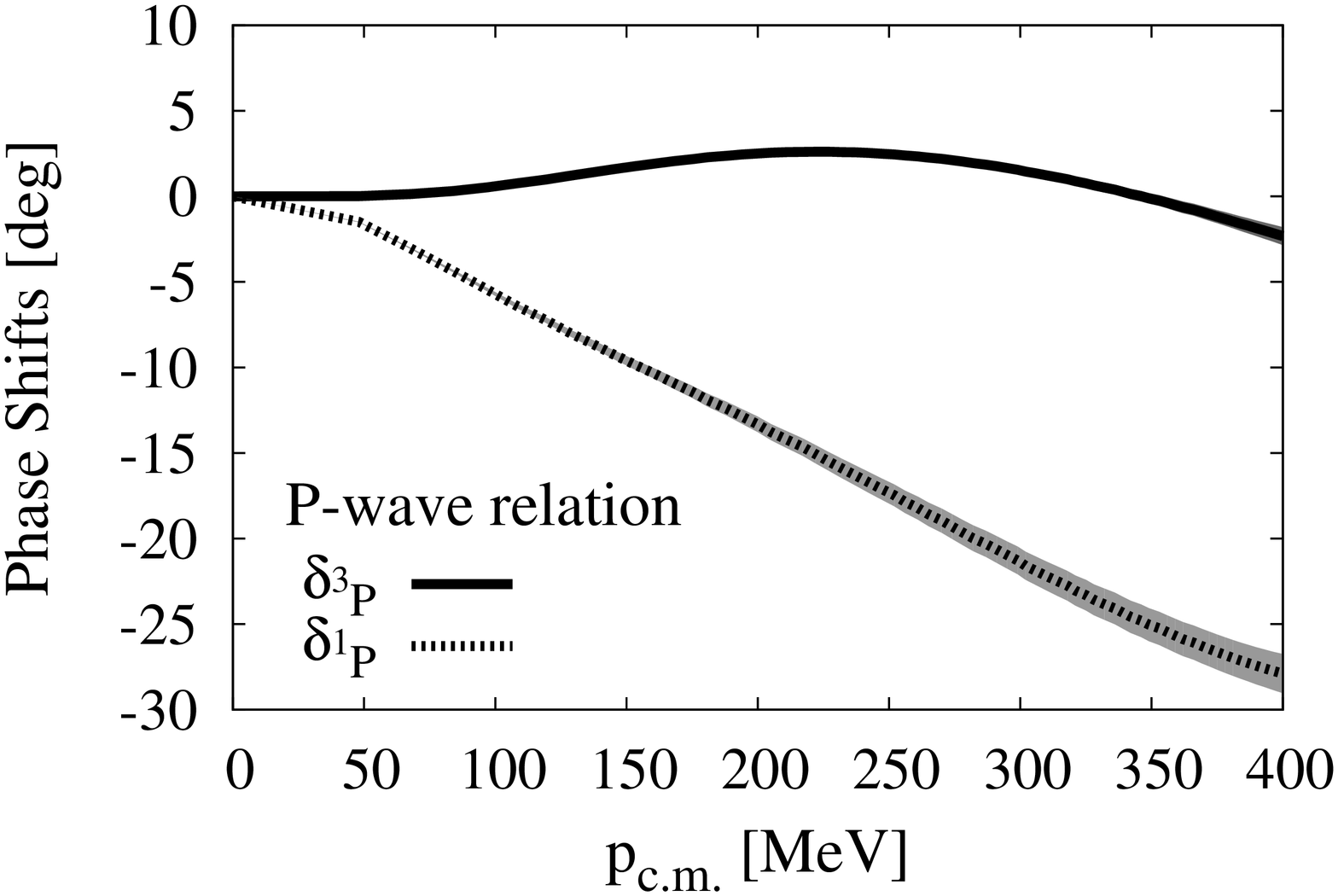}
\includegraphics[height=4cm,width=4cm,angle=270]{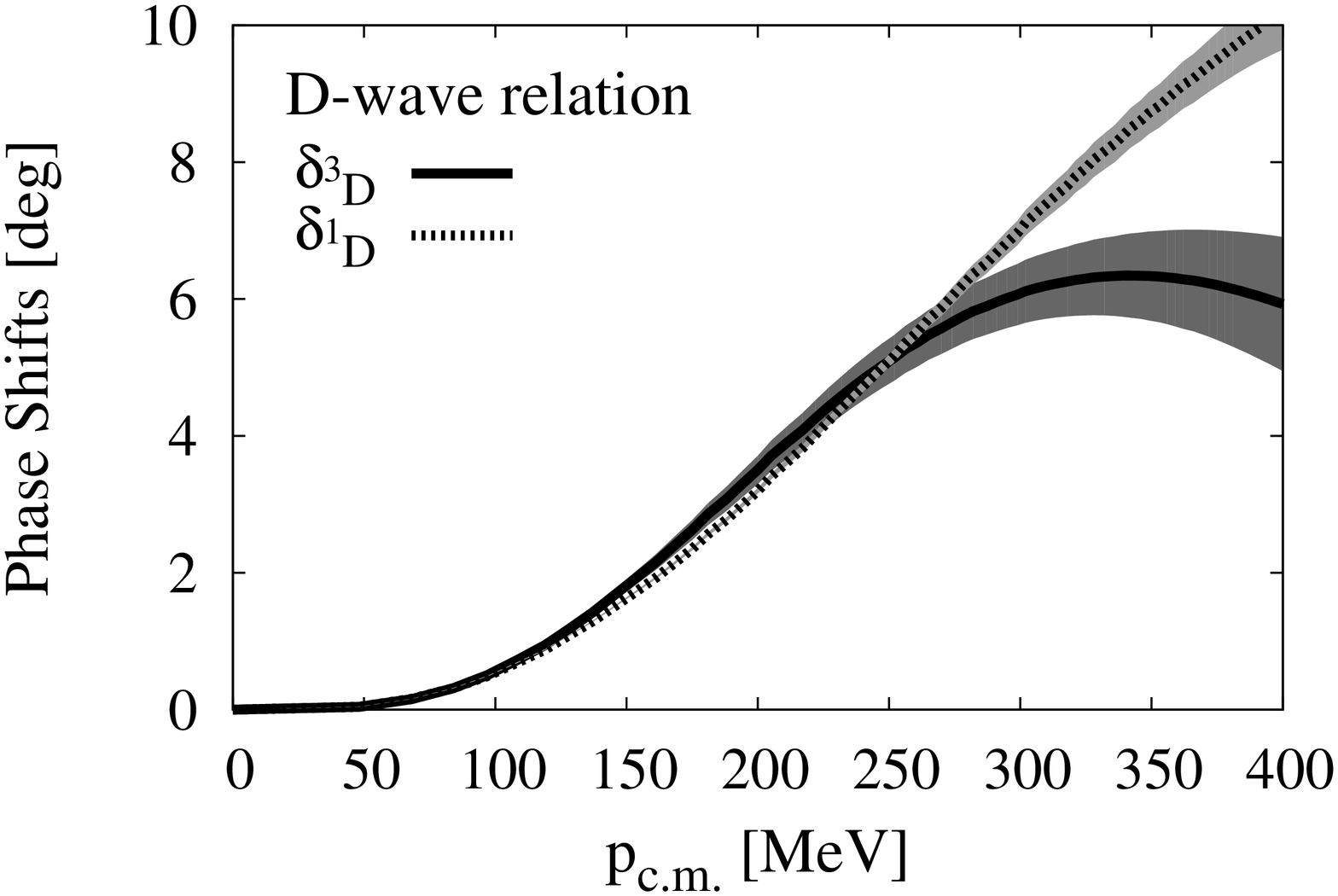} \\ 
\includegraphics[height=4cm,width=4cm,angle=270]{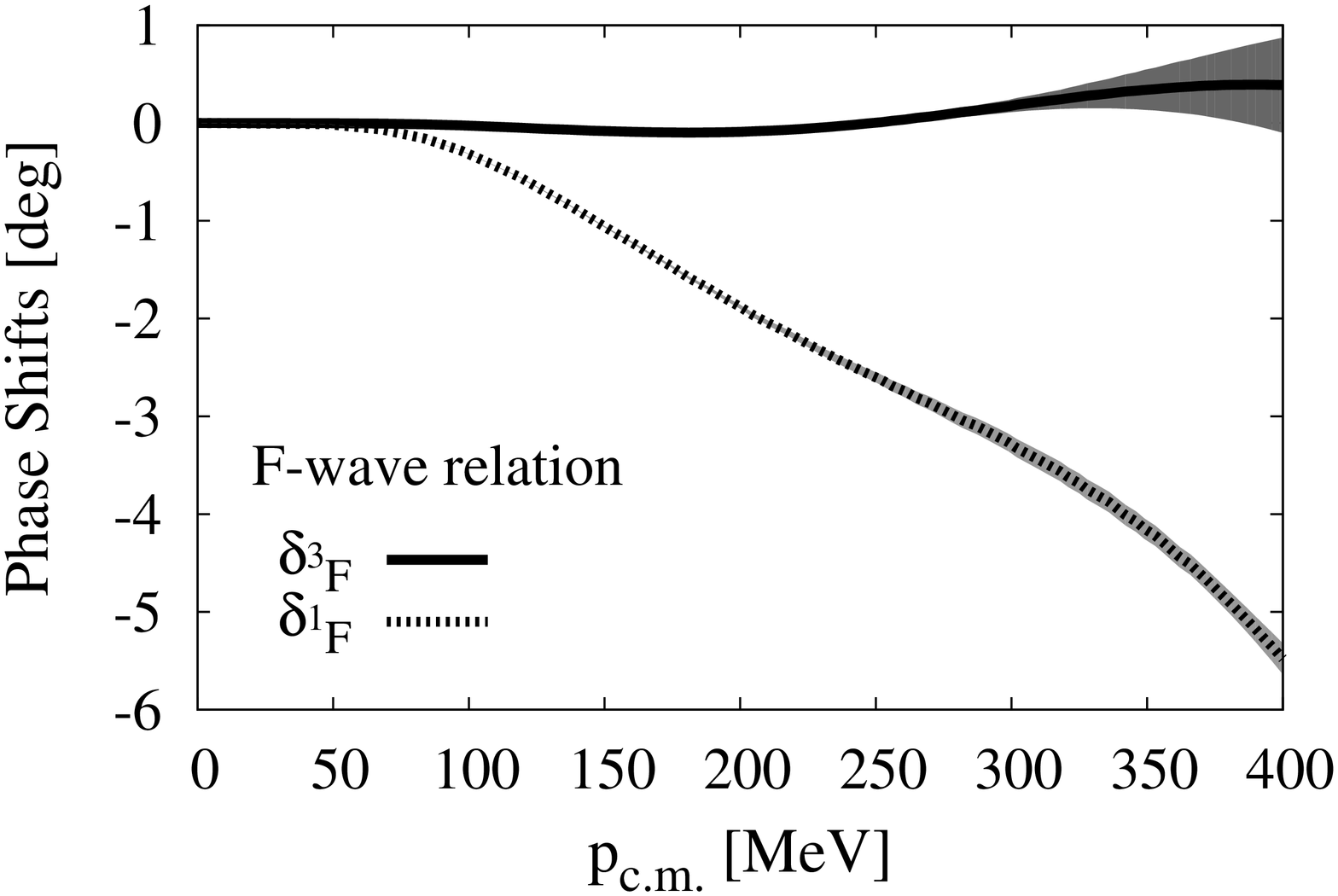}
\includegraphics[height=4cm,width=4cm,angle=270]{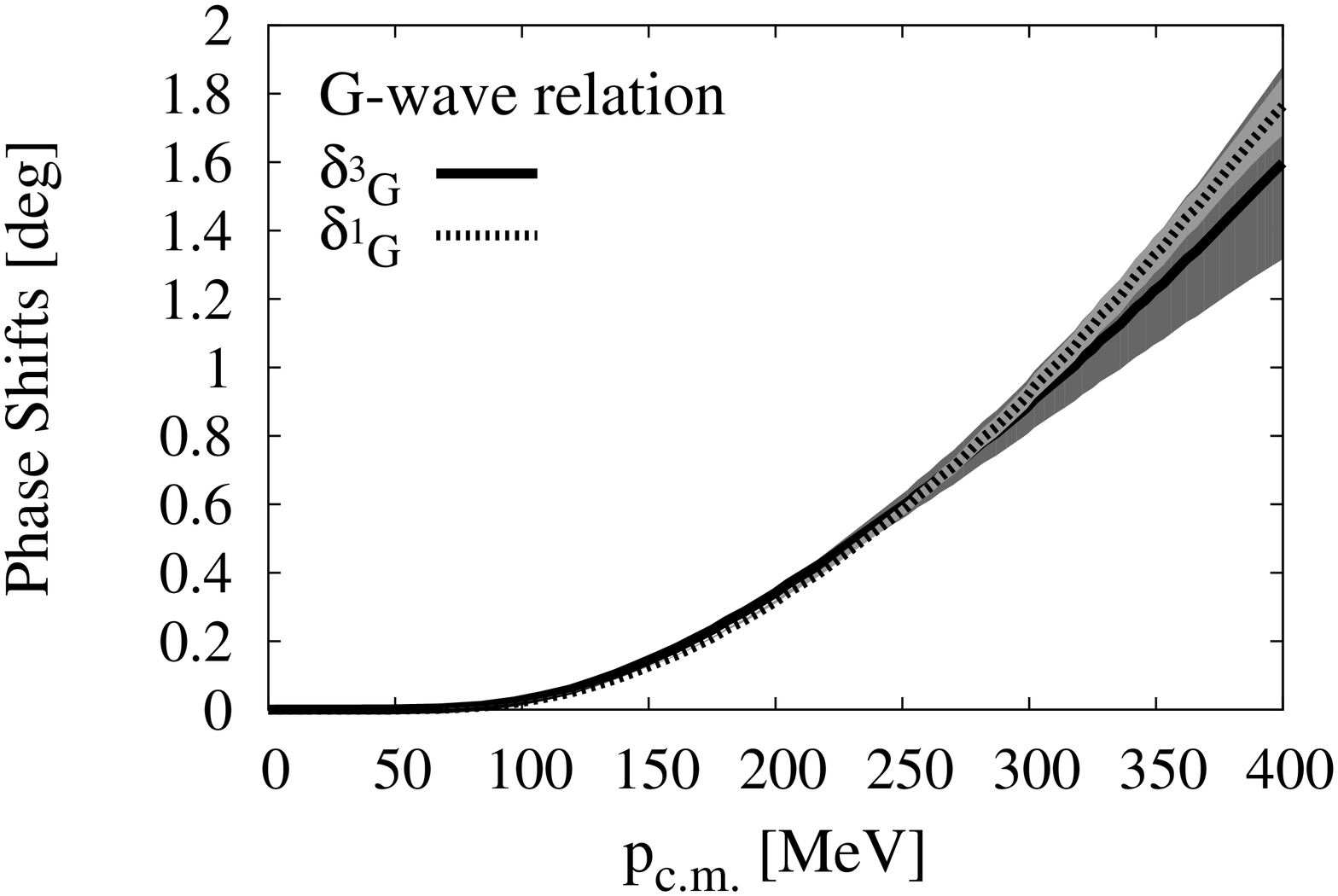}
\caption{Average values of the phase shifts~\cite{Stoks:1993tb} (in
  degrees) as a function of the CM momentum (in MeV).  (Upper left panel)
  P-waves. (Upper right panel) D-waves. (Lower left panel)
  F-waves. (Lower right panel) G-waves. According to the Wigner
  symmetry $\delta_{^1L}=\delta_{^3L}$. Serber symmetry implies
  $\delta_{^3L}=0$ for odd-L. One sees that L-even waves satisfy
  Wigner symmetry while L-odd waves satisfy Serber symmetry.}
\label{fig:higher}
\end{figure}

We proceed in first order perturbation theory, by using the central
symmetric distorted waves as the unperturbed states. Note that $\tau=
\tau_1 \cdot \tau_2 = 2 T(T+1)-3 $ and $\sigma= \sigma_1 \cdot
\sigma_2 = 2 S (S+1)-3 $ while Pauli principle requires
$(-)^{S+T+L}=-1$. The corresponding zeroth order wave function is of
the form
\begin{eqnarray}
\Psi ( \vec x ) = \frac{u_{L}^{ST} (r)}{r} Y_{L M_L}(\hat x) \chi^{S M_S}
\phi^{T M_T} \, ,  
\end{eqnarray}
with $\chi^{S M_S}$ and $\phi^{T M_T}$ spinors and isospinors with
good total spin $S=0,1$ and isospin $T=0,1$ respectively. The radial
wave functions satisfy the asymptotic boundary conditions
\begin{eqnarray}
u_{L}^{ST} (r) \to \sin \left( k r - \frac{L\pi}{2} + \delta_L^{ST}
\right) \, , 
\end{eqnarray}
where $k$ is the CM momentum. From the central potential assumption it
is clear that partial waves do not depend on the total angular
momentum and so we would have e.g.
$\delta_{^3P_0}=\delta_{^3P_1}=\delta_{^3P_2}$ and so on, in complete
contradiction to the data. As it is well-known the spin-orbit
interaction lifts the independence on the total angular momentum, via
the operator $\vec L \cdot \vec S$. Moreover, the tensor coupling
operator, $S_{12}$, mixes states with different orbital angular
momentum, so to account for the $J$-dependence we proceed in first
order perturbation theory in the spin-orbit and tensor potentials
using the orbital symmetric distorted waves as the unperturbed
states. Note that this is {\it not} the standard Born approximation
where all components of the potential are treated
perturbatively. According to a previous calculation~( see Appendix D
of Ref.~\cite{CalleCordon:2008cz}) the correction to the phase shift
to first order reads
 \begin{eqnarray}
\Delta \delta_{JL}^{ST} = - \frac{M}{p} 
\int_{0}^\infty dr \,
u_L^{ST} (r)^\dagger \Delta V  u_L^{ST} (r) \, ,
\end{eqnarray}
so that the perturbed eigenphases become
\begin{eqnarray}
\delta_{JL}^{ST} = \delta_{L}^{ST} + \Delta \delta_{JL}^{ST} \, . 
\end{eqnarray}
Note that to this order the mixing phases vanish, $\Delta
\epsilon_J=0$, and there is no difference between the eigen
phase shifts or the nuclear bar phase shifts. The spin-orbit interaction
lifts the independence on the total angular momentum, via the operator
$\vec L \cdot \vec S$. Further, the tensor coupling operator,
$S_{12}$, mixes states with different orbital angular
momentum. Nonetheless, these two perturbations leave the center of the
orbital multiplets unchanged.  Actually, since
\begin{eqnarray}
\sum_{J=L-1}^{L+1} (2J+1)
(\Delta V_{J}^{10})_{L,L}  = 0  \, , 
\label{eq:ps-lande''}
\end{eqnarray}
one has 
\begin{eqnarray}
\sum_{J=L-1}^{L+1} (2J+1) \Delta \delta_{LJ}^{10} = 0 \, . 
\label{eq:ps-lande'}
\end{eqnarray}
As a consequence 
\begin{eqnarray}
\bar \delta_{L}^{ST} &=& \frac{\sum_{J=L-1}^{L+1} (2J+1)
\delta_{LJ}^{ST}}{(2L+1)3} = \delta_{L}^{ST}
\, , \nonumber \\ 
\label{eq:ps-lande}
\end{eqnarray}
Thus, to first order we may define a common {\it mean} phase obtained
as the one obtained from a {\it mean} potential
\begin{eqnarray}
\bar V_{^3L} (r) =  \frac{\sum_{J=L-1}^{L+1} (2J+1)
V_{^3L_J} (r)}{3(2L+1)} \, .
\end{eqnarray} 
It is in terms of these potentials where we expect to formulate the
verification of a given symmetry. This is nothing but the standard
procedure of verifying a symmetry between multiplets by defining first
the center of the multiplet~\footnote{A familiar example is provided
  by the verification of SU(3) in the baryon spectrum. While the
  symmetry is rough the Gell-Mann-Okubo formula works rather well
  after it has been broken by a symmetry breaking term.}.
Now, Serber symmetry requires 
\begin{eqnarray} 
V_{^1L}(r)= V_{^3L}(r)=0 \qquad {\rm odd}-L \, ,  
\end{eqnarray} 
while Wigner symmetry requires 
\begin{eqnarray} 
V_{^{3}L}(r)=V_{^{1}L}(r)  \qquad {\rm all}-L  \, .  
\end{eqnarray} 
Clearly these two requirements are incompatible except when all
potentials vanish. In Fig.~\ref{fig:wigner-serber} we plot the Argonne
V-18 potentials~\cite{Wiringa:1994wb} for the center of the orbital
multiplets.  Thus the potentials suggest instead that for $r > 1.5
{\rm fm}$
\begin{eqnarray} 
V_{^3L} (r) &\ll & V_{^1L} (r) \, \, \qquad {\rm odd}-L \\ V_{^3L} (r)
& \sim & V_{^1L} (r) \qquad {\rm even}-L \, , 
\end{eqnarray} 
i.e. Wigner symmetry is fulfilled for {\it even}-L states while Serber
symmetry holds for odd-L {\it triplet} states at distances above
$1.5{\rm fm}$ in agreement with the expectations spelled out at the
beginning of this section.  The parallel statements for phase shifts
have been developed in detail in Ref.~\cite{CalleCordon:2008cz} (see
Fig.~\ref{fig:higher}) where the relation of long distance symmetry
and renormalization has been stressed.  The remarkable aspect, already
discussed there, is that the symmetry pattern while incompatible with
Wigner symmetry for odd-L states is fully compatible with large $N_c$
expectations~\cite{Kaplan:1996rk}.  It does not explain, however, {\it
  why} Serber symmetry is a good one.

\section{Searching the symmetry}
\label{sec:searching}

Most modern potential models of the NN interaction include OPE as the
dominant longest range contribution. However, they differ at short
distances where many effects compete and even are written in quite
different forms (energy dependent, momentum dependent, angular
momentum dependent, etc.). These ambiguities are of course compatible
with the inverse scattering problem and manifest mainly in the
off-shell behaviour of the NN forces. The relevant issue within the
present context and which we analyze below regards the {\it range} and
{\it form} of current NN interactions from the view point of long
distance symmetries.  Any potential fitting the elastic scattering
data must posses the symmetries displayed by the phase-shifts sum
rules as we see in Fig.~\ref{fig:higher}. However, it is not obvious 
that potentials display the symmetry explicitly.  



\subsection{One Pion Exchange}

The OPE potential reads 
\begin{eqnarray}
V^\pi (r) = \tau \left( \sigma W_S^\pi(r) + S_{12} W_T^\pi (r) \right) \, . 
\end{eqnarray} 
While OPE complies to the Wigner symmetry it does not embody exactly
the Serber symmetry.  Actually we get for even-L waves
\begin{eqnarray}
V_{^1S}^\pi (r) &=& V_{^1D}^\pi (r) = V_{^1G}^\pi (r) = 
-3 W_S^\pi  (r) \, , \\  
V_{^3S}^\pi (r) &=& V_{^3D}^\pi (r) = V_{^3G}^\pi (r) = 
-3 W_S^\pi  (r) \, ,
\end{eqnarray}
while for odd-L waves we have 
\begin{eqnarray}
V_{^1P}^\pi (r) &=& V_{^1F}^\pi (r) = V_{^1H}^\pi (r) = 
9 W_S^\pi  (r) \, , \\  
V_{^3P}^\pi (r) &=& V_{^3F}^\pi (r) = V_{^3H}^\pi (r) = 
\, \,  W_S^\pi  (r) \, .
\end{eqnarray}
The factor $9$ for the singlet to triplet ratio is nonetheless a close
approximation to the Serber limit in a region where the potential is
anyhow small.  These OPE relations are verified in practice for
distances above $3-4 {\rm fm}$. As we see from
Fig.~\ref{fig:wigner-serber} the vanishing of $^3P$ potential happens
down to the region around $1.5 {\rm fm}$. For lower distances,
potential models start deviating from each other (see
e.g. \cite{Stoks:1994wp}) but this vanishing of $^3P$ potential is a
common feature which occurs {\it beyond} the validity of OPE.

\subsection{Boundary conditions (alias $V_{\rm high \, R}$)}
\label{sec:boundary}

We now analyze the symmetry issue for the highly successful
PWA~\cite{Stoks:1993tb} of the Nijmegen group.  There, a OPE potential
is used down to $r_c=1.4 {\rm fm}$ and the interaction below that
distance is represented by a boundary condition determined by a square
well potential with an energy dependent height,
\begin{eqnarray} 
2 \mu  V_{S,\beta}(k^2) = \sum_{n=0}^N a_ {n,\beta} k^{2n} \, ,
\end{eqnarray} 
where $\beta$ stands for the corresponding channel, so that the total
potential reads
\begin{eqnarray}
V_\beta (r) &=& \left[ V_\beta^\pi (r) + V_\beta^{\rm int} (r) \right]
\theta (r -r_c) + V_{S,\beta} (k^2) \theta (r_c -r) \, , \nonumber \\ 
\end{eqnarray} 
where $V_\beta^{\rm int} (r) $ is a phenomenological intermediate
range potential which acts in the region $1.4 {\rm fm} \le r \le 2.0
{\rm fm}$. Then,  for the center of the
L-multiplets ( V in MeV and k in fm ) we have
\begin{eqnarray}
V_{S,^1P} ( k^2 ) &=& 139.448 - 23.417 k^2 + 2.479 k^4 \, , 
\\ 
V_{S,^3P} ( k^2 ) &=& \, 14.666 + \, 0.92 k^2 + 0.029 k^4 \, , 
\\ 
V_{S,^1F} ( k^2 ) &=& 248.73 \, , 
\\ 
V_{S,^3F} ( k^2 ) &=& -33.08 + 5.90 k^2 \, , 
\end{eqnarray} 
where, again, we see that Serber symmetry takes place since $V_{S,^3P}
( k^2 ) \ll V_{S,^1P} ( k^2 )$ and $V_{S,^3F} ( k^2 ) \ll V_{S,^1F} (
k^2 )$. Actually, the factor is strikingly similar to the $1/9$ of the
OPE interaction which in the analysis of holds up to $r_c=1.4 {\rm
  fm}$.  Thus, in the Nijmegen PWA decomposition of the interaction we
find the remarkable relation
\begin{eqnarray}
V_{^3L} (r) \ll  V_{^1L} (r)  \qquad {\rm odd}-L , \qquad {\rm all} \, \, r 
\end{eqnarray} 
showing that there is Serber symmetry in the short range piece of the
potential. On the other hand, the even partial waves yield
\begin{eqnarray}
V_{S,^1S} ( k^2 ) &=& -17.813 - 1.016 k^2 + 2.564 k^4 \, , 
\\ 
V_{S,^3S} ( k^2 ) &=& -40.955 + 4.714 k^2 + 1.779 k^4 \, , 
\\ 
V_{S,^1D} ( k^2 ) &=& 61.42-15.678 k^2 \, , 
\\ 
V_{S,^3D} ( k^2 ) &=& 28.869 - 3.579 k^2 \,  , 
\\ 
V_{S,^1G} ( k^2 ) &=& 466.566 \, , 
\\ 
V_{S,^3G} ( k^2 ) &=& 0 \, , 
\end{eqnarray} 
where we clearly see the violations of Wigner symmetry at short distances,
i.e.  we only have 
\begin{eqnarray}
V_{^3L} (r) \sim  V_{^1L} (r)  \qquad {\rm even}-L \qquad  \, r \ge r_c \, . 
\end{eqnarray} 
This simple analysis suggests that Serber symmetry, when it works,
holds to shorter distances than the Wigner symmetry. Our previous
analysis in terms of mean phases~\cite{CalleCordon:2008cz} fully supports this
fact. Indeed, higher partial waves with angular momentum $l$ are
necessarily small at small momenta due to the well known $\delta_l(p)
\sim -\alpha_l p^{2l+1}$ threshold behaviour. In fact, this is the
case for $\delta_{^1P}$ and $\delta_{^1F}$. However, Serber symmetry
implies that $\delta_{^3P}$ and $\delta_{^3F}$ are rather small {\it
  not only} in the threshold region but also in the entire elastic
region as can be clearly seen from Fig.~\ref{fig:higher}.

\subsection{Potentials and $V_{\rm low \, k}$}
\label{sec:Vlow}

\begin{figure}
\includegraphics[height=4cm,width=4cm,angle=270]{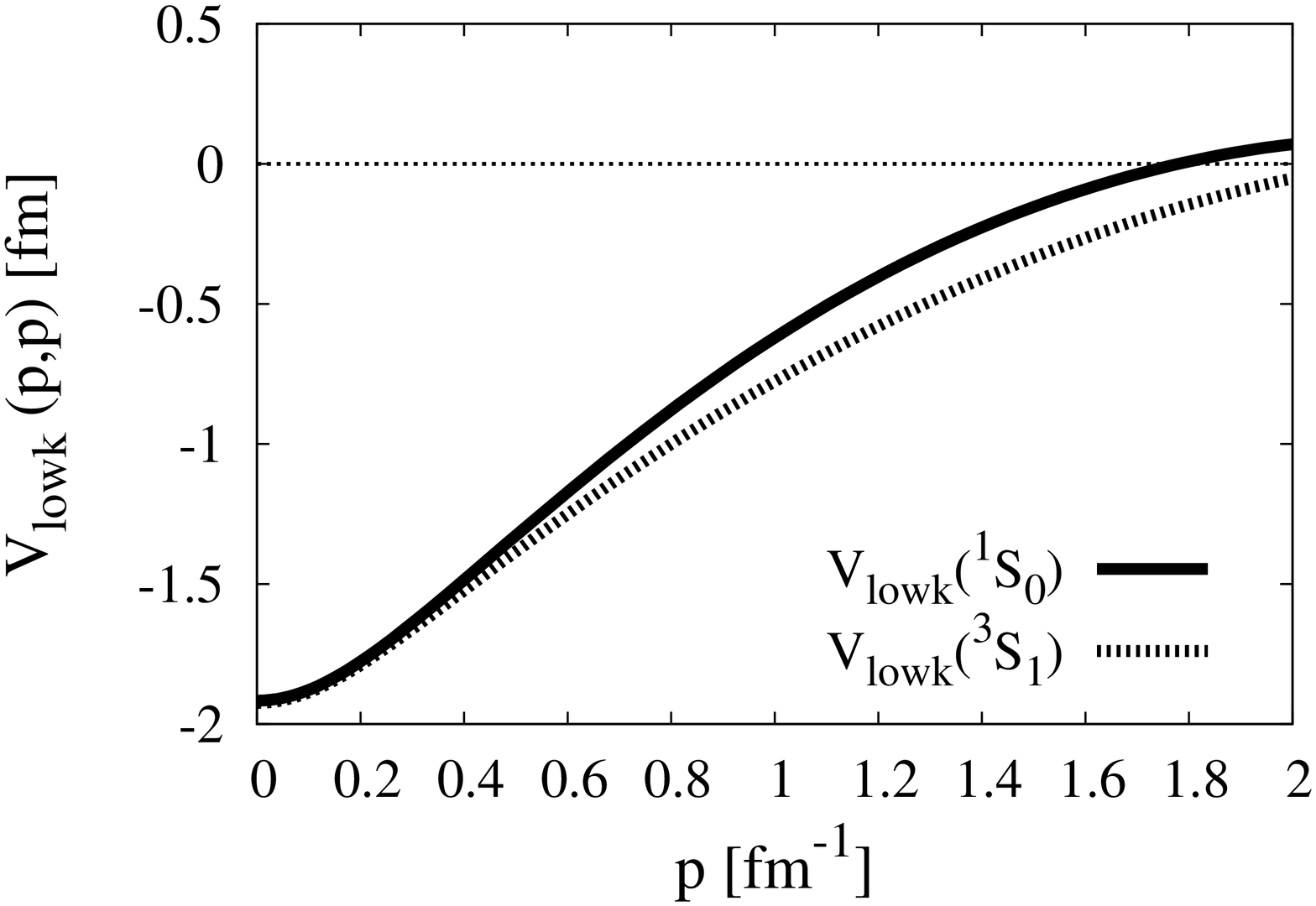}
\includegraphics[height=4cm,width=4cm,angle=270]{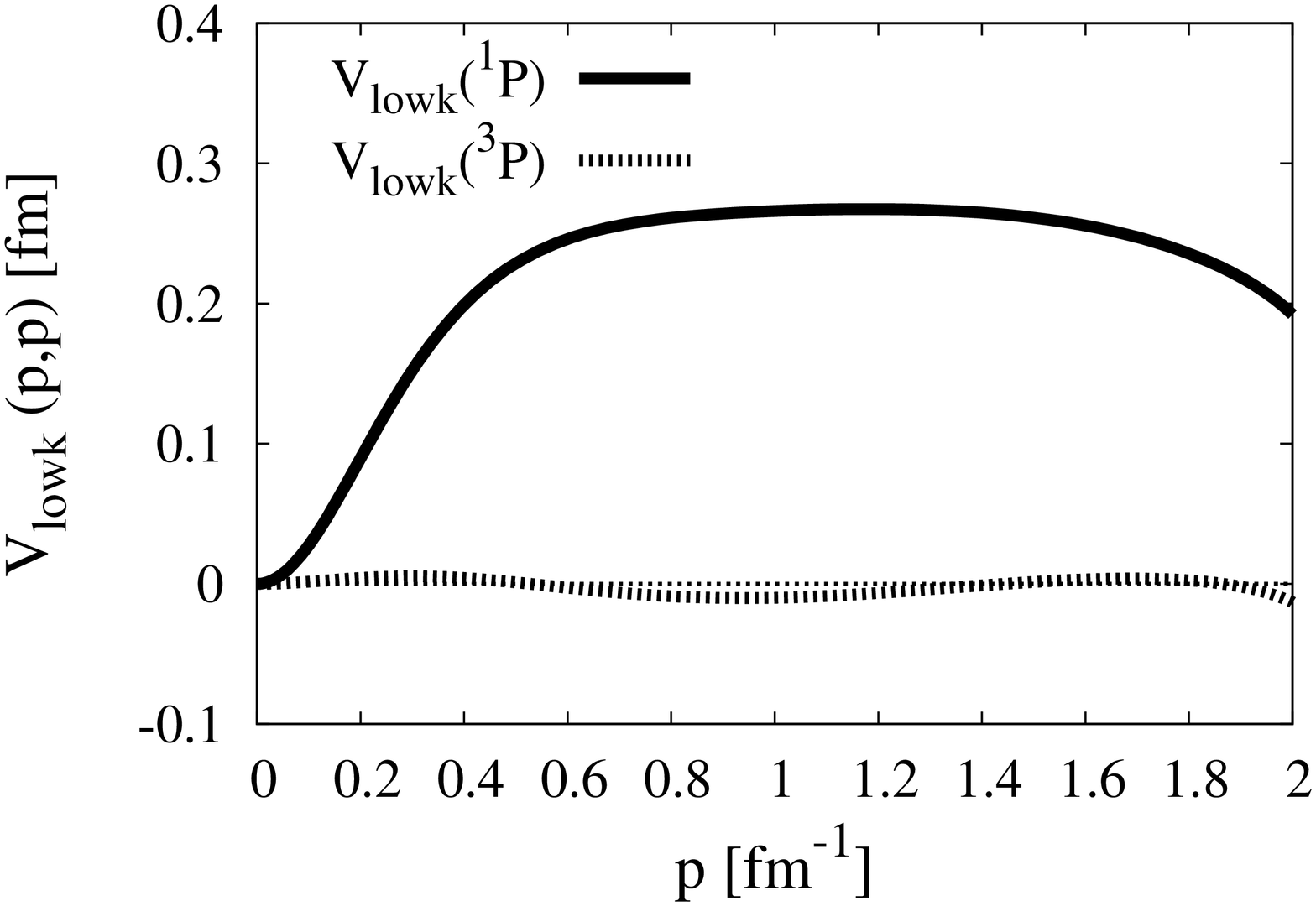}\\ \includegraphics[height=4cm,width=4cm,angle=270]{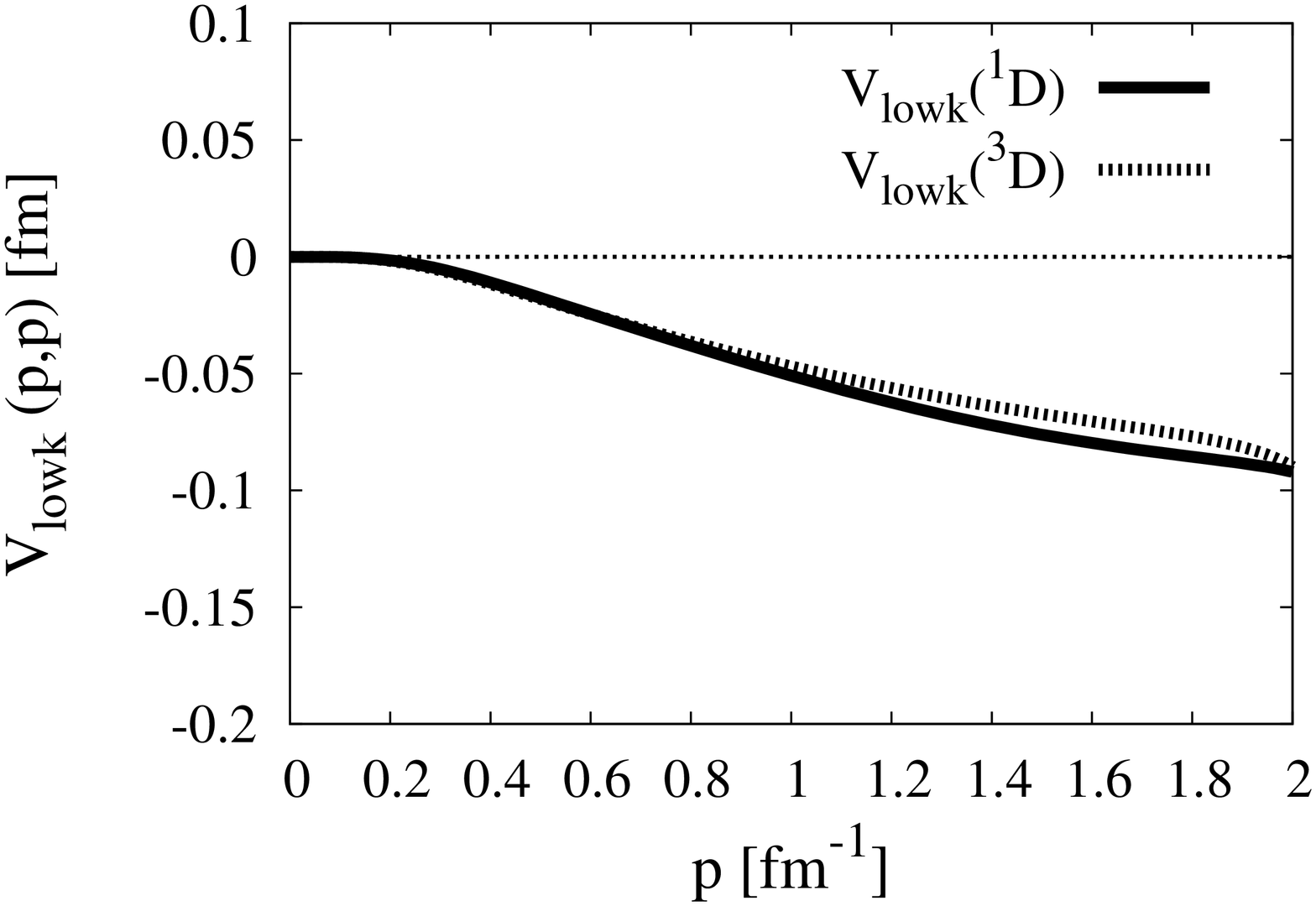}
\includegraphics[height=4cm,width=4cm,angle=270]{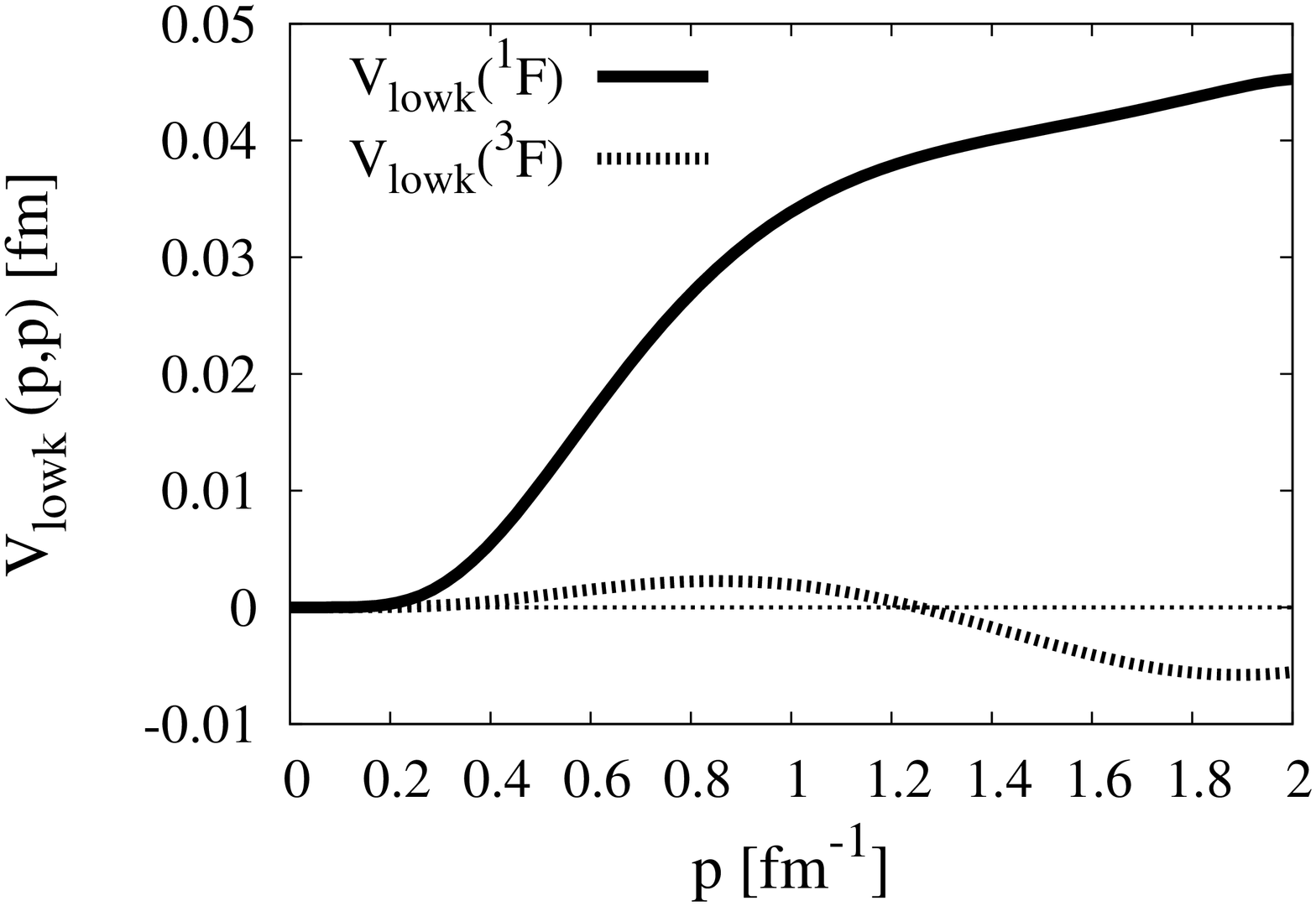}
\caption{Diagonal $V_{\rm low k} (p,p)$ potentials (in fm) as a
  function of the momentum p (in ${\rm fm}^{-1}$) for the
  Argonne-V18~\cite{Wiringa:1994wb}, for the center of the
  Serber-Wigner multiplets. Wigner symmetry requires singlet and
  triplet potentials to coincide. Serber symmetry implies vanishing
  odd-L partial waves.  Even-L waves posses Wigner symmetry while
  odd-L triplet waves exhibit Serber symmetry.}
\label{fig:vlowk}
\end{figure}

A somewhat different perspective arises from a Wilsonian analysis of
the NN interaction which corresponds to a coarse graining of the
potential. This viewpoint was implemented in Ref.~\cite{Bogner:2003wn}
where the so-called $V_{\rm low k}$ approach has been pursued, and
corresponds to integrating out high momentum modes below a given
cut-off $ k \le \Lambda$ from the Lippmann-Schwinger equation. It was
found that high quality potential models, i.e. fitting the NN data to
high accuracy and also incorporating OPE, collapse into a unique
self-adjoint nonlocal potential for $\Lambda \sim 400 {\rm MeV}$. This
is a not a unreasonable result since all the potentials provide a
rather satisfactory description of elastic NN scattering data up to $p
\sim 400 {\rm MeV}$. Moreover, the potential which comes out from
eliminating high energy modes can be accurately represented as the sum
of the truncated original potential and a polynomial in the
momentum~\cite{Holt:2003rj},
\begin{eqnarray}
V_{\rm low k} ( k' , k) = V_{\rm NN} ( k' , k) + V_{\rm CT}^\Lambda ( k' , k)
\, , \qquad (k,k') \le \Lambda \, , \nonumber \\ 
\label{eq:vlowk} 
\end{eqnarray} 
where $V_{\rm NN} ( k' , k)$ is the original potential in momentum
space for a given partial wave channel and $V_{\rm CT}^\Lambda ( k' ,
k)$ is the effect of the high energy states, 
\begin{eqnarray}
V_{\rm CT}^\Lambda ( k' , k) = k^l k'^{l'} \left[ C_0^{l l'}
  (\Lambda)+ C_2^{l l'}(\Lambda) ( k^2+ k'^2 ) + \dots \right] \, ,
\end{eqnarray} 
where the coefficients $C_n^{l l'}(\Lambda) $ play the role of
counterterms.  It should be noted that here $ V_{\rm NN} ( k' , k) $
is cut-off {\it independent} whereas $V_{\rm CT}^\Lambda ( k' , k)$
does depend on $\Lambda$.  When the potential given by
Eq.~(\ref{eq:vlowk}) is plugged into the truncated Lippmann-Schwinger
equation, i.e. intermediate states $q \le \Lambda$, the phase shifts
corresponding to the full original potential $V_{\rm NN} ( k' , k)$
are reproduced. In Fig.~\ref{fig:vlowk} the corresponding diagonal
$V_{\rm low k} (p,p)$ mean potentials are plotted for the Argonne-V18
force~\cite{Wiringa:1994wb}~\footnote{We thank Scott K. Bogner for
  kindly providing the numbers of Ref.~\cite{Holt:2003rj}.}. As we see
both Wigner and Serber symmetries are, again, vividly seen. The
important observation here is that the separation assumed by
Eq.~(\ref{eq:vlowk}) does not manifestly display the
symmetry. Actually, a more convenient representation would be to
separate off all polynomial dependence explicitly from the original
potential
\begin{eqnarray}
V_{\rm low k} ( k' , k) = \bar V_{\rm NN} ( k' , k) + \bar V_{\rm CT}^\Lambda ( k' , k)\, , 
\qquad (k,k') \le \Lambda \, , \nonumber \\ 
\label{eq:vlowk2} 
\end{eqnarray} 
so that if $\bar V_{\rm CT}^\Lambda ( k' , k)$ contains up to ${\cal
  O} (p^n)$ then $\bar V_{\rm NN} ( k' , k)$ starts off at ${\cal O}
(p^{n+1})$, i.e. the next higher order. This way the departures from a
pure polynomial may be viewed as true and explicit effects due to the
potential. In terms of these polynomials, Wigner and Serber symmetries
are formulated  from the coefficients
\begin{eqnarray}
\bar C_0 = C_0 + C^{\rm high}_0 (\Lambda)
\end{eqnarray}
constructed from the sum of the potential and the integrated out
contribution below a cut-off $\Lambda$, namely 
\begin{eqnarray}
\bar C_{0,^1L} &=& \bar C_{0,^3L} \, , \qquad {\rm even}-L \, , \nonumber \\ 
\bar C_{0,^3L} &=& 0 \qquad \, ,\qquad {\rm odd}-L    \, . 
\label{eq:c's-s-w}
\end{eqnarray}   
It should be noted that the $V_{\rm low k}$ approach is in
spirit nothing but the momentum space version of the PWA of the
Nijmegen group in coordinate space where short distances, $r \le r_c$,
are integrated out and parameterized by means of an energy dependent
boundary condition. From this viewpoint the similarities as regards
the Wigner and Serber symmetries are not surprising. This is why the
standard boundary condition approach might be also denominated $V_{\rm
  high \, R}$ (see also Ref.~\cite{Entem:2007jg} for further
discussions).

\subsection{Chiral two pion exchange}
\label{sec:chiral}

\begin{figure}
\includegraphics[height=4cm,width=4cm,angle=270]{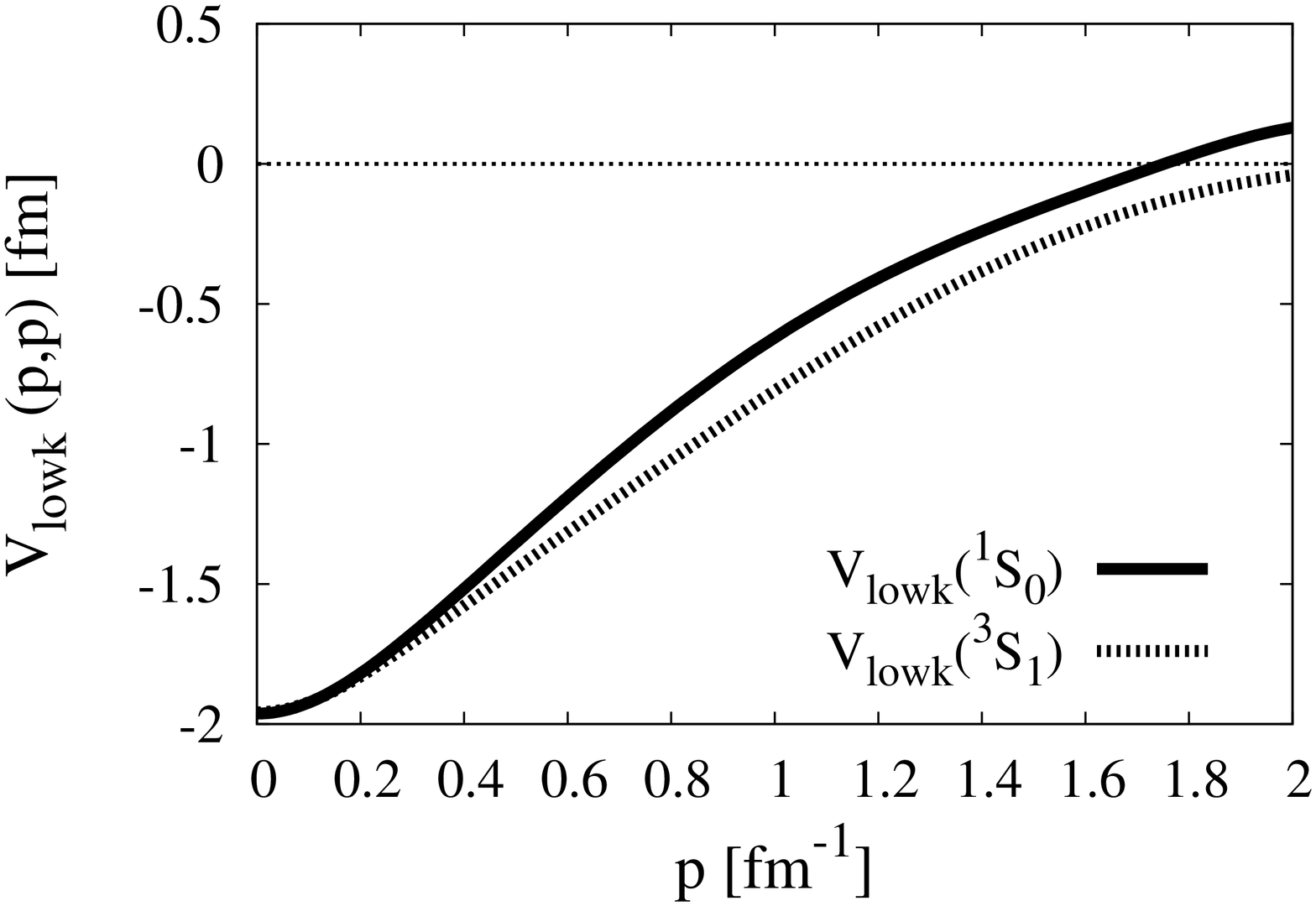}
\includegraphics[height=4cm,width=4cm,angle=270]{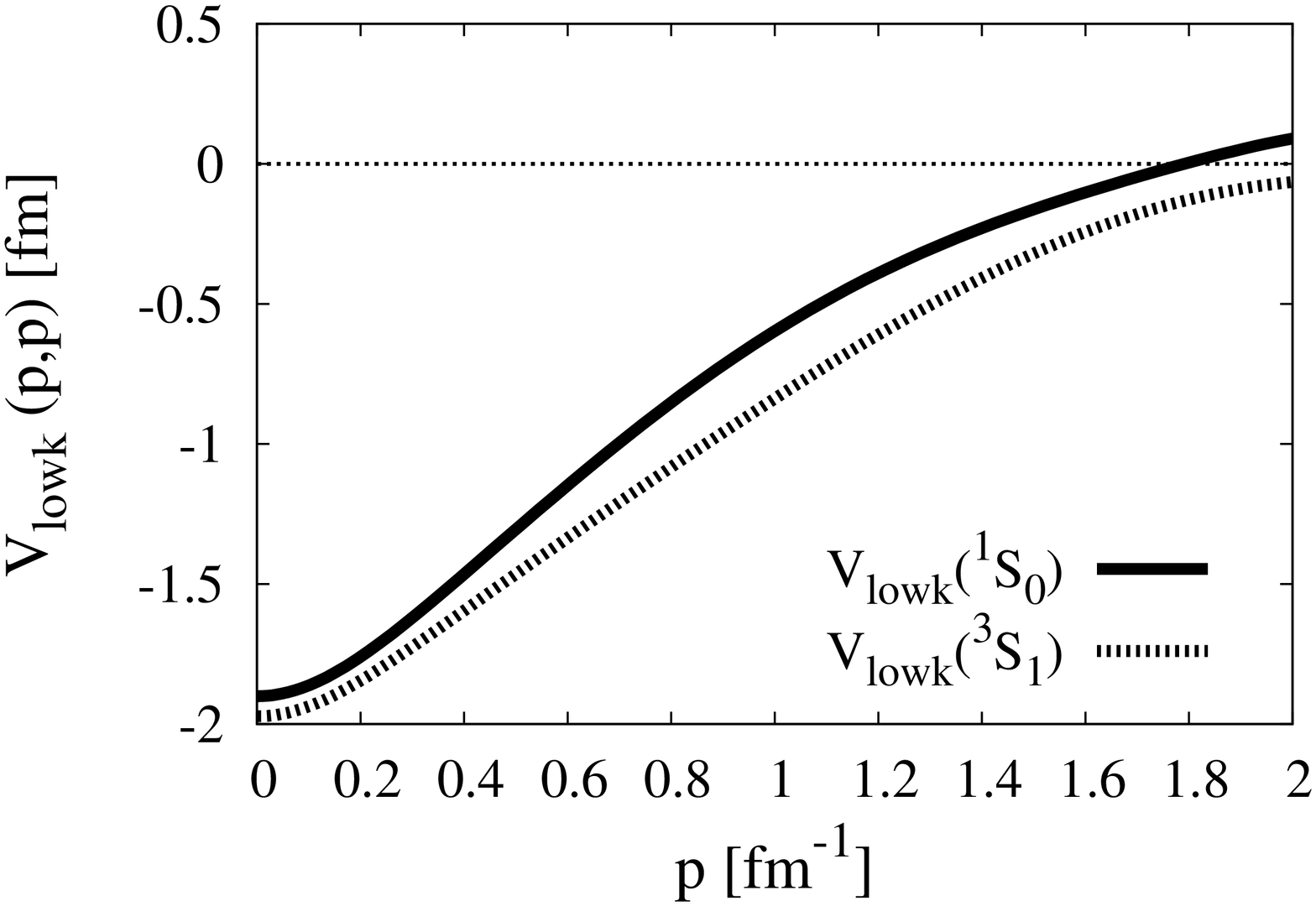}
\caption{Diagonal $V_{\rm low k} (p,p)$ potentials (in fm) as a
  function of the momentum p (in ${\rm fm}^{-1}$) for the N$^3$LO-chiral
  potentials~\cite{Entem:2003ft} for the $S$-wave states when the
  chiral cut-offs $\Lambda_\chi=500 {\rm MeV}$ and $\Lambda_\chi=600
  {\rm MeV}$ are used.  As we see there is a $5\%$ violation of Wigner
  symmetry in the second case.}
\label{fig:vlowk-chiral}
\end{figure}

The chiral Two Pion Exchange (TPE) potentials computed in
Ref.~\cite{Kaiser:1998wa} are understood as direct consequences of the
spontaneous chiral symmetry breaking in QCD. Actually, the TPE
contribution takes over the OPE one at about $r=2{\rm fm}$. At very
long distances one has
\begin{eqnarray}
V_{2 \pi}^{\rm ChPT} (r) = ( 1 + 2 \vec \tau_1 \cdot \vec \tau_2 )
\frac{e^{-2m_\pi r}}{r} \frac{3 g_A^4 m_\pi^5}{1024 f_\pi^4 M_N \pi^2}
+ \dots \, ,
\end{eqnarray} 
where $m_\pi$ and $M_N$ are the pion and nucleon masses respectively,
$g_A$ the axial coupling constant and $f_\pi$ the pion weak decay
constant. As we see Serber symmetry is broken already at long
distances. Generally, these chiral potentials are supplemented by
counterterms or equivalently boundary conditions when discussing NN
scattering and generating phase shifts (see
e.g. Ref.~\cite{Rentmeester:1999vw}). Given that these NN phase shifts
do fulfill the symmetry (see Fig.~\ref{fig:higher}) we expect that the
breaking of the symmetry at long distances must be compensated by the
counterterms which encode the unknown short distance
physics~\cite{Rentmeester:1999vw}. This can be verified by looking
e.g. at the $V_{\rm low \, k}$ potential corresponding to the
Next-to-next-to-next-to-leading order (N$^3$LO) chiral potential which
contains its own cut-off parameter of $\Lambda_\chi=500 {\rm
  MeV}$~\cite{Entem:2003ft}. This potential contains OPE and describes
successfully the data and hence falls into the universality class of
high-quality potentials~\cite{Bogner:2006vp} when the common $V_{\rm
  low k}$ cut-off scale $\Lambda=400 {\rm MeV}$ is used. If the chiral
potential is slightly detuned by taking $\Lambda_\chi=600 {\rm MeV}$
one sees a low momentum violation of the Wigner symmetry in
Fig.~\ref{fig:vlowk-chiral} in total contradiction with the fact that
one expects that asymptotically OPE should dominate. This shows that
regarding the symmetry $\Lambda_\chi$ is fine-tuned. A more complete
account of these issues will be presented
elsewhere~\cite{PavonWSchiral}.

\section{Are counterterms fingerprints of long distance symmetries ?}
\label{sec:ct}

Given the fact that both Wigner and Serber symmetries can be
interpreted as long distance symmetries which roughly materialize at
low energies in the potentials (see Fig.~\ref{fig:wigner-serber}), the
phase shifts (see Fig.~\ref{fig:higher}) and the $V_{\rm low k}$
potentials (see Fig.~\ref{fig:vlowk}) we find it appropriate to
discuss how these results fit into renormalization ideas and the role
played by the corresponding counterterms. 

\subsection{The perturbative point of view}

As we have mentioned in the previous section, chiral potentials are
generally used to describe NN scattering with the additional
implementation of counterterms which cannot directly be determined
from chiral symmetry alone. On the other hand, one expects these
counterterms to encode short distance physics and hence to be related
to the exchange of heavier mesonic degrees of freedom alike those
employed in the One Boson Exchange (OBE)
potentials~\cite{Machleidt:1987hj}. The idea is quite naturally based
on the resonance saturation hypothesis of the exchange forces (see
e.g. \cite{Ecker:1988te} for a discussion in the $\pi\pi$ scattering
case). This is achieved by integrating out the heavy fields using
their classical equations of motion, and expanding the exchanged
momentum between the nucleons as compared to the resonance mass
case~\cite{Epelbaum:2001fm,Epelbaum:2003xx}. Schematically it
corresponds to power expand the Yukawa-like NN potentials (we ignore
spin and isospin for simplicity),
\begin{eqnarray}
\frac{g_M^2}{{\bf q}^2+M^2} &=& \frac{g_M^2}{M^2} - \frac{ g_M^2 {\bf
    q^2}}{M^4} + \dots \nonumber \\ &=& C_0 + C_2 \left({\bf p}^2+
     {\bf p}'^2 \right) + C_1 {\bf p} \cdot {\bf p}' + \dots \, , 
\label{eq:low-q}
\end{eqnarray} 
where we are working in the CM system and we take the momentum
transfer as ${\bf q}={\bf p}-{\bf p'}$. The meaning of the terms above
is as follows: $C_0$ is an s-wave zero range, $C_2$ is an s-wave
finite range, $C_1$ is a p-wave etc. . More generally,
Eq.~(\ref{eq:low-q}) corresponds to a power series expansion of the
potential in momentum space. Obviously, we expect such a procedure to
be meaningful whenever the scattering process can be treated
perturbatively, like e.g.  the case of peripheral waves. As is well
known, central s-waves cannot be treated perturbatively as the
corresponding scattering amplitudes have poles very close to threshold
corresponding a virtual state in the $^1S_0$ channel and the deuteron
in the $^3S_1-^3D_1$ channel. This does not mean, however, that the
potential cannot be represented in the polynomial form of
Eq.~(\ref{eq:low-q}), but rather that the coefficients cannot be
computed {\it directly} as the Fourier components of the potential.

\subsection{The Wilsonian point of view}

The momentum space $V_{\rm low k}$ approach~\cite{Holt:2003rj} makes
clear that the long distance behaviour of the theory is not determined
by the low momentum components of the original potential {\it only},
one has to add virtual high energy states which also contribute to the
interaction at low energies in the form of counterterms, as outlined
by Eqs.~(\ref{eq:vlowk}) and (\ref{eq:vlowk2}). Alternatively the more
conventional coordinate space boundary condition (alias $V_{\rm high
  R}$) method shows that the low energy behaviour of the theory is not
determined {\it only} by the long distance behaviour of the potential,
one has to include the contribution from integrated out short
distances in the form of boundary conditions. A true statement is that
the low momentum features of the interaction in the $V_{\rm low
  k}(p,p)$ potential can be mapped into long distance characteristics
of the potential $V(r)$. The symmetries are formulated in terms of the
conditions in Eq.~(\ref{eq:c's-s-w}).

\subsection{Long distance symmetries in Nuclear Potentials}

In order to substantiate our points further, let us note that in
Ref.~\cite{Holt:2003rj} it was suggested that the $V_{\rm low k}$ was
a viable way of determining the effective interactions which could be
further used in shell model calculations for finite nuclei. Actually,
this interpretation when combined with our observation of
Fig.~\ref{fig:vlowk} that Serber symmetry shows up quite universally
has interesting consequences. Schematically, this can be implemented
as a Skyrme type effective (pseudo)potential~\cite{Skyrme:1959zz}
\begin{eqnarray} 
V (\vec r) &=& t_0 (1 + x_0 P_\sigma ) \delta^{(3)} (\vec r) + t_1 (1
+ x_1 P_\sigma ) \left\{ -\nabla^2 , \delta^{(3)} (\vec r) \right\}
\nonumber \\ &-& t_2 (1 + x_2 P_\sigma ) \nabla \delta^{(3)} (\vec r)
\nabla + \dots 
\label{eq:skyrme}
\end{eqnarray} 
where $P_\sigma = (1+ \sigma_1 \cdot \sigma_2)/2$ is the spin exchange
operator with $P_\sigma=1$ for spin single $S=0$ and $P_\sigma=1$ for
spin triplet $S=1$ states. The dots stand for spin-orbit, tensor
interaction, etc. It should be noted the close resemblance of the
momentum space version of this potential
\begin{eqnarray} 
V ({\bf p}',{\bf p}) &=&   
t_0 (1 + x_0 P_\sigma ) + t_1 (1 + x_1 P_\sigma )
  ({\bf p}'^2 + {\bf p}^2) \nonumber \\ &+& t_2 (1 + x_2 P_\sigma )
        {\bf p}' \cdot {\bf p} + \dots
\label{eq:skyrme2}
\end{eqnarray} 
with Eq.~(\ref{eq:vlowk2}) after projection onto partial waves, where
only S- and P-waves have been retained.  Traditionally, binding
energies have been used to determine the parameters $t_i$ and $x_i$
within a mean field approach and many possible fits arise depending on
the chosen observables (see e.g. Ref.~\cite{Chabanat:1997qh}) possibly
displaying some spurious short distance sensitivity beyond the range
of applicability of Eq.~(\ref{eq:skyrme}). The low momentum character
of the Skyrme force, suggests using the longest possible wavelength
properties.  Actually, inclusion of tensor force and a new fitting
strategy to single particle energies~\cite{Zalewski:2008is} yields
$x_2 = -0.99 $ which is an almost perfect Serber force for
spin-triplets ($P_\sigma=1$) and reproduces the so-called SLy4 form of
the Skyrme functional~\cite{Chabanat:1997qh}.  On the light of our
discussion this result seems quite natural as single particle energies
place attention in long wavelength states, a situation where $V_{\rm
  low k}$ can be described by a pure polynomial in momentum (see
Eq.~(\ref{eq:vlowk2})) and Serber symmetry becomes manifest {\it
  directly} from a coarse graining of the NN-interaction.

\subsection{Matching OBE potentials to chiral potentials}

In Ref.~\cite{Epelbaum:2001fm} a systematic determination of
counterterms has been carried out for a variety of realistic
potentials which successfully fit the NN data by reading off the
Fourier components of the potential (see e.g. Eq.~(\ref{eq:low-q})).
The counterterms so obtained are then compared to those determined
from direct fits to the NN data when the chiral potential is added.
The spread of values for these counterterms found in
Ref.~\cite{Epelbaum:2001fm} for realistic potentials, however, does
not comply to the fact that all those potentials provide a quite
satisfactory description of the phase shifts. Moreover, in
Ref.~\cite{Epelbaum:2001fm} it is found that for the OBE Bonn
potential~\cite{Machleidt:1987hj}
\begin{eqnarray}
C_{^1P}&=& +0.454 \times 10^{-4} {\rm GeV}^{-4} \, ,  \\ 
C_{^3P} &\equiv& \frac{1}{9} \left( C_{^3P_0} + 3 C_{^3P_1}+ 5C_{^3P_2} \right) \nonumber \\  
&=& -0.140 \times 10^{-4} {\rm GeV}^{-4}  \, .
\end{eqnarray} 
Thus, the triplet to singlet ratio is $C_{^3P} / C_{^1P} \sim -0.33 $
in this case. For the CD Bonn potential~\cite{Machleidt:2000ge} one
has $C_{^3P} / C_{^1P} \sim -0.7$ whereas Argonne
AV-18~\cite{Wiringa:1994wb} yields $C_{^3P} / C_{^1P} \sim -0.54$.
These large factors contrast with the much smaller factor $V_{S,^3P} (
k^2 ) / V_{S,^1P} ( k^2 ) \sim 1/10 $ of the PWA sketched above in
Sect.~\ref{sec:boundary}. They also disagree with the almost
vanishing ratio $V_{^3P} (p,p ) / V_{^1P} ( p,p ) $ found for the
$V_{\rm low k}$ potentials described in Section~\ref{sec:Vlow} which
yield a universal result (see also Fig.~\ref{fig:vlowk} for the
particular AV-18 potential). The reason is that the correct
formulation of the symmetry conditions is given by
Eq.~(\ref{eq:c's-s-w}) which are made up from the potential plus the
contribution of the high energy tail. Thus, it appears that in the
approach of Ref.~\cite{Epelbaum:2001fm} Serber symmetry is more
strongly violated at short distances than expected from other
means. In our view the spread of values found in
Ref.~\cite{Epelbaum:2001fm} might possibly reflect an inadequacy of
the method used to characterize the long distance coarse grained NN
dynamics where, as we have shown, Serber symmetry becomes quite
accurate. Actually, the matching of counterterms between, say, the OBE
potential and the chiral potentials is done in terms of objects which
have a radically different large $N_c$ behaviour (see Section
\ref{sec:serber-largeN} for further details). For instance, while
$C^{\rm OBE}_0 \sim g_M^2 /m_M^2 \sim N_c$ because $g_M \sim
N_c^{\frac12} $ and $m_M \sim N_c^0$ one has $C_0^{\rm chiral} \sim
g_A^4 m_\pi^2/f_\pi^4 \sim N_c^2$ since $m_\pi \sim N_c^0$, $f_\pi
\sim N_c^{\frac12} $ and $g_A \sim N_c$. In fact, the value of the
counterterms determined from resonance exchange is generally {\it not}
simply determined by the coefficients of the power series expansion of
potential in momentum space, as schematically given  by Eq.~(\ref{eq:low-q}),
since they undergo renormalization and hence run with the scale.

\subsection{Long distance symmetry and off-shellness}

The previous analysis shows that nothing forbids to have a potential
which breaks the symmetry strongly on the one hand and being able to
{\it simultaneously} fit the scattering data which manifestly display
the symmetry on the other hand. Actually, this can only be achieved by
some degree of compensating symmetry violation between long and short
distances~\footnote{This is the case for instance of chiral TPE
  potentials, see Section~\ref{sec:chiral}, where the
  potential~\cite{Kaiser:1998wa} breaks the symmetry above $1.6 {\rm
    fm }$ but the data can be described~\cite{Rentmeester:1999vw} with
  this truncated potential plus suitable energy dependent boundary
  conditions.}.  However, it is somewhat unnatural as it does not
reflect the true character of the theory and relegates the role of the
symmetry to be an accidental one. As it is widely accepted, unveiling
symmetries is not mandatory but makes life much easier~\footnote{The
  above discussion is somewhat similar to the use of regularization
  schemes in EFT; while it is possible to break the symmetry by all
  allowed counterterms, final physical results will depend on
  redundant combinations of parameters expressing the symmetry. In
  practice it is far more convenient to use a regularization scheme
  where the symmetry is preserved manifestly.}. 

Of course, these observations are true as long as we restrict to
on-shell properties, such as NN scattering. However, would these
symmetries have any consequence for off-shell nucleons ?.  One may
clearly have arbitrary short distance parameterizations of the force
without a sizeable change of the phase shifts. However, the
universality of long distance potentials above $\sim 1.5 {\rm fm}$ or,
equivalently, a coarse graining of the interaction with the given
length scale $\sim \pi/\Lambda $ such as $V_{\rm low k}$ is {\it by
  definition} based on insensitivity to shorter wavelengths. Our
discussion above on effective forces illustrates the fact that these
redefinitions of the potential in the NN scattering problem {\it
  cannot} affect the effective force and so a violation of the Serber
symmetry has a physical significance for wavelengths larger than the
coarse graining scale.

\section{Serber force from a large $N_c$ perspective} 
\label{sec:serber-largeN}

Up till now, in this paper we have provided evidence that long
distance symmetries such as Wigner's and Serber's do really take place
in the two nucleon system. From now on we are concerned with trying to
determine whether those symmetries are purely accidental ones or obey
some pattern following more closely from QCD. Actually, we
found~\cite{CalleCordon:2008cz} that large $N_c$
limit~\cite{'tHooft:1973jz,Witten:1979kh} (for comprehensive reviews
see e.g. \cite{Manohar:1998xv,Jenkins:1998wy,Lebed:1998st}) provides a
rationale for Wigner symmetry. The fact that Serber symmetry holds
when Wigner symmetry fails suggests analyzing the large $N_c$
consequences more thoroughly. While we do not find a completely unique
answer regarding the origin of Serber symmetry, the analysis does show
interesting features, as will be discussed.

\subsection{The large $N_c$ expansion}

In this section we want to analyze these long distance Serber and
Wigner symmetries within the two nucleon system from the large $N_c$
expansion~\cite{'tHooft:1973jz,Witten:1979kh} (for comprehensive
reviews see e.g. \cite{Manohar:1998xv,Jenkins:1998wy,Lebed:1998st}).
One feature of large $N_c$ which becomes relevant for the NN problem
is that is does not specially hold for long or short distances. This
allows in particular to switch from perturbative quarks and gluons at
short distances to the non-perturbative hadrons, the degrees of
freedom of interest to nuclear physics. This quark-hadron duality
makes possible the applicability of large $N_c$ counting rules
directly to baryon-meson interactions, at distances where explicit
quark-gluon effects are not expected to be crucial. The procedure
provides utterly a set of consistency conditions from which the
contracted SU(4) symmetry is
deduced~\cite{Manohar:1998xv,Jenkins:1998wy,Lebed:1998st}.  Thus,
while the large $N_c$ scaling behaviour and spin-flavour structure of
the NN potential, Eq.~(\ref{eq:pot-largeN}), is directly established
in terms of quarks and gluons~\cite{Kaplan:1996rk}, quark-hadron
duality at distances larger than the confinement scale requires an
identification of the corresponding exchanged mesons, and hence a link
to the OBE potentials is provided. However, for internal consistency
of the hadronic version of the large $N_c$ expansion, these counting
rules should hold regardless of the number of exchanged mesons between
the nucleons. Actually, naive power counting suggests huge violations
of the NN counting rules.  The issue has been clarified after the work
of Banerjee, Cohen and Gelman for all meson exchange cases with spin 0
and spin 1~\cite{Banerjee:2001js} where the necessary cancellations
between meson retardation in direct box diagrams and crossed box
diagrams was indeed shown to take place. In the TPE case the
$\Delta$-isobar embodying the contracted SU(4) symmetry was explicitly
needed. Although the exchange of three and higher mesons appeared
initially to present puzzling inconsistencies~\cite{Belitsky:2002ni} a
possible outcome was outlined in Ref.~\cite{Cohen:2002im} by noting
that large $N_c$ counting rules apply to energy independent and hence
self-adjoint potentials.

\subsection{Large $N_c$ potentials}

Based on the contracted SU(4) large $N_c$ symmetry the spin-flavour
structure of the NN interaction was analyzed by Kaplan, Savage and
Manohar~\cite{Kaplan:1995yg,Kaplan:1996rk} who found that the leading
$N_c$ nucleon-nucleon potential indeed scales as $N_c$ and has the
structure
\begin{eqnarray}
V (r) = V_C (r) + \sigma_1 \cdot \sigma_2 \tau_1 \cdot \tau_2 W_S (r)
+ S_{12} \tau_1 \cdot \tau_2 W_T(r) \, .  \nonumber \\ 
\label{eq:pot-largeN}
\end{eqnarray} 
It is noteworthy that
the tensor force appears at the leading order in the large $N_c$
expansion. From the large $N_c$ potential, Eq.~(\ref{eq:pot-largeN}), 
we have for the center of multiplet potentials the sum rules 
\begin{eqnarray}
V_{^1L} = V_{^3L} &=&   
V_C (r) - 3 W_S (r)  \, , \quad (-1)^L=+1 \nonumber  \\  \\ 
V_{^1L} &=&   V_C (r) + 9 W_S (r)  \, , \quad  (-1)^L=-1 
 \nonumber  \\ \\ 
V_{^3L} &=&   V_C (r) +  \, \, \,W_S (r) \, ,  \quad (-1)^L=-1  \nonumber  \\  
\end{eqnarray} 
where as we see $ V_{^1L} \neq V_{^3L}$ for odd-L. Thus, large $N_c$
{\it implies} Wigner symmetry in even-L channels and {\it allows} a
violation of Wigner symmetry in odd-L partial waves while it {\it
  allows} a violation of Serber symmetry in spin singlet channels. The
question is whether or not large $N_c$ {\it implies} Serber symmetry
in spin triplet channels as we observe both for the potentials in
Fig.~\ref{fig:wigner-serber} as well as for the phase shifts in
Fig.~\ref{fig:wigner-serber}.  On the other hand, from the odd-waves
we see from Fig.~\ref{fig:higher} that the mean triplet phase is close
to null, thus one might attribute this feature to an accidental
symmetry where the odd-waves potentials are likewise negligible. In
the large $N_c$ limit this means $V_C + 9 W_S \gg V_C + W_S$, a fact
which should be verified.

\subsection{OBE large $N_c$ potentials}

According to Refs.~\cite{Kaplan:1995yg,Kaplan:1996rk} in the leading
large $N_c$ one has $V_C \sim W_S \sim N_c$ while $V_S \sim W_C \sim
1/N_c$. To proceed further and gain some insight we write the
potentials in terms of leading single meson exchanges (see also
Ref.~\cite{Banerjee:2001js}) one has Yukawa like potentials (we use
the notation of Ref.~\cite{Machleidt:1987hj})
\begin{eqnarray}
V_C (r) &=& - \frac{g_{\sigma NN}^2}{4 \pi} \frac{e^{-m_\sigma r}}{r}
+ \frac{g_{\omega NN}^2}{4 \pi} \frac{e^{-m_\omega r}}{r} \, , \nonumber 
\\ 
W_S(r)
&=& \frac1{12} \frac{g_{\pi NN}^2}{4\pi} \frac{m_\pi^2}{\Lambda_N^2}
\frac{e^{-m_\pi r}}{r} + \frac1{6} \frac{f_{\rho NN}^2}{4
\pi}\frac{m_\rho^2}{\Lambda_N^2} \frac{e^{-m_\rho r}}{r} \, , 
\nonumber 
\\ 
\end{eqnarray} 
where $\Lambda_N = 3M_N/N_c$ is a scale which is numerically equal to
the nucleon mass and is ${\cal O} (N_c^0)$. All meson couplings scale
as $g_{\sigma NN} , g_{\pi NN}, g_{\omega NN}, f_{\rho NN} \sim
\sqrt{N_c}$ whereas all meson masses scale as
$m_\pi,m_\sigma,m_\rho,m_\omega \sim N_c^0$.  In principle there would
be infinitely many contributions but we stop at the vector mesons. A
relevant question which will be postponed to the next Section regards
{\it what} values of Yukawa masses should one take. This is
particularly relevant for the $m_\sigma$ case.  Note that the
tensorial structure of the potential Eq.~(\ref{eq:pot-largeN}) is
complete to ${\cal O} (N_c^{-1})$. This leaves room for 
${\cal O}(N_c^0)$ corrections to the NN potential {\it without} generating
  new dependences triggered by sub-leading mass shifts $\Delta m_\sigma
  ={\cal O} (N_c^{-1})$ and sub-leading vertex corrections $\Delta
  g_{\sigma NN} = {\cal O} (N_c^{-1/2})$.

As we have mentioned above, to obtain Serber symmetry we must get a
large cancellation.  At short distances the Yukawa OBE potentials have
Coulomb like behavior $V \to C/(4 \pi r)$ with the dimensionless
combinations
\begin{eqnarray}
C_{V_C+W_S}&=&-g_{\sigma NN}^2 +g_{\omega NN}^2 + 
 \frac{ f_{\rho NN}^2 m_\rho^2 }{6 M_N^2}
 \nonumber \\
C_{V_C+9W_S}&=&-g_{\sigma NN}^2 +g_{\omega NN}^2 + 
 \frac{ 3 f_{\rho NN}^2 m_\rho^2 }{2 M_N^2}
 \nonumber \\
\end{eqnarray}
where the small OPE contribution has been dropped. To resemble Serber
symmetry we should have $C_{V_C+W_S} \ll C_{V_C+9W_S} $. There are
several scenarios where this can be achieved. For instance, if we
impose the OPE 1/9-rule for the full potential we have $g_{\sigma
  NN}^2 = g_{\omega NN}^2 $. Using SU(3), $3 g_{\rho NN}= g_{\omega NN}
$, Sakurai's universality $g_{\rho NN}= g_{\rho \pi\pi}/2$, the
current-algebra KSFR relation, $\sqrt{2} g_{\rho \pi\pi} f_\pi=
m_\rho$, and the scalar Goldberger-Treiman relation, $g_{\sigma NN}
f_\pi = M_N$, one would get $ M_N = N_c m_\rho/(2 \sqrt{2}) $ a
not unreasonable result.  This only addresses the cancellation at short
distances. The cancellation would be more effective at intermediate
distances if $m_\rho $ and $m_\sigma$ would be numerically closer. In
this regard, let us note that there are various schemes where an
identity between scalar and vector meson masses are explicitly
verified~\cite{Weinberg:1990xn,Svec:1996xp,Megias:2004uj}. In reality,
however, the scalar and vector masses are sufficiently different
$m_\sigma = 444 {\rm MeV}$ vs $m_\rho = 770 {\rm MeV}$. In the next
Section we want to analyze this apparent contradiction.

\section{From $\pi\pi$ resonances to $NN$ Yukawa potentials}
\label{sec:yukawa}

\subsection{Correlated two pion exchange}

As we have already mentioned TPE is a genuine test of chiral
symmetry. On the other hand, it is well known that the iterated
exchange of two pions may become in the $t$-channel either a $\sigma$
or a $\rho$ resonance for isoscalar and isovector states
respectively. While the interactions leading to this collectiveness
are controlled to a great extent by chiral
symmetry~\cite{Oller:1997ng,Nieves:1999bx,Nieves:2001de}, the
resulting contributions to the NN potential in terms of boson
exchanges bear a very indirect relation to it. The relation of the
ubiquitous scalar meson in nuclear physics and NN forces in terms of
correlated two pion exchange has been pointed out many
times~\cite{Partovi:1969wd,Machleidt:1987hj,Lin:1990cx,Kim:1994ce}
(see
e.g. Refs.~\cite{Kaiser:1998wa,Oset:2000gn,Kaskulov:2004kr,Donoghue:2006rg}
for a discussion in a chiral context). Attempts to introduce chiral
Lagrangeans in nuclear physics have been
numerous~\cite{Stoks:1996yj,Furnstahl:1996wv,Papazoglou:1998vr} but
the implications for the OBE potential are meager despite the fact
that useful relations among couplings can be deduced~\footnote{We
  should mention the Goldberger-Treimann relation for pions $g_A M_N =
  g_{\pi NN} f_\pi $ and scalars $ M_N = g_{\sigma NN} f_\pi$ which
  yields $g_{\pi NN} = 12.8 $ and $g_{\sigma NN}=10.1$ and the
  KSFR-universality relation which yields $ g_{\rho NN} = g_{\rho \pi
    \pi}/2 = m_\rho / f_\pi /\sqrt{8} = 2.9$}. As we will see, they
are complementary information to the large $N_c$ requirements.

Note that the leading term generating the scalar meson is $ g_A^4 /
f_\pi^6 \sim N_c$ but occurs first at N$^3$LO in the chiral counting. The
central potential reads~\cite{Kaiser:1998wa,Oset:2000gn,Kaskulov:2004kr,Donoghue:2006rg},
\begin{eqnarray}
V_{NN}^C (r) = - \frac{32 \pi}{3 m_\pi^4}\int \frac{d^3 q}{(2 \pi)^3}
e^{i q \cdot x} \left[\sigma_{\pi N} (-q^2) \right]^2 t_{00} (-q^2) \, \nonumber \\
\label{eq:Vsig-piN} 
\end{eqnarray} 
where $\sigma_{\pi N} (s) $ is the $\pi N$ sigma term and $t_{00} (s)
$ the $\pi\pi$ scattering amplitude in the $I=J=0$ channel as a
function of the $\pi\pi$ CM energy $\sqrt{s}$ (see also
Eq.~(\ref{eq:tIJ})). Under the inclusion of $\Delta$ resonance
contributions Eq.~(\ref{eq:Vsig-piN}) is modified by an aditive
redefinition of $\sigma_{\pi N}$ to include those
$\Delta$-states~\cite{Oset:2000gn}. In the large $N_c$ limit, $ t_{\pi
  \pi} (s) \sim 1/N_c $ while $\sigma_{\pi N} (s) \sim N_c $ yielding
$V_{NN} \sim N_c $ as expected~\cite{Kaplan:1996rk}. Actually, at the
sigma pole
\begin{eqnarray}
\frac{32 \pi}{3 m_\pi^4}\left[\sigma_{\pi N} (s) \right]^2 t_{\pi
\pi}^{II} (s) &\to& \frac{g_{\sigma NN}^2}{s - (m_\sigma- i
\Gamma_\sigma/2 )^2} \nonumber \\ 
&\to& \frac{g_{\sigma NN}^2}{s - m_\sigma^2} \, , 
\end{eqnarray} 
where in the second step we have taken the large $N_c$ limit. This
yields $g_{\sigma \pi\pi} \sim 1/ \sqrt{N_c}$, provided $m_\sigma \sim
N_c^0$ and $\Gamma_\sigma \sim 1 / N_c$. The ``fictitious'' narrow
$\sigma$ exchange has been attributed to $N\Delta+ \Delta \Delta$
intermediate states~\cite{Kaiser:1998wa}, to $2 \pi$ iterated
pions~\cite{Donoghue:2006rg} or both~\cite{Oset:2000gn}. This
identification is based on fitting the resulting r-space potentials to
a Yukawa function in a {\it reasonable} distance range.

\subsection{Exchange of Pole Resonances}

In this section we separate the resonance contribution to the $NN$
potential from the background, neglecting for simplicity the vertex
correction in Eq.~(\ref{eq:Vsig-piN}). The most obvious definition of
the $\sigma$ or $\rho$ propagator is via the $\pi\pi$ scattering
amplitude in the scalar-isoscalar and vector-isovector channels,
$(J,I)=(0,0)$ and $(J,I)=(1,1)$ respectively. Using the definition
\begin{eqnarray}
t_{IJ}(s) = \frac1{2 i \rho_{\pi\pi}(s)}\left( e^{2 i \delta_{IJ} (s)}
-1 \right) \, ,
\label{eq:tIJ}
\end{eqnarray} 
with $\rho_{\pi \pi} (s)=\sqrt{1-4 m_\pi^2/s}$ the phase space in our
notation.  Taking into account the fact that on the second Riemann
sheet (taking $\sigma$ as an example) the amplitude has a pole 
\begin{eqnarray}
t^{\rm II}_{00} (s) \to  \frac{R_\sigma}{s-s_\sigma}  \, , 
\end{eqnarray} 
with $\sqrt{s_\sigma}= M_\sigma - i \Gamma_\sigma/2 $ the pole
position and $R_\sigma $ the corresponding residue. Here we define, as
usual, the analytical continuation as 
\begin{eqnarray}
t^{\rm II}_{00} (z)^{-1} - t^{\rm I}_{00} (z)^{-1} = 2 i \rho_{\pi\pi}
(z) \, . 
\label{eq:analytic}
\end{eqnarray} 
By continuity $t_{00} (s) \equiv t_{00}^{\rm I}(s \pm i0^+)= t_{00}^{\rm II}(s \mp
i0^+)$ and thus unitarity requires $ \rho_{\pi\pi} ( s+ i
0^+)=-\rho_{\pi\pi}(s-i0^+)$. One has for the (suitably normalized)
scalar propagator
\begin{eqnarray}
D_S (s) = \frac{t_{00} (s)}{|R_\sigma|} \, ,  
\end{eqnarray} 
in the whole complex plane. In particular
\begin{eqnarray}
D_S^{\rm II}(s) = \frac{t_{00}^{\rm II} (s)}{|R_\sigma|} \to
\frac{e^{i\varphi_\sigma}}{s-(M_\sigma - i \Gamma_\sigma /2)^2} \, .
\end{eqnarray} 
where the phase $\varphi_\sigma$ is defined as $e^{i\varphi_\sigma
}=R_\sigma/|R_\sigma|$ and is related to the background, i.e. the
non-pole contribution.  In Appendix~\ref{sec:toy} we discuss a toy
model for $\pi\pi$ scattering~\cite{Binstock:1972gx} which proves
quite useful to fix ideas. The function $D_S (s)$ is analytic in the
complex s-plane with a $2\pi$ right cut along the $( 4 m_\pi^2,
\infty)$ line stemming from unitarity in $\pi\pi$ scattering and a
left cut running from $(-\infty,0)$ due to particle exchange in the
$u$ and $t$ channels. Assuming the scattering amplitude to be
proportional to this propagator the corresponding $\pi\pi$ phase shift
is then given by
\begin{eqnarray}
e^{2 i \delta_{00}(s)}&=& \frac{t_{00}(s+i0^+)}{t_{00}(s-i0^+)}
=\frac{D_S(s+i0^+)}{D_{S}(s-i0^+)}  \, . 
\end{eqnarray}
The propagator satisfies the unsubtracted dispersion
relation~\cite{Ericson:1988gk},
\begin{eqnarray}
D_S (q^2) = \int_{4m_\pi^2}^\infty d \mu^2 \frac{\rho_S (\mu^2)}{\mu^2-q^2} \, , 
\label{eq:disp-rel}
\end{eqnarray} 
where the spectral function is related to the discontinuity across
the unitarity branch cut~\footnote{Defined as ${\rm Disc} \, t
  (s+i0^+) = t (s+i0^+) - t (s-i0^+) = 2 i {\rm Im}\, t(s)$ for a real
  function below $\pi\pi$ threshold, $0< s< 4m^2$.}
\begin{eqnarray}
\rho_S ( s) &=& \frac1{2 i \pi} {\rm Disc} D_S (s+i0^+) \\ 
&=& \frac1{\pi |R_\sigma|} \rho_{\pi\pi} (s) | t_{00}(s)|^2 \, , 
\label{eq:spectral-t}
 \end{eqnarray} 
which satisfies the normalization condition 
\begin{eqnarray}
\int_{4m_\pi^2}^\infty d \mu^2 \rho_S ( \mu^2) = Z_\sigma \, . 
\label{eq:norm-spec}
\end{eqnarray} 
where $Z_\sigma$ is the integrated strength. Thus, the Fourier
transformation of the propagator is
\begin{eqnarray}
D_S(r)&=&\int \frac{d^3 q}{(2 \pi)^3} e^{i {\vec q} \cdot \vec x} D_S(-\vec q^2) \nonumber \\ 
&=& - \frac{1}{4\pi r} \int_{4 m_\pi^2}^\infty d \mu^2 \rho_S(\mu^2)
e^{-\mu r} \, . 
\label{eq:dis-coor}
\end{eqnarray} 
According to Eq.~(\ref{eq:norm-spec}), $D_S(r) \sim -Z_\sigma /(4\pi
r)$ for small distances. We define the analytic function $\rho_S(z)
e^{-\sqrt{z} r}$ for $r> 0$ in the cut plane without $(-\infty,0)$ and
$(4 m_\pi^2, \infty)$ where
\begin{eqnarray}
\rho_S ( z) &=& \frac1{\pi |R_\sigma|} \rho_{\pi\pi} (z) t_{00}^{\rm
  I}(z) t_{00}^{\rm II}(z) \, ,  
\label{eq:spectral-t2}
 \end{eqnarray} 
\begin{figure}
\includegraphics[height=4cm,width=5.5cm]{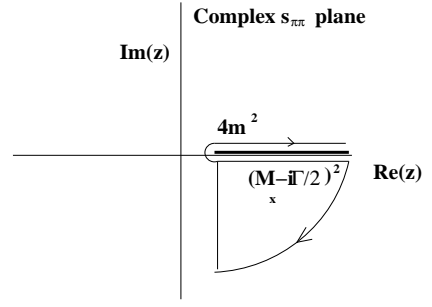} 
\caption{$\pi\pi$ complex squared CM energy plane. The contour used
  in the main text yielding the pole+background decomposition for the
  coordinate space scalar-isoscalar propagator in
  Eq.~(\ref{eq:sig+pipi}).}
\label{fig:contour}
\end{figure}
and fulfilling the boundary value condition $\rho_S (s) \equiv \rho_S
(s+i0^+)$.  This function has a pole at
the complex point $z = s_\sigma = ( M_\sigma - i \Gamma_\sigma/2)^2$ and a square
root branch cut at $z= 4m_\pi^2$ triggered by the phase space factor
 only since
$t_{00}^{\rm I}(z) t_{00}^{\rm II}(z)$ is continuous, so that $\rho_S
( s+ i 0^+)=-\rho_S(s-i0^+)$. Thus, we can write the spectral
integral, Eq.~(\ref{eq:dis-coor}) as running below the unitarity cut
and by suitably deforming the contour in the fourth quadrant in the
second Riemann Sheet, as shown in Fig.~\ref{fig:contour}, we can
separate explicitly the contribution from the pole and the $2\pi$
background yielding
\begin{eqnarray}
D_S (r)= D_\sigma(r) + D_{2 \pi} (r) \, .  
\label{eq:sig+pipi}
\end{eqnarray} 
While, in principle, both contributions are complex, the total result
must be real and their imaginary parts cancel identically (see
Appendix~\ref{sec:toy} for a specific example). 
Using that Eq.~(\ref{eq:analytic}) implies $2 i t_{00}^I ( s_\sigma)
\rho_{\pi\pi} (s_\sigma) =1$ the $\sigma$-pole contribution is
effectively given by
\begin{eqnarray}
{\rm Re} D_\sigma(r) &=&- \frac{Z_\sigma e^{-M_\sigma r}}{4 \pi r} \\  
&\times& \left[ \cos
  \left( \frac{r \Gamma_\sigma}2 \right) - \tan \varphi_\sigma \sin \left(
  \frac{r\Gamma_\sigma} 2 \right) \right] \, , \nonumber 
\end{eqnarray} 
which is an oscillating function damped by an exponential. In the
narrow resonance limit, $\Gamma_\sigma\to 0$,  one has $\varphi_\sigma =
{\cal O} (\Gamma_\sigma)$ yielding 
\begin{eqnarray}
{\rm Re} D_\sigma(r) \sim - \frac{Z_\sigma e^{-M_\sigma r}}{4 \pi r} \left[ 1 +
  {\cal O} ( \Gamma_\sigma^2) \right] \, ,
\label{eq:yuk-narrow}
\end{eqnarray} 
which is a Yukawa potential.  
The $2\pi$ background reads
\begin{eqnarray}
{\rm Re} D_{2\pi}(r) &=& -\frac1{4 \pi r} \frac{i}2 \Big[\int_{0}^\infty d y
  \rho_S( 4m_\pi^2-iy) e^{-r \sqrt{4m_\pi^2-iy} } \nonumber \\ &&-
   \int_{0}^\infty d y \rho_S( 4m_\pi^2+iy) e^{-r
    \sqrt{4m_\pi^2+iy} } \, \, \Big] \, .
\end{eqnarray}
At large distances the integral is dominated by the small $y$ region,
and we get the distinct TPE behaviour $\sim e^{- 2 m_\pi r}$. The
pre-factor is obtained by expanding at small $y$ and using that
unitarity imposes the spectral density to be proportional to the phase
space factor, Eq.~(\ref{eq:spectral-t}).  Close to threshold, $s \to 4
m_\pi^2$, involves the $\pi\pi$ scattering length $a_{00}$ defined as
$\delta_{00} (s) \sim a_{00} \sqrt{s - 4 m_\pi^2}$ yielding 
\begin{eqnarray} 
\rho_S (s ) =   \frac{ 2 m_\pi a_{00}^2
}{\pi |R_\sigma|} \, \sqrt{s - 4 m_\pi^2} + \dots \, .  
\end{eqnarray} 
We therefore get 
\begin{eqnarray}
D_{2\pi}(r) &=&  - \frac{K_2 ( 2 m_\pi r)}{r^2} \frac{ 4 m_\pi^3 a_{00}^2
}{\pi^2 |R_\sigma|} + \dots \nonumber \\ 
&\sim& -\frac{e^{-2
      m_\pi r}}{r^{\frac52}} \frac{2 a_{00}^2 m_\pi^\frac52}{\pi |R_\sigma|} \, . 
\end{eqnarray}
In Appendix~\ref{sec:toy} the pole-background decomposition in
Eq.~(\ref{eq:sig+pipi}) is checked explicitly in a toy model. The
resonance contribution saturates the normalization completely, the
$2\pi$ continuum background yielding a vanishing contribution to the
integrated strength. On the other hand, the resonance produces a
Yukawa tail with an oscillatory modulation which alternates between
attraction and repulsion, although the region where the oscillation is
relevant depends largely on $\varphi_\sigma$.

\subsection{$\pi\pi$ resonances at large $N_c$}

The large $N_c$ analysis also opens up the possibility to a better
understanding of the role played by the ubiquitous scalar meson. This
is an essential ingredient accounting phenomenologically for the
mid range nuclear attraction and which, with a mass of $ \sim 500{\rm
  MeV}$, was originally proposed in the fifties~\cite{PhysRev.98.783}
to provide saturation and binding in nuclei. Along the years, there
has always been some arbitrariness on the ``effective'' or
"fictitious'' scalar meson mass and coupling constant to the nucleon,
partly stimulated by lack of other sources of
information~\footnote{For instance, in the very successful Charge
  Dependent (CD) Bonn potential~\cite{Machleidt:2000ge} any partial
  wave $^{2S+1} L_J$-channel is fitted with a different scalar meson
  mass and coupling.}. During the last decade, the situation has
steadily changed, and insistence and efforts of
theoreticians~\cite{Tornqvist:1995ay}, have finally culminated with
the inclusion of the $0^{++}$ resonance (commonly denoted by $\sigma$)
in the PDG~\cite{Yao:2006px} as the $f_0 (600)$ seen as a $\pi\pi$
resonance, with a wide spread of values ranging from $400-1200 {\rm
  MeV}$ for the mass and a $600-1200 {\rm MeV}$ for the width are
displayed~\cite{vanBeveren:2002mc}. These uncertainties have recently
been sharpened by a benchmark determination based on Roy equations and
chiral symmetry~\cite{Caprini:2005zr} yielding the value $m_\sigma - i
\Gamma_\sigma /2 = 441^{+16}_{-8} - i 272^{+9}_{-12} {\rm MeV}$. Once
the formerly fictitious sigma has become a real and well determined
lowest resonance of the QCD spectrum it is mandatory to analyze its
consequences all over. Clearly, these numbers represent the value for
$N_c=3$, but large $N_c$ counting requires that for mesons $m_\sigma
\sim N_c^0$ and $\Gamma_\sigma \sim 1/N_c$.

\begin{figure}
\includegraphics[height=4cm,width=5.5cm]{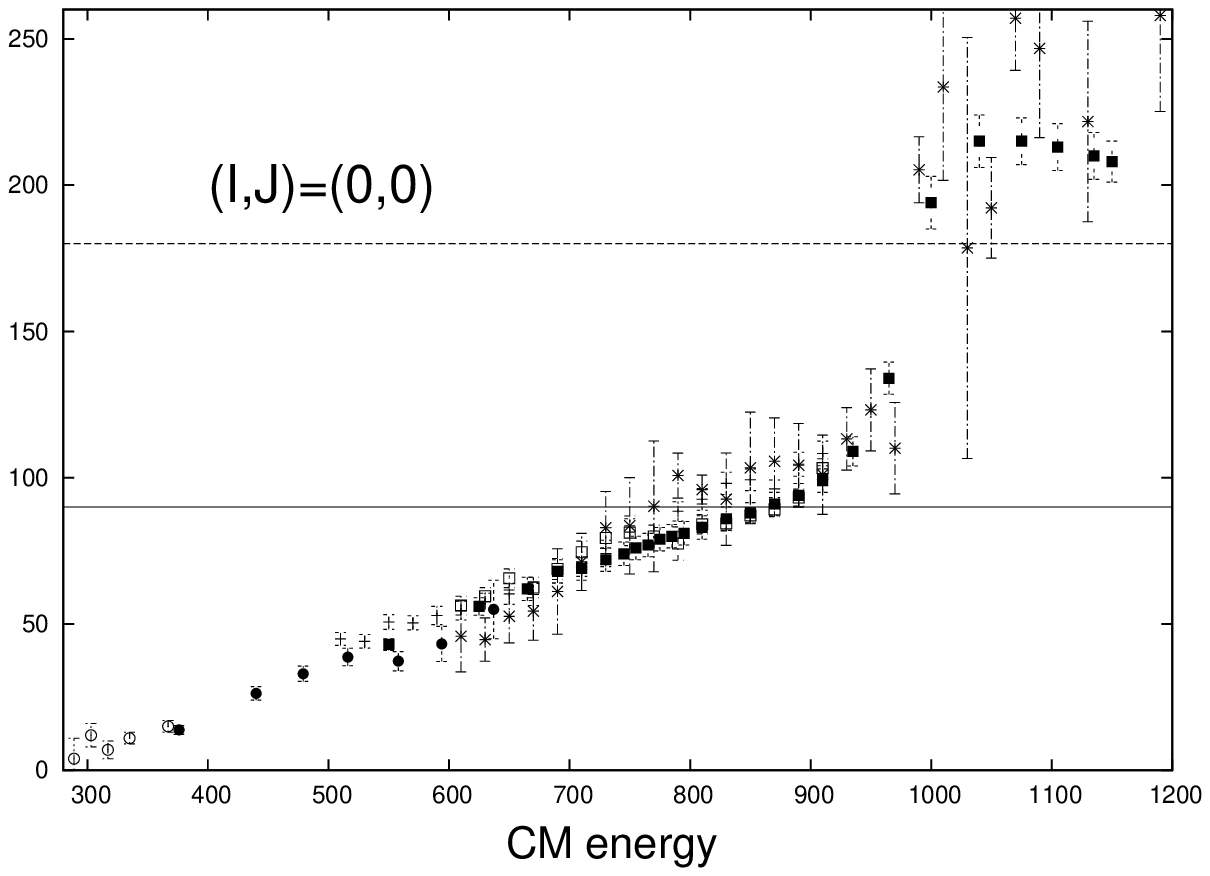} \\ 
\includegraphics[height=4cm,width=5.5cm]{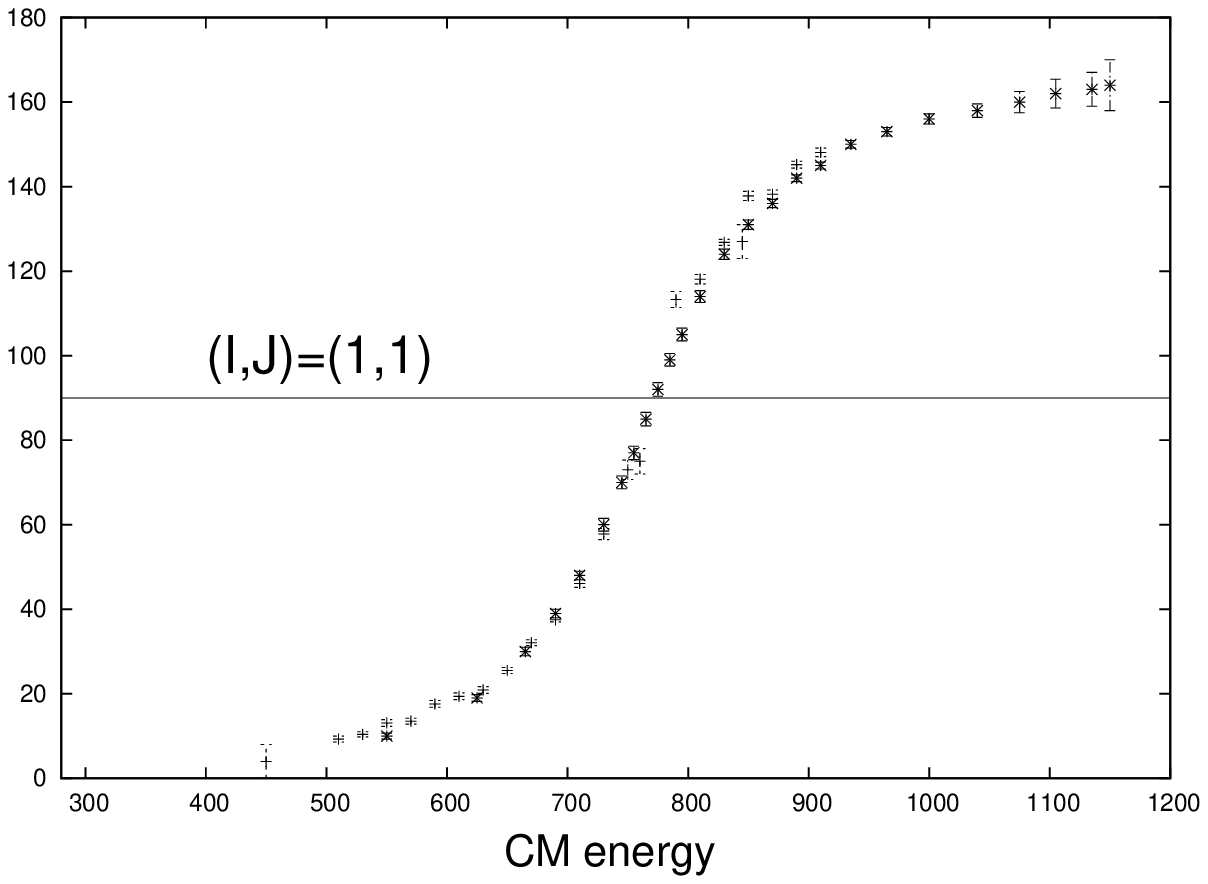}
\caption{$\pi\pi$ scattering phase shifts (in degrees) as a function
  of the CM energy $\sqrt{s}$. Horizontal lines mark the position of
  the Breit-Wigner resonances. Data are from
  \cite{Protopopescu:1973sh,Hyams:1973zf,Estabrooks:1974vu,Srinivasan:1975tj,Kaminski:1996da,Hoogland:1977kt,Froggatt:1977hu}.} 
\label{fig:pipi}
\end{figure}

In this regard the large $N_c$ analysis may provide a clue of {\it
  what} value should be taken for the $\sigma$ mass
~\cite{CalleCordon:2008eu}~\footnote{Large $N_c$ studies in $\pi\pi$
  scattering based on scaling and unitarization with the Inverse
  Amplitude Method (IAM) of ChPT amplitudes provide results which
  regarding the troublesome scalar meson depend on details of the
  scheme used. While the one loop coupled channel
  approach~\cite{Pelaez:2003dy} yields any possible $m_\sigma$ and a
  large width (in apparent contradiction with standard large $N_c$
  counting~\cite{'tHooft:1973jz,Witten:1979kh}), the (presumably more
  reliable) two loop approach~\cite{Pelaez:2006nj}, yields a large
  mass shift (a factor of 2) for the scalar meson when going from
  $N_c=3$ to $N_c = \infty$ yielding $m_\sigma \to 900 {\rm MeV}$, but
  a small shift in the case of the $\rho$ meson. One should note the
  large uncertainties of the two loop IAM method documented in
  Ref.~\cite{Nieves:2001de}. Based on the Bethe-Salpeter approach to
  lowest order~\cite{Nieves:1999bx} we have estimated $m_\sigma \to
  500 {\rm MeV}$~\cite{CalleCordon:2008eu}.}. Of course, similar
remarks apply to the width of other mesons, such as $\rho$, as well.
If we make use of the large $N_c$ expansion according to the standard
assumption ($M^{(k)} \sim N_c^{-k}$)
\begin{eqnarray}
M_\sigma  &=& M_\sigma^{(0)}  + M_\sigma^{(1)} +  {\cal O} (N_c^{-2})  \, ,\\ 
\Gamma_\sigma &=& \Gamma_\sigma^{(1)} +  {\cal O} (N_c^{-2})  \, ,
\end{eqnarray} 
the pole contribution becomes 
\begin{eqnarray}
D_\sigma(r) = - \frac{e^{-m_\sigma r}}{4 \pi r} + {\cal O} (N_c^{-2})
 \, , 
\end{eqnarray} 
where $m_\sigma= M_\sigma^{(0)} + M_\sigma^{(1)} $, representing the
resonance mass to NLO in the $1/N_c$ expansion, should be used. Note
that the width does not contribute to this order.  Thus, for all
purposes we may use a Yukawa potential to represent the exchange of a
resonance. However, what is the {\it numerical} value of this
$m_\sigma$ one should use for the NN problem?. Model calculations
based on $N_c$ scaling of $\pi\pi$ chiral unitary phase shifts for
$N_c=3$ suggest sizeable modifications as compared with the accurately
determined pole position when $N_c$ is varied but the numerical
results are not very robust~\cite{JuanEnrique}~\footnote{Actually,
  according to Ref.~\cite{Flambaum:2007xj} the effect of a meson width
  in the Yukawa-like potential is
$$V(r) = - \frac{g^2}{4 \pi} \left( 1-\frac{\Gamma_\sigma}{m_\sigma
    \pi} \right) e^{- (m_\sigma + \Gamma_\sigma /\pi) r } $$ which
  corresponds to a NLO large $N_c$ renormalization of the mass and
  coupling and providing a ${\cal O}(N_c^ 0) $ correction to the
  central potential. The analysis is based on separating the integrand
  into different intervals which become dominant at large
  distances. Our analysis separates first the pole contribution form
  the background and then studies each contribution separately. We
  note however, that one can extract a Yukawa potential of the meson
  {\it even} for the large and physical width in the region where the
  potential is operating with quite sensible
  values~\cite{Binstock:1972gx}. In Appendix \ref{sec:pipi-realistic}
  we update this analysis using recent parameterizations of $\pi\pi$
  scattering provided in Refs.~\cite{Yndurain:2007qm,Caprini:2008fc}
  confirming the Yukawa behaviour. The main reason is that the the
  potential is being probed for space-like exchanged four momentum,
  while the resonance behaviour takes place in the time-like region
  corresponding to the crossed process $\bar N N \to 2\pi$.}. An
alternative viewpoint where, to the same accuracy, the large $N_c$-NLO
pole contribution could be replaced by the equivalent Breit-Wigner
resonance mass to the same approximation, since according
to~\cite{JuanEnrique} we may take
\begin{eqnarray}
\delta_{00}(m_\sigma^2) &=& \frac{\pi}{2} + {\cal O} (1/N_c^3) 
\end{eqnarray}
Thus, at LO and NLO in the large $N_c$ limit the exchange of a
resonance between nucleons can be represented at long distances as a
Yukawa potential with the Breit-Wigner mass to ${\cal O} (N_c^{-2})$.
The vertex correction $\sigma_{\pi N}$, see
e.g. Eq.~(\ref{eq:Vsig-piN}), just adds a coupling constant yielding
\begin{eqnarray}
V_\sigma (r) = -\frac{g_{\sigma NN}^2}{4\pi}\frac{e^{- m_\sigma r}}{r}
+ {\cal O} (1/N_c) \, .
\end{eqnarray} 
Of course, the same type of arguments apply to the $\rho$-meson
exchange, with the only modification 
\begin{eqnarray} 
\delta_{11}(m_\rho^2)= \frac{\pi}2 + {\cal O}(1/N_c^3) \, , 
\end{eqnarray} 
where now $m_\rho= M_\rho^{(0)} + M_\rho^{(1)} $. In
Fig.~\ref{fig:pipi} we show the data for $\pi\pi$ phase shifts, where
we see that the {\it true} Breit-Wigner masses or not very different.
Of course, these arguments do not imply that the Yukawa masses should
{\it exactly} coincide, but at least suggest that one should expect a
large shift for the $\sigma$ mass from the pole position and a very
small one for the $\rho$ meson mass when the next-to-leading $1/N_c$
correction to the pole masses are considered. The identity of scalar
and vector masses has been deduced from several scenarios based on
algebraic chiral
symmetry~\cite{Weinberg:1969hw,Weinberg:1990xn}. Actually, it has been
shown that $m_\rho=m_\sigma$ {\it without} appealing to the strict
large $N_c$ limit but assuming the narrow resonance approximation (See
also Ref.~\cite{Weinberg:1994tu}).

\section{conclusions} 
\label{sec:conclusions}

Serber symmetry seems to be an evident but puzzling symmetry of the NN
system. Since it was proposed more than 70 years ago no clear
explanation based on the more fundamental QCD Lagrangean has been put
forward. 

In the present paper we have analyzed the problem from the viewpoint
of long distance symmetries, a concept which has proven useful in the
study of Wigner $SU(4)$ spin-flavour symmetry.  Actually, Serber
symmetry is clearly seen in the np differential cross section implying
a set of sum rules for the partial wave phase shifts which are well
verified to a few percent level in the entire elastic region. While
this situation corresponds to scattering of on-shell nucleons, it
would be rather interesting to establish the symmetry beyond this
case. Therefore, we have formulated these sum rules at the level of
high quality potentials, i.e. potentials which describe elastic NN
scattering with $\chi^2 /{\rm DOF} \sim 1$ which are also well
verified at distances above $1 {\rm fm}$. This suggests that a coarse
graining of the NN interaction might also display the symmetry. The
equivalent momentum space Wilsonian viewpoint is implemented
explicitly by the $V_{\rm low k}$ approach by integrating all modes
below a certain cut-off $\Lambda \sim {\rm 400}$. By analyzing
existing $V_{\rm low k}$ calculations for high quality potentials we
have shown that Serber symmetry is indeed fulfilled to a high
degree. We remind that within the $V_{\rm low k}$ approach this
symmetry has direct implications in shell model calculations for
finite nuclei since the $V_{\rm low k}$ potential corresponds to the
effective nuclear interaction.

A surprising finding of the present paper is that chiral potentials,
while implementing extremely important QCD features, do not fulfill
the symmetry to the same degree as current high quality
potentials. This effect must necessarily be compensated by similar
symmetry violations in the counterterms encoding the non-chiral and
unknown short distance interaction and needed to describe NN phase
shifts where the symmetry does indeed happen. While this is not
necessarily a deficiency of the chiral approach it is disturbing that
the symmetry does not manifest at long distances, unlike high quality
potentials. This may be a general feature of chiral potentials which
requires further investigation~\cite{PavonWSchiral}.

In an attempt to provide a more fundamental understanding of the
striking but so far accidental Serber symmetry, we have also
speculated how it might arise from QCD within the large $N_c$
perspective on the second part of the paper. The justification for
advocating such a possible playground is threefold. Firstly, the NN
potential tensorial structure is determined with a relative $1/N_c^2$
accuracy, which naively suggests a bold $10\%$. In the second place,
the meson exchange picture is justified. Finally, we have found
previously that such an expansion provides a rationale for the equally
accidental and pre-QCD Wigner $SU(4)$ symmetry. Actually, we found
that large $N_c$ predicts the NN channels where Wigner symmetry indeed
works and fails phenomenologically. The intriguing point is that when
Wigner symmetry fails, as allowed by large $N_c$ considerations,
Serber symmetry holds instead. In the present paper we have verified
the previous statements at the level of potentials at large distances
or using $V_{\rm low k}$ potentials, reinforcing our previous
conclusions based on just a pure phase shift analysis. Under those
circumstances, it is therefore natural to analyze to what extent and
even if Serber symmetry could be justified at all from a large $N_c$
viewpoint.  In practical terms we have shown that within a One Boson
Exchange framework, the fulfillment of the symmetry at the potential
level is closely related to having not too dissimilar values of
$\sigma$ and $\rho$ meson masses as they appear in Yukawa
potentials. Actually, these $\sigma$ and $\rho$ states are associated
to resonances which are seen in $\pi\pi$ scattering and can be
uniquely defined as poles in the second Riemann sheet of the
scattering amplitude at the invariant mass $\sqrt{s}= M_R - i
\Gamma_R/2$. We have therefore analyzed the meaning of those
resonances within the large $N_c$ picture, by assuming the standard
mass $m_R \sim N_c^0 $ and width $\Gamma_R \sim 1/N_c$ scaling. We
have found that, provided we keep terms in the potential to NLO, meson
widths do not contribute to the NN potential, as they are ${\cal O}
(N_c^ {-1})$, i.e. a relative $1/N_c^2$ correction to the LO
contribution. This justifies using a Yukawa potential where the mass
corresponds to an approximation to the pole mass $m_R = M_R^{(0)}+
M_R^{(1)}$ which cannot be distinguished from the Breit-Wigner mass up
to ${\cal O} (N_c^{-2})$. This suggests that the masses $m_\sigma$ and
$m_\rho$ which appear in the OBE potential could be interpreted as an
approximation to the pole mass rather than its exact value. This
supports the customary two-Yukawa representation of complex-pole
resonances pursued in phenomenological approaches since it was first
proposed ~\cite{Binstock:1972gx}, since in practice only the lowest
Yukawa mass contributes significantly. The question on {\it what}
numerical value should be used for the Yukawa mass is a difficult one,
and at present we know of no other direct way than NN scattering fits
for which $m_\sigma=520(40) {\rm MeV}$ might be
acceptable~\cite{Alvaro2009} when the uncorrelated $2 \pi$
contribution is disregarded.

On a more fundamental level, however, lattice QCD calculations at
variable $N_c$ values~(see e.g. Ref.~\cite{Teper:2008yi} for a review)
might reliably determine the Yukawa mass parameters appearing in the
large $N_c$ potential. A recent quenched QCD lattice calculation
yields~\cite{Bali:2008an} $m_\rho/ \sqrt{\sigma}=
1.670(24)-0.22(23)/N_c^2 $ with $\sqrt{\sigma}$ the string tension,
which for $\sqrt{\sigma}=444 {\rm MeV} $ yields $m_\rho = 740 {\rm
  MeV}$ for $m_\pi=0$ (see also Ref.~\cite{DelDebbio:2007wk}). The
extension of those calculations to compute the needed $1/N_c$ mass
shift would be most welcome and would require full dynamical quarks.
Of course, one should not forget that Serber symmetry holds to great
accuracy in the real $N_c=3$ world, and in this sense it represents a
stringent test to lattice QCD calculations in $P$-waves. Amazingly,
the only existing of S-wave potential calculation\cite{Ishii:2006ec}
displays Wigner symmetry quite accurately.

In
any case the large $N_c$ form of the potential
Eq.~(\ref{eq:pot-largeN}) can be retained with relative $1/N_c^2$
accuracy since meson widths enter beyond that accuracy as sub-leading
corrections, on equal footing with many other effects (spin-orbit,
relativistic and other mesons), independently on how large
the $\sigma$ width is in the real $N_c=3$ world. In our view this
paves the way for further investigations on the relevance of large
$N_c$ based ideas for the two nucleon system.

\begin{acknowledgements} 

{\it We gratefully acknowledge Manuel Pav\'on Valderrama for a
  critical reading of the manuscript, Juan Nieves, Lorenzo Luis
  Salcedo and Jos\'e Ram\'on Pel\'aez for discussions and Scott
  K. Bogner for kindly providing the data corresponding to the V18
  potentials of Ref.~\cite{Bogner:2003wn}. We also thank Jacek
  Dobaczewski and Rupp Machleidt for useful communications.  This work
  has been partially supported by the Spanish DGI and FEDER funds with
  grant FIS2008-01143/FIS, Junta de Andaluc{\'\i}a grant FQM225-05,
  and EU Integrated Infrastructure Initiative Hadron Physics Project
  contract RII3-CT-2004-506078.}
\end{acknowledgements}

\appendix 

\section{Toy model for $\pi\pi$ scattering}
\label{sec:toy} 

\begin{figure}
\includegraphics[height=7.5cm,width=5cm,angle=270]{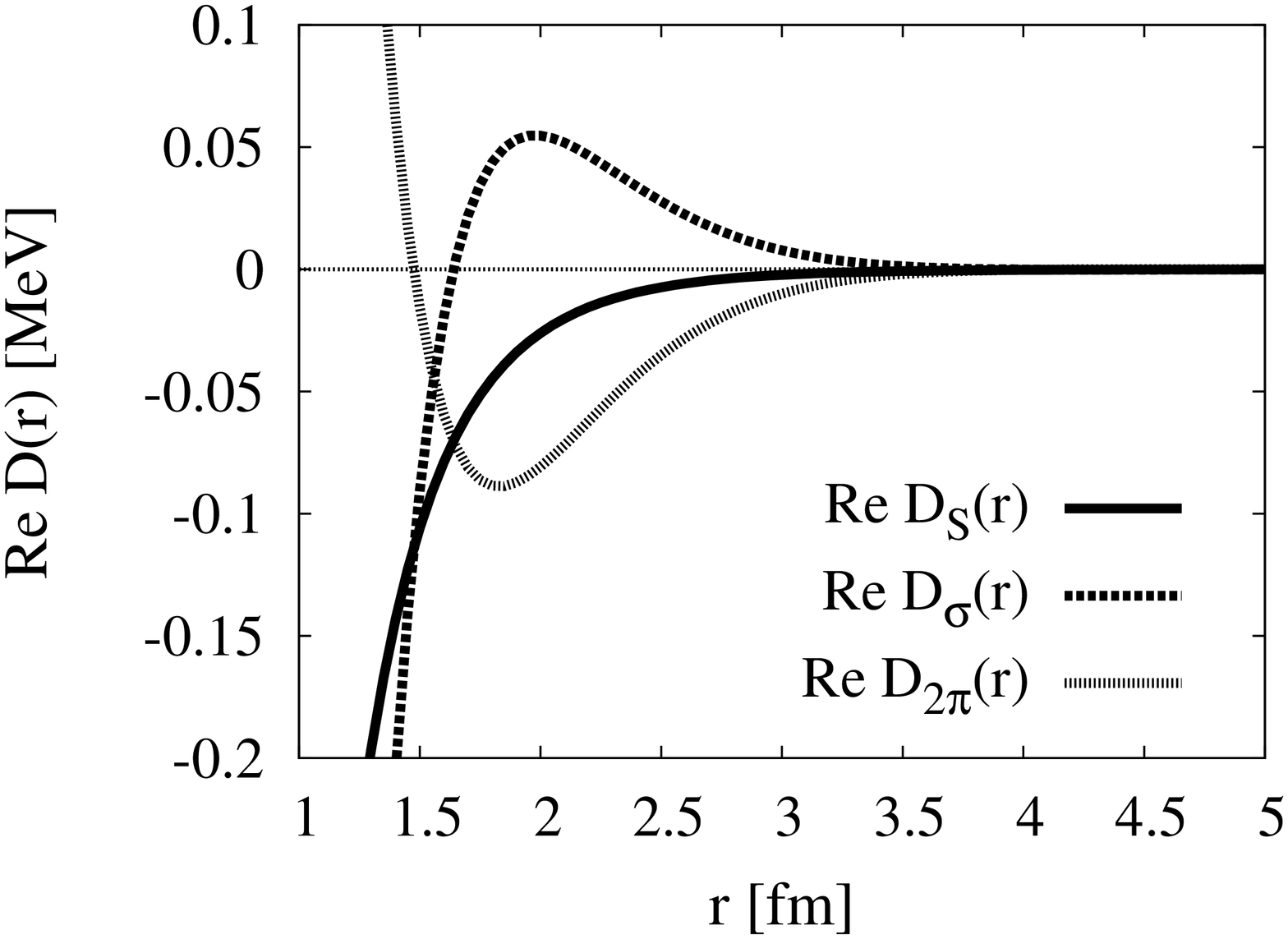}  
\includegraphics[height=7.5cm,width=5cm,angle=270]{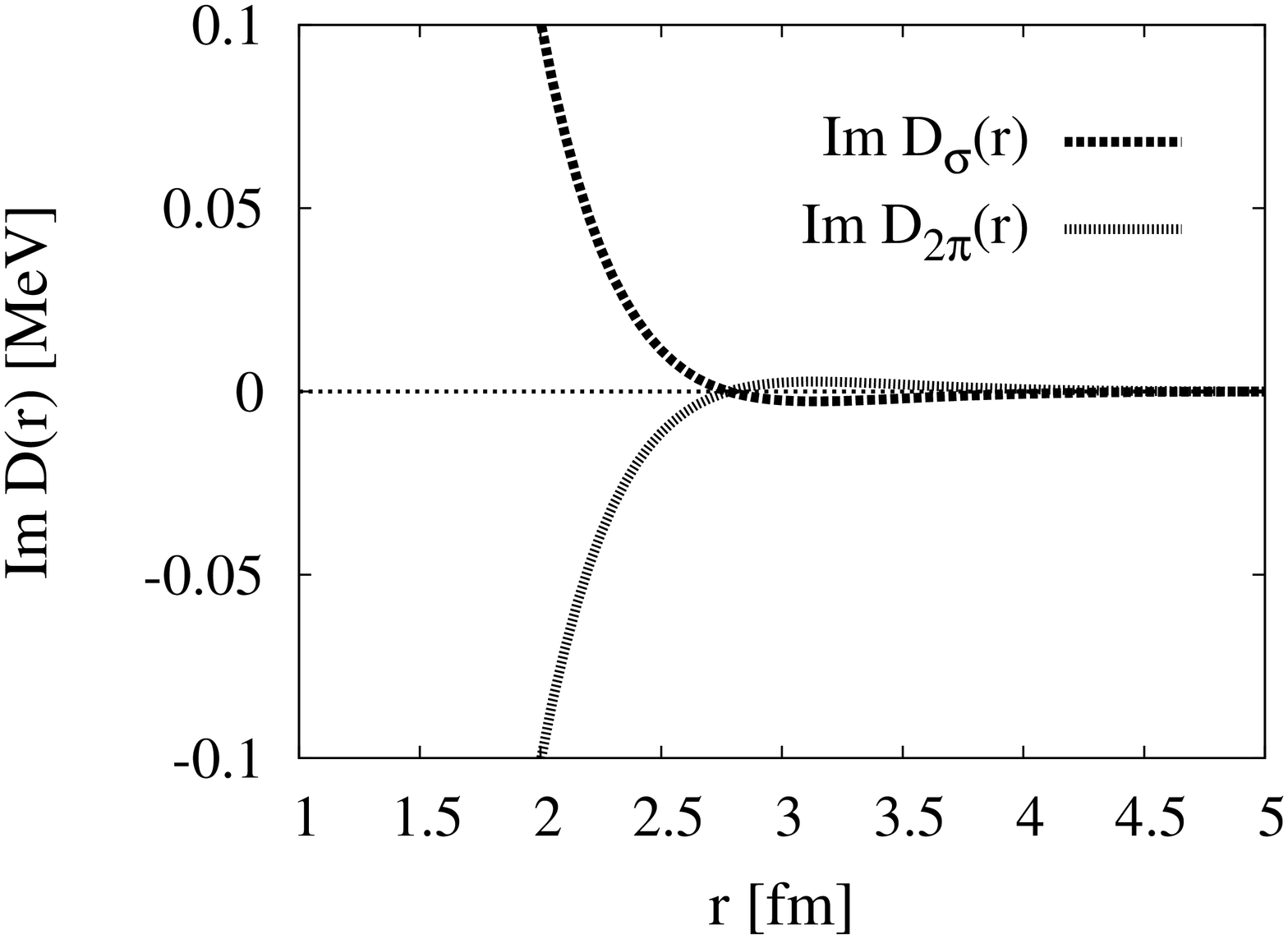}  
\caption{Correlated $\pi\pi$ coordinate space propagator $D (r)$ (in
  MeV) as a function of the distance (in fm) for the $\pi\pi$
  scattering toy model. In the upper panel we draw the real part.  We
  separate the pole contribution $D_\sigma(r)$ (dashed-dotted line)
  from the continuum contribution $D_{2\pi}(r)$ (dotted-line) and the
  total result $D(r)$ (solid line). The identity $D(r) = D_\sigma(r) +
  D_{2\pi}(r)$ is verified. In the lower panel we show the
  cancellation of imaginary parts, ${\rm Im} D_\sigma(r) = -{\rm Im}
  D_{2\pi}(r)$.}
\label{fig:Dpi}
\end{figure}

In this appendix we illustrate with a specific example our discussion
of Section~\ref{sec:yukawa} and in particular the pole-background
decomposition of Eq.~(\ref{eq:sig+pipi}). According to
Ref.~\cite{Binstock:1972gx} the finite width of the scalar meson can
be modelled by the propagator
\begin{eqnarray}
D_S (s) = \frac1{s- m_\sigma^2- i m_\sigma \gamma_\sigma
  \sqrt{\frac{s-4 m_\pi^2}{m_\sigma^2-4 m_\pi^2}}} \, , 
\end{eqnarray} 
for $t \ge 4 m_\pi^2$. Below the elastic scattering threshold we use
the standard definition $\sqrt{t-4 m_\pi^2}= -i \sqrt{|t-4 m_\pi^2|}
e^{i \theta}$ where $0 \le \theta= {\rm Arg}(t-4m_\pi^2) < 2
\pi$. This defines the propagator in the first Riemann sheet, the
second Riemann sheet is determined from the usual continuity equation
$D^{\rm II}_S (s+i0^+)= D^{\rm I}_S (s-i0^+)$. The pole position is given by 
\begin{eqnarray}
s_\sigma &=& (M_\sigma-i \Gamma_\sigma/2)^2 = m_\sigma^2 - 
  \frac{ \gamma_\sigma^2  m_\sigma^2 }{2 m_\sigma^2 - 8 m_\pi^2} \nonumber \\ &-& i \, 
  \frac{\gamma_\sigma m_\sigma \sqrt{4 (m_\sigma^2-4 m_\pi^2)^2- \gamma_\sigma^2 m_\sigma^2} }{2 m_\sigma^2-8m_\pi^2} \, .   
\label{eq:width}
\end{eqnarray} 
In the small width limit the position of the pole and width are 
\begin{eqnarray}
M_\sigma &=& m_\sigma - \frac{\gamma_\sigma^2}{8
  m_\sigma}\frac{m_\sigma^2+4 m_\pi^2}{m_\sigma^2-4 m_\pi^2}+ {\cal
  O}(\gamma_\sigma^4) \\ \Gamma_\sigma &=& \gamma_\sigma + {\cal
  O}(\gamma_\sigma^3) \, . 
\label{eq:width-pert}
\end{eqnarray} 
Despite the large $\sigma$-width $m_\sigma \sim \gamma_\sigma$ 
this expansion works because the $1/8$-factor yields (the next correction has a numerical $1/128$, see below).
Assuming the
scattering amplitude to be proportional to this propagator the
corresponding $\pi\pi$ phase shift is then given by
\begin{eqnarray}
e^{2 i \delta_{00}(s)}
&=&\frac{s-m_\sigma^2- i m_\sigma \gamma_\sigma \sqrt{\frac{s-4
      m_\pi^2}{m_\sigma^2-4 m_\pi^2}} }{s-m_\sigma^2+ i m_\sigma
  \gamma_\sigma \sqrt{\frac{s-4 m_\pi^2}{m_\sigma^2-4 m_\pi^2}}}
\end{eqnarray} 
The parameterization is such that the standard Breit-Wigner definition
of the resonance is fulfilled for the bare parameters, 
\begin{eqnarray}
\delta_{00}(m_\sigma^2) = \frac{\pi}{2} \, ,  \qquad \gamma_\sigma =
\frac{1}{m_\sigma \delta_{00}'(m_\sigma^2)}
\end{eqnarray}
Of course, in the limit of narrow resonances both definitions are
indistinguishable and we have $M_\sigma \to m_\sigma$ and
$\Gamma_\sigma \to \gamma_\sigma$.  If we use the pole position in the
second Riemann sheet of the S-matrix or equivalently the zero in the
first Riemann sheet from~\cite{Caprini:2005zr} yielding the value
$M_\sigma - i \Gamma_\sigma /2 = 441^{+16}_{-8} - i 272^{+9}_{-12}
{\rm MeV}$ we get
\begin{eqnarray}
m_\sigma= 567(10) {\rm MeV} \qquad \gamma_\sigma/2= 276(10) {\rm MeV} \, . 
\end{eqnarray}
From the small width expansion, Eq.~(\ref{eq:width-pert}), one gets
$m_\sigma= 554(10){\rm MeV}$, despite the apparent large width.  From
Ref.~\cite{Leutwyler:2008xd} one has the magnitude of the residue
$|R_\sigma|=0.218^{+0.023}_{-0.012} {\rm GeV}^2$ whereas we get
$|R_\sigma|=0.430{\rm GeV}^2$. Note the $120(20) {\rm MeV}$ shift
between the Breit-Wigner and the pole position. With the above
parameters the scattering length is $a_{00} m_\pi = 0.36 $ which is
clearly off the value $a_{00} m_\pi = 0.220(2)$ deduced from ChPT.
The propagator satisfies the unsubtracted dispersion in
Eq.~(\ref{eq:disp-rel}) where the spectral function is given by
\begin{eqnarray}
\rho( \mu^2) = \frac1{\pi}\frac{\gamma_\sigma m_\sigma
  \sqrt{m_\sigma^2-4m_\pi^2}\sqrt{\mu^2-4 m_\pi^2}}{
  (m_\sigma^2-4m_\pi^2)(\mu^2-m_\sigma^2)^2+\gamma_\sigma^2 m_\sigma^2
  (\mu^2-4 m_\pi^2)} \, , \nonumber \\
\end{eqnarray} 
and satisfies the normalization condition given by
Eq.~(\ref{eq:norm-spec}) with $Z_\sigma=1$. Thus, using the Fourier
transformation of the propagator and separating explicitly the
contribution from the poles $D_\sigma(r) $ and the $2\pi$ background
$D_{2 \pi} (r)$ in Eq.~(\ref{eq:sig+pipi}). This yields the result
depicted in Fig.~(\ref{fig:Dpi}) which illustrates and checks the
pole-background decomposition and shows that the {\it total}
contribution, although describable by a Yukawa shape {\it does not}
correspond to the pole piece. In addition the cancellation of
imaginary parts, ${\rm Im} D_\sigma(r) = -{\rm Im} D_{2\pi}(r)$, is
explicitly verified.
Using the inverse relations of Eq.~(\ref{eq:width-pert}), in the
narrow width approximation the pole contribution becomes
\begin{eqnarray}
{\rm Re} D_\sigma  (r)   &=&  -  \frac{e^{-M_\sigma  r}}{4  \pi   r}   \\ &\times& 
\left[  1 + \frac{r \Gamma_\sigma^2}{8}
  \left(\frac{2  M_\sigma}{ M_\sigma^2-4  m_\pi^2} -
    r\right) + \dots \right] \, , \nonumber 
\end{eqnarray}
in qualitative agreement with Eq.~(\ref{eq:yuk-narrow}). On the other
hand the $2 \pi$ contribution at long distances becomes 
\begin{eqnarray}
D_{2\pi}(r) &=& - \frac{K_2 ( 2 m_\pi r)}{r^2}\frac{\gamma_\sigma m_\pi^2 m_\sigma  
}{\pi^2 (m_\sigma^2- 4 m_\pi^2)^\frac52}  + \dots \nonumber \\ 
&=& -\frac{e^{- 2 m_\pi r}}{r^\frac52} \frac{\gamma_\sigma m_\pi^\frac32 m_\sigma}{\pi^\frac32 (m_\sigma^2- 4 m_\pi^2)^\frac52} + \dots 
\end{eqnarray} 
The asymptotic form in the first lime reproduces $95\%$ accuracy the
full result Eq.~(\ref{eq:sig+pipi}) for $r > 5 {\rm fm}$.

Finally, the $\rho$ meson propagator and the associated $(I,J)=(1,1)$
phase shift can be dealt with {\it mutatis mutandis} by using
\begin{eqnarray}
[D_V (s)]^{-1} = s- m_\rho^2- i m_\rho \gamma_\rho \left[\frac{s-4
      m_\pi^2}{m_\rho^2-4 m_\pi^2}\right]^{\frac32} 
\end{eqnarray} 
where the $p-$wave character of the $\rho \to 2 \pi $ decay can be
recognized in the phase space factor.

\section{Realistic scalar-isoscalar $\pi\pi$ scattering }
\label{sec:pipi-realistic}

Realistic parameterizations of the $\pi\pi$ scattering data have been
proposed based on the conformal
mappings~\cite{Yndurain:2007qm,Caprini:2008fc} with several
variations. Our results show little dependence on those and we show 
here the ghost-full version and the Adler zero located at the lowest
order ChPT $s_A=m^2/2$~\cite{Yndurain:2007qm} which reads
\begin{eqnarray}
\rho (s) \cot \delta_{00}(s) = \frac{m^2}{s-m^2/2} 
\left[
  \frac{m}{\sqrt{s}}+ B_0 + B_1 w + B_2 w^2 \right] \, ,  \nonumber \\ 
\label{eq:conformal} 
\end{eqnarray} 
where the conformal mapping is 
\begin{eqnarray}
w(s)= \frac{\sqrt{s}-\sqrt{4m_K^2-s}}{\sqrt{s}+\sqrt{4m_K^2-s}} \, .
\end{eqnarray} 
For the three sets of $B_{0,1,2}$ parameters discussed in
Ref.~\cite{Yndurain:2007qm} the resulting complex pole position is
slightly higher than the Roy equation value $\sqrt{s_\sigma}=
441^{+16}_{-8} - i 272^{+9}_{-12} {\rm MeV}$~\cite{Caprini:2005zr}. If
we use the dispersive representation for $D(r)$ given in
Eq.~(\ref{eq:dis-coor}) and cut the integral at the $\bar K K$
threshold $\mu = 2 m_K$ we get a function which can be fitted in the
range $1 {\rm fm} \le r \le 5 {\rm fm} $ by a Yukawa potential with
$m_\sigma= 600(50) {\rm MeV}$.  The uncertainty corresponds to
changing the $B_{0,1,2}$ parameters within
errors~\cite{Yndurain:2007qm} as well as the varying the fitting
interval. This is the modern version of the result found long
ago~\cite{Binstock:1972gx} using a relativistic Breit-Wigner form (see
Appendix~\ref{sec:toy}).


\begin{thebibliography}{93}
\expandafter\ifx\csname natexlab\endcsname\relax\def\natexlab#1{#1}\fi
\expandafter\ifx\csname bibnamefont\endcsname\relax
  \def\bibnamefont#1{#1}\fi
\expandafter\ifx\csname bibfnamefont\endcsname\relax
  \def\bibfnamefont#1{#1}\fi
\expandafter\ifx\csname citenamefont\endcsname\relax
  \def\citenamefont#1{#1}\fi
\expandafter\ifx\csname url\endcsname\relax
  \def\url#1{\texttt{#1}}\fi
\expandafter\ifx\csname urlprefix\endcsname\relax\def\urlprefix{URL }\fi
\providecommand{\bibinfo}[2]{#2}
\providecommand{\eprint}[2][]{\url{#2}}

\bibitem[{\citenamefont{Epelbaum et~al.}(2008)\citenamefont{Epelbaum, Hammer,
  and Mei{\ss}ner}}]{Epelbaum:2008ga}
\bibinfo{author}{\bibfnamefont{E.}~\bibnamefont{Epelbaum}},
  \bibinfo{author}{\bibfnamefont{H.-W.} \bibnamefont{Hammer}},
  \bibnamefont{and} \bibinfo{author}{\bibfnamefont{U.-G.}
  \bibnamefont{Mei{\ss}ner}} (\bibinfo{year}{2008}), \eprint{0811.1338}.

\bibitem[{\citenamefont{Serber}(1938)}]{Serber}
\bibinfo{author}{\bibfnamefont{R.}~\bibnamefont{Serber}},
  \bibinfo{journal}{Phys. Rev.} \textbf{\bibinfo{volume}{53}},
  \bibinfo{pages}{211} (\bibinfo{year}{1938}).

\bibitem[{\citenamefont{Ashkin and Wu}(1948)}]{PhysRev.73.973}
\bibinfo{author}{\bibfnamefont{J.}~\bibnamefont{Ashkin}} \bibnamefont{and}
  \bibinfo{author}{\bibfnamefont{T.-Y.} \bibnamefont{Wu}},
  \bibinfo{journal}{Phys. Rev.} \textbf{\bibinfo{volume}{73}},
  \bibinfo{pages}{973} (\bibinfo{year}{1948}).

\bibitem[{\citenamefont{{Christian}}(1952)}]{1952RPPh...15...68C}
\bibinfo{author}{\bibfnamefont{R.~S.} \bibnamefont{{Christian}}},
  \bibinfo{journal}{Reports on Progress in Physics}
  \textbf{\bibinfo{volume}{15}}, \bibinfo{pages}{68} (\bibinfo{year}{1952}).

\bibitem[{\citenamefont{Blatt and Weisskopf}(1952)}]{TheoLib:NUC05}
\bibinfo{author}{\bibfnamefont{J.}~\bibnamefont{Blatt}} \bibnamefont{and}
  \bibinfo{author}{\bibfnamefont{V.}~\bibnamefont{Weisskopf}},
  \emph{\bibinfo{title}{Theoretical Nuclear Physics}} (\bibinfo{publisher}{John
  Wiley \& Sons}, \bibinfo{year}{1952}).

\bibitem[{\citenamefont{Christian and Hart}(1950)}]{PhysRev.77.441}
\bibinfo{author}{\bibfnamefont{R.~S.} \bibnamefont{Christian}}
  \bibnamefont{and} \bibinfo{author}{\bibfnamefont{E.~W.} \bibnamefont{Hart}},
  \bibinfo{journal}{Phys. Rev.} \textbf{\bibinfo{volume}{77}},
  \bibinfo{pages}{441} (\bibinfo{year}{1950}).

\bibitem[{\citenamefont{Gerjuoy}(1950)}]{PhysRev.77.568}
\bibinfo{author}{\bibfnamefont{E.}~\bibnamefont{Gerjuoy}},
  \bibinfo{journal}{Phys. Rev.} \textbf{\bibinfo{volume}{77}},
  \bibinfo{pages}{568} (\bibinfo{year}{1950}).

\bibitem[{\citenamefont{Nakabayasi and Sato}(1952)}]{PhysRev.88.144}
\bibinfo{author}{\bibfnamefont{K.}~\bibnamefont{Nakabayasi}} \bibnamefont{and}
  \bibinfo{author}{\bibfnamefont{I.}~\bibnamefont{Sato}},
  \bibinfo{journal}{Phys. Rev.} \textbf{\bibinfo{volume}{88}},
  \bibinfo{pages}{144} (\bibinfo{year}{1952}).

\bibitem[{\citenamefont{Jastrow}(1951)}]{PhysRev.81.165}
\bibinfo{author}{\bibfnamefont{R.}~\bibnamefont{Jastrow}},
  \bibinfo{journal}{Phys. Rev.} \textbf{\bibinfo{volume}{81}},
  \bibinfo{pages}{165} (\bibinfo{year}{1951}).

\bibitem[{\citenamefont{Hull et~al.}(1961)\citenamefont{Hull, Lassila, Ruppel,
  McDonald, and Breit}}]{PhysRev.122.1606}
\bibinfo{author}{\bibfnamefont{M.~H.} \bibnamefont{Hull}},
  \bibinfo{author}{\bibfnamefont{K.~E.} \bibnamefont{Lassila}},
  \bibinfo{author}{\bibfnamefont{H.~M.} \bibnamefont{Ruppel}},
  \bibinfo{author}{\bibfnamefont{F.~A.} \bibnamefont{McDonald}},
  \bibnamefont{and} \bibinfo{author}{\bibfnamefont{G.}~\bibnamefont{Breit}},
  \bibinfo{journal}{Phys. Rev.} \textbf{\bibinfo{volume}{122}},
  \bibinfo{pages}{1606} (\bibinfo{year}{1961}).

\bibitem[{\citenamefont{de~la Ripelle et~al.}(2005)\citenamefont{de~la Ripelle,
  Sofianos, and Adam}}]{delaRipelle:2004ms}
\bibinfo{author}{\bibfnamefont{M.~F.} \bibnamefont{de~la Ripelle}},
  \bibinfo{author}{\bibfnamefont{S.~A.} \bibnamefont{Sofianos}},
  \bibnamefont{and} \bibinfo{author}{\bibfnamefont{R.~M.} \bibnamefont{Adam}},
  \bibinfo{journal}{Ann. Phys.} \textbf{\bibinfo{volume}{316}},
  \bibinfo{pages}{107} (\bibinfo{year}{2005}), \eprint{nucl-th/0410016}.

\bibitem[{\citenamefont{Fetter and Walecka}(1971)}]{Fetter}
\bibinfo{author}{\bibfnamefont{A.}~\bibnamefont{Fetter}} \bibnamefont{and}
  \bibinfo{author}{\bibfnamefont{J.}~\bibnamefont{Walecka}},
  \emph{\bibinfo{title}{Quantum Theory of Many-Panicle Systems}}
  (\bibinfo{publisher}{McGraw–Hill, New York}, \bibinfo{year}{1971}).

\bibitem[{\citenamefont{Bethe and Longmire}(1950)}]{PhysRev.77.647}
\bibinfo{author}{\bibfnamefont{H.~A.} \bibnamefont{Bethe}} \bibnamefont{and}
  \bibinfo{author}{\bibfnamefont{C.}~\bibnamefont{Longmire}},
  \bibinfo{journal}{Phys. Rev.} \textbf{\bibinfo{volume}{77}},
  \bibinfo{pages}{647} (\bibinfo{year}{1950}).

\bibitem[{\citenamefont{{Lashko} and {Filippov}}(2007)}]{2007PAN....70.1440L}
\bibinfo{author}{\bibfnamefont{Y.~A.} \bibnamefont{{Lashko}}} \bibnamefont{and}
  \bibinfo{author}{\bibfnamefont{G.~F.} \bibnamefont{{Filippov}}},
  \bibinfo{journal}{Physics of Atomic Nuclei} \textbf{\bibinfo{volume}{70}},
  \bibinfo{pages}{1440} (\bibinfo{year}{2007}).

\bibitem[{\citenamefont{Ali et~al.}(1985)\citenamefont{Ali, Ahmad, and
  Ferdous}}]{RevModPhys.57.923}
\bibinfo{author}{\bibfnamefont{S.}~\bibnamefont{Ali}},
  \bibinfo{author}{\bibfnamefont{A.~A.~Z.} \bibnamefont{Ahmad}},
  \bibnamefont{and} \bibinfo{author}{\bibfnamefont{N.}~\bibnamefont{Ferdous}},
  \bibinfo{journal}{Rev. Mod. Phys.} \textbf{\bibinfo{volume}{57}},
  \bibinfo{pages}{923} (\bibinfo{year}{1985}).

\bibitem[{\citenamefont{Skyrme}(1959)}]{Skyrme:1959zz}
\bibinfo{author}{\bibfnamefont{T.}~\bibnamefont{Skyrme}},
  \bibinfo{journal}{Nucl. Phys.} \textbf{\bibinfo{volume}{9}},
  \bibinfo{pages}{615} (\bibinfo{year}{1959}).

\bibitem[{\citenamefont{Chabanat et~al.}(1997)\citenamefont{Chabanat, Meyer,
  Bonche, Schaeffer, and Haensel}}]{Chabanat:1997qh}
\bibinfo{author}{\bibfnamefont{E.}~\bibnamefont{Chabanat}},
  \bibinfo{author}{\bibfnamefont{J.}~\bibnamefont{Meyer}},
  \bibinfo{author}{\bibfnamefont{P.}~\bibnamefont{Bonche}},
  \bibinfo{author}{\bibfnamefont{R.}~\bibnamefont{Schaeffer}},
  \bibnamefont{and} \bibinfo{author}{\bibfnamefont{P.}~\bibnamefont{Haensel}},
  \bibinfo{journal}{Nucl. Phys.} \textbf{\bibinfo{volume}{A627}},
  \bibinfo{pages}{710} (\bibinfo{year}{1997}).

\bibitem[{\citenamefont{Zalewski et~al.}(2008)\citenamefont{Zalewski,
  Dobaczewski, Satula, and Werner}}]{Zalewski:2008is}
\bibinfo{author}{\bibfnamefont{M.}~\bibnamefont{Zalewski}},
  \bibinfo{author}{\bibfnamefont{J.}~\bibnamefont{Dobaczewski}},
  \bibinfo{author}{\bibfnamefont{W.}~\bibnamefont{Satula}}, \bibnamefont{and}
  \bibinfo{author}{\bibfnamefont{T.~R.} \bibnamefont{Werner}},
  \bibinfo{journal}{Phys. Rev.} \textbf{\bibinfo{volume}{C77}},
  \bibinfo{pages}{024316} (\bibinfo{year}{2008}), \eprint{0801.0924}.

\bibitem[{\citenamefont{Stoks et~al.}(1993)\citenamefont{Stoks, Kompl,
  Rentmeester, and de~Swart}}]{Stoks:1993tb}
\bibinfo{author}{\bibfnamefont{V.~G.~J.} \bibnamefont{Stoks}},
  \bibinfo{author}{\bibfnamefont{R.~A.~M.} \bibnamefont{Kompl}},
  \bibinfo{author}{\bibfnamefont{M.~C.~M.} \bibnamefont{Rentmeester}},
  \bibnamefont{and} \bibinfo{author}{\bibfnamefont{J.~J.}
  \bibnamefont{de~Swart}}, \bibinfo{journal}{Phys. Rev.}
  \textbf{\bibinfo{volume}{C48}}, \bibinfo{pages}{792} (\bibinfo{year}{1993}).

\bibitem[{\citenamefont{Stoks et~al.}(1994)\citenamefont{Stoks, Klomp,
  Terheggen, and de~Swart}}]{Stoks:1994wp}
\bibinfo{author}{\bibfnamefont{V.~G.~J.} \bibnamefont{Stoks}},
  \bibinfo{author}{\bibfnamefont{R.~A.~M.} \bibnamefont{Klomp}},
  \bibinfo{author}{\bibfnamefont{C.~P.~F.} \bibnamefont{Terheggen}},
  \bibnamefont{and} \bibinfo{author}{\bibfnamefont{J.~J.}
  \bibnamefont{de~Swart}}, \bibinfo{journal}{Phys. Rev.}
  \textbf{\bibinfo{volume}{C49}}, \bibinfo{pages}{2950} (\bibinfo{year}{1994}),
  \eprint{nucl-th/9406039}.

\bibitem[{\citenamefont{Ishii et~al.}(2007)\citenamefont{Ishii, Aoki, and
  Hatsuda}}]{Ishii:2006ec}
\bibinfo{author}{\bibfnamefont{N.}~\bibnamefont{Ishii}},
  \bibinfo{author}{\bibfnamefont{S.}~\bibnamefont{Aoki}}, \bibnamefont{and}
  \bibinfo{author}{\bibfnamefont{T.}~\bibnamefont{Hatsuda}},
  \bibinfo{journal}{Phys. Rev. Lett.} \textbf{\bibinfo{volume}{99}},
  \bibinfo{pages}{022001} (\bibinfo{year}{2007}), \eprint{nucl-th/0611096}.

\bibitem[{\citenamefont{Beane et~al.}(2006)\citenamefont{Beane, Bedaque,
  Orginos, and Savage}}]{Beane:2006mx}
\bibinfo{author}{\bibfnamefont{S.~R.} \bibnamefont{Beane}},
  \bibinfo{author}{\bibfnamefont{P.~F.} \bibnamefont{Bedaque}},
  \bibinfo{author}{\bibfnamefont{K.}~\bibnamefont{Orginos}}, \bibnamefont{and}
  \bibinfo{author}{\bibfnamefont{M.~J.} \bibnamefont{Savage}},
  \bibinfo{journal}{Phys. Rev. Lett.} \textbf{\bibinfo{volume}{97}},
  \bibinfo{pages}{012001} (\bibinfo{year}{2006}), \eprint{hep-lat/0602010}.

\bibitem[{\citenamefont{Calle~Cordon and
  Ruiz~Arriola}(2008{\natexlab{a}})}]{CalleCordon:2008cz}
\bibinfo{author}{\bibfnamefont{A.}~\bibnamefont{Calle~Cordon}}
  \bibnamefont{and}
  \bibinfo{author}{\bibfnamefont{E.}~\bibnamefont{Ruiz~Arriola}},
  \bibinfo{journal}{Phys. Rev.} \textbf{\bibinfo{volume}{C78}},
  \bibinfo{pages}{054002} (\bibinfo{year}{2008}{\natexlab{a}}),
  \eprint{0807.2918}.

\bibitem[{\citenamefont{Wigner}(1937)}]{Wigner:1936dx}
\bibinfo{author}{\bibfnamefont{E.}~\bibnamefont{Wigner}},
  \bibinfo{journal}{Phys. Rev.} \textbf{\bibinfo{volume}{51}},
  \bibinfo{pages}{106} (\bibinfo{year}{1937}).

\bibitem[{\citenamefont{{Hund}}(1937)}]{1937ZPhy..105..202H}
\bibinfo{author}{\bibfnamefont{F.}~\bibnamefont{{Hund}}},
  \bibinfo{journal}{Zeitschrift fur Physik} \textbf{\bibinfo{volume}{105}},
  \bibinfo{pages}{202} (\bibinfo{year}{1937}).

\bibitem[{\citenamefont{'t~Hooft}(1974)}]{'tHooft:1973jz}
\bibinfo{author}{\bibfnamefont{G.}~\bibnamefont{'t~Hooft}},
  \bibinfo{journal}{Nucl. Phys.} \textbf{\bibinfo{volume}{B72}},
  \bibinfo{pages}{461} (\bibinfo{year}{1974}).

\bibitem[{\citenamefont{Witten}(1979)}]{Witten:1979kh}
\bibinfo{author}{\bibfnamefont{E.}~\bibnamefont{Witten}},
  \bibinfo{journal}{Nucl. Phys.} \textbf{\bibinfo{volume}{B160}},
  \bibinfo{pages}{57} (\bibinfo{year}{1979}).

\bibitem[{\citenamefont{Manohar}(1998)}]{Manohar:1998xv}
\bibinfo{author}{\bibfnamefont{A.~V.} \bibnamefont{Manohar}}
  (\bibinfo{year}{1998}), \eprint{hep-ph/9802419}.

\bibitem[{\citenamefont{Jenkins}(1998)}]{Jenkins:1998wy}
\bibinfo{author}{\bibfnamefont{E.~E.} \bibnamefont{Jenkins}},
  \bibinfo{journal}{Ann. Rev. Nucl. Part. Sci.} \textbf{\bibinfo{volume}{48}},
  \bibinfo{pages}{81} (\bibinfo{year}{1998}), \eprint{hep-ph/9803349}.

\bibitem[{\citenamefont{Lebed}(1999)}]{Lebed:1998st}
\bibinfo{author}{\bibfnamefont{R.~F.} \bibnamefont{Lebed}},
  \bibinfo{journal}{Czech. J. Phys.} \textbf{\bibinfo{volume}{49}},
  \bibinfo{pages}{1273} (\bibinfo{year}{1999}), \eprint{nucl-th/9810080}.

\bibitem[{\citenamefont{Kaplan and Savage}(1996)}]{Kaplan:1995yg}
\bibinfo{author}{\bibfnamefont{D.~B.} \bibnamefont{Kaplan}} \bibnamefont{and}
  \bibinfo{author}{\bibfnamefont{M.~J.} \bibnamefont{Savage}},
  \bibinfo{journal}{Phys. Lett.} \textbf{\bibinfo{volume}{B365}},
  \bibinfo{pages}{244} (\bibinfo{year}{1996}), \eprint{hep-ph/9509371}.

\bibitem[{\citenamefont{Kaplan and Manohar}(1997)}]{Kaplan:1996rk}
\bibinfo{author}{\bibfnamefont{D.~B.} \bibnamefont{Kaplan}} \bibnamefont{and}
  \bibinfo{author}{\bibfnamefont{A.~V.} \bibnamefont{Manohar}},
  \bibinfo{journal}{Phys. Rev.} \textbf{\bibinfo{volume}{C56}},
  \bibinfo{pages}{76} (\bibinfo{year}{1997}), \eprint{nucl-th/9612021}.

\bibitem[{\citenamefont{Banerjee et~al.}(2002)\citenamefont{Banerjee, Cohen,
  and Gelman}}]{Banerjee:2001js}
\bibinfo{author}{\bibfnamefont{M.~K.} \bibnamefont{Banerjee}},
  \bibinfo{author}{\bibfnamefont{T.~D.} \bibnamefont{Cohen}}, \bibnamefont{and}
  \bibinfo{author}{\bibfnamefont{B.~A.} \bibnamefont{Gelman}},
  \bibinfo{journal}{Phys. Rev.} \textbf{\bibinfo{volume}{C65}},
  \bibinfo{pages}{034011} (\bibinfo{year}{2002}), \eprint{hep-ph/0109274}.

\bibitem[{\citenamefont{Caprini et~al.}(2006)\citenamefont{Caprini, Colangelo,
  and Leutwyler}}]{Caprini:2005zr}
\bibinfo{author}{\bibfnamefont{I.}~\bibnamefont{Caprini}},
  \bibinfo{author}{\bibfnamefont{G.}~\bibnamefont{Colangelo}},
  \bibnamefont{and}
  \bibinfo{author}{\bibfnamefont{H.}~\bibnamefont{Leutwyler}},
  \bibinfo{journal}{Phys. Rev. Lett.} \textbf{\bibinfo{volume}{96}},
  \bibinfo{pages}{132001} (\bibinfo{year}{2006}), \eprint{hep-ph/0512364}.

\bibitem[{\citenamefont{Amsler et~al.}(2008)}]{Amsler:2008zz}
\bibinfo{author}{\bibfnamefont{C.}~\bibnamefont{Amsler}} \bibnamefont{et~al.}
  (\bibinfo{collaboration}{Particle Data Group}), \bibinfo{journal}{Phys.
  Lett.} \textbf{\bibinfo{volume}{B667}}, \bibinfo{pages}{1}
  (\bibinfo{year}{2008}).

\bibitem[{\citenamefont{Wiringa et~al.}(1995)\citenamefont{Wiringa, Stoks, and
  Schiavilla}}]{Wiringa:1994wb}
\bibinfo{author}{\bibfnamefont{R.~B.} \bibnamefont{Wiringa}},
  \bibinfo{author}{\bibfnamefont{V.~G.~J.} \bibnamefont{Stoks}},
  \bibnamefont{and}
  \bibinfo{author}{\bibfnamefont{R.}~\bibnamefont{Schiavilla}},
  \bibinfo{journal}{Phys. Rev.} \textbf{\bibinfo{volume}{C51}},
  \bibinfo{pages}{38} (\bibinfo{year}{1995}), \eprint{nucl-th/9408016}.

\bibitem[{\citenamefont{Bogner et~al.}(2003)\citenamefont{Bogner, Kuo, and
  Schwenk}}]{Bogner:2003wn}
\bibinfo{author}{\bibfnamefont{S.~K.} \bibnamefont{Bogner}},
  \bibinfo{author}{\bibfnamefont{T.~T.~S.} \bibnamefont{Kuo}},
  \bibnamefont{and} \bibinfo{author}{\bibfnamefont{A.}~\bibnamefont{Schwenk}},
  \bibinfo{journal}{Phys. Rept.} \textbf{\bibinfo{volume}{386}},
  \bibinfo{pages}{1} (\bibinfo{year}{2003}), \eprint{nucl-th/0305035}.

\bibitem[{\citenamefont{Holt et~al.}(2004)\citenamefont{Holt, Kuo, Brown, and
  Bogner}}]{Holt:2003rj}
\bibinfo{author}{\bibfnamefont{J.~D.} \bibnamefont{Holt}},
  \bibinfo{author}{\bibfnamefont{T.~T.~S.} \bibnamefont{Kuo}},
  \bibinfo{author}{\bibfnamefont{G.~E.} \bibnamefont{Brown}}, \bibnamefont{and}
  \bibinfo{author}{\bibfnamefont{S.~K.} \bibnamefont{Bogner}},
  \bibinfo{journal}{Nucl. Phys.} \textbf{\bibinfo{volume}{A733}},
  \bibinfo{pages}{153} (\bibinfo{year}{2004}), \eprint{nucl-th/0308036}.

\bibitem[{\citenamefont{Entem et~al.}(2007)\citenamefont{Entem, Ruiz~Arriola,
  Pavon~Valderrama, and Machleidt}}]{Entem:2007jg}
\bibinfo{author}{\bibfnamefont{D.~R.} \bibnamefont{Entem}},
  \bibinfo{author}{\bibfnamefont{E.}~\bibnamefont{Ruiz~Arriola}},
  \bibinfo{author}{\bibfnamefont{M.}~\bibnamefont{Pavon~Valderrama}},
  \bibnamefont{and} \bibinfo{author}{\bibfnamefont{R.}~\bibnamefont{Machleidt}}
  (\bibinfo{year}{2007}), \eprint{0709.2770}.

\bibitem[{\citenamefont{Kaiser et~al.}(1998)\citenamefont{Kaiser,
  Gerstendorfer, and Weise}}]{Kaiser:1998wa}
\bibinfo{author}{\bibfnamefont{N.}~\bibnamefont{Kaiser}},
  \bibinfo{author}{\bibfnamefont{S.}~\bibnamefont{Gerstendorfer}},
  \bibnamefont{and} \bibinfo{author}{\bibfnamefont{W.}~\bibnamefont{Weise}},
  \bibinfo{journal}{Nucl. Phys.} \textbf{\bibinfo{volume}{A637}},
  \bibinfo{pages}{395} (\bibinfo{year}{1998}), \eprint{nucl-th/9802071}.

\bibitem[{\citenamefont{Rentmeester et~al.}(1999)\citenamefont{Rentmeester,
  Timmermans, Friar, and de~Swart}}]{Rentmeester:1999vw}
\bibinfo{author}{\bibfnamefont{M.~C.~M.} \bibnamefont{Rentmeester}},
  \bibinfo{author}{\bibfnamefont{R.~G.~E.} \bibnamefont{Timmermans}},
  \bibinfo{author}{\bibfnamefont{J.~L.} \bibnamefont{Friar}}, \bibnamefont{and}
  \bibinfo{author}{\bibfnamefont{J.~J.} \bibnamefont{de~Swart}},
  \bibinfo{journal}{Phys. Rev. Lett.} \textbf{\bibinfo{volume}{82}},
  \bibinfo{pages}{4992} (\bibinfo{year}{1999}), \eprint{nucl-th/9901054}.

\bibitem[{\citenamefont{Entem and Machleidt}(2003)}]{Entem:2003ft}
\bibinfo{author}{\bibfnamefont{D.~R.} \bibnamefont{Entem}} \bibnamefont{and}
  \bibinfo{author}{\bibfnamefont{R.}~\bibnamefont{Machleidt}},
  \bibinfo{journal}{Phys. Rev.} \textbf{\bibinfo{volume}{C68}},
  \bibinfo{pages}{041001} (\bibinfo{year}{2003}), \eprint{nucl-th/0304018}.

\bibitem[{\citenamefont{Bogner et~al.}(2007)\citenamefont{Bogner, Furnstahl,
  Ramanan, and Schwenk}}]{Bogner:2006vp}
\bibinfo{author}{\bibfnamefont{S.~K.} \bibnamefont{Bogner}},
  \bibinfo{author}{\bibfnamefont{R.~J.} \bibnamefont{Furnstahl}},
  \bibinfo{author}{\bibfnamefont{S.}~\bibnamefont{Ramanan}}, \bibnamefont{and}
  \bibinfo{author}{\bibfnamefont{A.}~\bibnamefont{Schwenk}},
  \bibinfo{journal}{Nucl. Phys.} \textbf{\bibinfo{volume}{A784}},
  \bibinfo{pages}{79} (\bibinfo{year}{2007}), \eprint{nucl-th/0609003}.

\bibitem[{\citenamefont{Pavon~Valderrama and
  Ruiz~Arriola}(2009)}]{PavonWSchiral}
\bibinfo{author}{\bibfnamefont{M.}~\bibnamefont{Pavon~Valderrama}}
  \bibnamefont{and}
  \bibinfo{author}{\bibfnamefont{E.}~\bibnamefont{Ruiz~Arriola}}
  (\bibinfo{year}{2009}), \eprint{In preparation}.

\bibitem[{\citenamefont{Machleidt et~al.}(1987)\citenamefont{Machleidt,
  Holinde, and Elster}}]{Machleidt:1987hj}
\bibinfo{author}{\bibfnamefont{R.}~\bibnamefont{Machleidt}},
  \bibinfo{author}{\bibfnamefont{K.}~\bibnamefont{Holinde}}, \bibnamefont{and}
  \bibinfo{author}{\bibfnamefont{C.}~\bibnamefont{Elster}},
  \bibinfo{journal}{Phys. Rept.} \textbf{\bibinfo{volume}{149}},
  \bibinfo{pages}{1} (\bibinfo{year}{1987}).

\bibitem[{\citenamefont{Ecker et~al.}(1989)\citenamefont{Ecker, Gasser, Pich,
  and de~Rafael}}]{Ecker:1988te}
\bibinfo{author}{\bibfnamefont{G.}~\bibnamefont{Ecker}},
  \bibinfo{author}{\bibfnamefont{J.}~\bibnamefont{Gasser}},
  \bibinfo{author}{\bibfnamefont{A.}~\bibnamefont{Pich}}, \bibnamefont{and}
  \bibinfo{author}{\bibfnamefont{E.}~\bibnamefont{de~Rafael}},
  \bibinfo{journal}{Nucl. Phys.} \textbf{\bibinfo{volume}{B321}},
  \bibinfo{pages}{311} (\bibinfo{year}{1989}).

\bibitem[{\citenamefont{Epelbaum et~al.}(2002)\citenamefont{Epelbaum, Meissner,
  Gloeckle, and Elster}}]{Epelbaum:2001fm}
\bibinfo{author}{\bibfnamefont{E.}~\bibnamefont{Epelbaum}},
  \bibinfo{author}{\bibfnamefont{U.~G.} \bibnamefont{Meissner}},
  \bibinfo{author}{\bibfnamefont{W.}~\bibnamefont{Gloeckle}}, \bibnamefont{and}
  \bibinfo{author}{\bibfnamefont{C.}~\bibnamefont{Elster}},
  \bibinfo{journal}{Phys. Rev.} \textbf{\bibinfo{volume}{C65}},
  \bibinfo{pages}{044001} (\bibinfo{year}{2002}), \eprint{nucl-th/0106007}.

\bibitem[{\citenamefont{Epelbaum et~al.}(2004)\citenamefont{Epelbaum, Gloeckle,
  and Meissner}}]{Epelbaum:2003xx}
\bibinfo{author}{\bibfnamefont{E.}~\bibnamefont{Epelbaum}},
  \bibinfo{author}{\bibfnamefont{W.}~\bibnamefont{Gloeckle}}, \bibnamefont{and}
  \bibinfo{author}{\bibfnamefont{U.-G.} \bibnamefont{Meissner}},
  \bibinfo{journal}{Eur. Phys. J.} \textbf{\bibinfo{volume}{A19}},
  \bibinfo{pages}{401} (\bibinfo{year}{2004}), \eprint{nucl-th/0308010}.

\bibitem[{\citenamefont{Machleidt}(2001)}]{Machleidt:2000ge}
\bibinfo{author}{\bibfnamefont{R.}~\bibnamefont{Machleidt}},
  \bibinfo{journal}{Phys. Rev.} \textbf{\bibinfo{volume}{C63}},
  \bibinfo{pages}{024001} (\bibinfo{year}{2001}), \eprint{nucl-th/0006014}.

\bibitem[{\citenamefont{Belitsky and Cohen}(2002)}]{Belitsky:2002ni}
\bibinfo{author}{\bibfnamefont{A.~V.} \bibnamefont{Belitsky}} \bibnamefont{and}
  \bibinfo{author}{\bibfnamefont{T.~D.} \bibnamefont{Cohen}},
  \bibinfo{journal}{Phys. Rev.} \textbf{\bibinfo{volume}{C65}},
  \bibinfo{pages}{064008} (\bibinfo{year}{2002}), \eprint{hep-ph/0202153}.

\bibitem[{\citenamefont{Cohen}(2002)}]{Cohen:2002im}
\bibinfo{author}{\bibfnamefont{T.~D.} \bibnamefont{Cohen}},
  \bibinfo{journal}{Phys. Rev.} \textbf{\bibinfo{volume}{C66}},
  \bibinfo{pages}{064003} (\bibinfo{year}{2002}), \eprint{nucl-th/0209072}.

\bibitem[{\citenamefont{Weinberg}(1990)}]{Weinberg:1990xn}
\bibinfo{author}{\bibfnamefont{S.}~\bibnamefont{Weinberg}},
  \bibinfo{journal}{Phys. Rev. Lett.} \textbf{\bibinfo{volume}{65}},
  \bibinfo{pages}{1177} (\bibinfo{year}{1990}).

\bibitem[{\citenamefont{Svec}(1997)}]{Svec:1996xp}
\bibinfo{author}{\bibfnamefont{M.}~\bibnamefont{Svec}}, \bibinfo{journal}{Phys.
  Rev.} \textbf{\bibinfo{volume}{D55}}, \bibinfo{pages}{5727}
  (\bibinfo{year}{1997}), \eprint{hep-ph/9607297}.

\bibitem[{\citenamefont{Megias et~al.}(2004)\citenamefont{Megias, Ruiz~Arriola,
  Salcedo, and Broniowski}}]{Megias:2004uj}
\bibinfo{author}{\bibfnamefont{E.}~\bibnamefont{Megias}},
  \bibinfo{author}{\bibfnamefont{E.}~\bibnamefont{Ruiz~Arriola}},
  \bibinfo{author}{\bibfnamefont{L.~L.} \bibnamefont{Salcedo}},
  \bibnamefont{and}
  \bibinfo{author}{\bibfnamefont{W.}~\bibnamefont{Broniowski}},
  \bibinfo{journal}{Phys. Rev.} \textbf{\bibinfo{volume}{D70}},
  \bibinfo{pages}{034031} (\bibinfo{year}{2004}), \eprint{hep-ph/0403139}.

\bibitem[{\citenamefont{Oller et~al.}(1998)\citenamefont{Oller, Oset, and
  Pelaez}}]{Oller:1997ng}
\bibinfo{author}{\bibfnamefont{J.~A.} \bibnamefont{Oller}},
  \bibinfo{author}{\bibfnamefont{E.}~\bibnamefont{Oset}}, \bibnamefont{and}
  \bibinfo{author}{\bibfnamefont{J.~R.} \bibnamefont{Pelaez}},
  \bibinfo{journal}{Phys. Rev. Lett.} \textbf{\bibinfo{volume}{80}},
  \bibinfo{pages}{3452} (\bibinfo{year}{1998}), \eprint{hep-ph/9803242}.

\bibitem[{\citenamefont{Nieves and Ruiz~Arriola}(2000)}]{Nieves:1999bx}
\bibinfo{author}{\bibfnamefont{J.}~\bibnamefont{Nieves}} \bibnamefont{and}
  \bibinfo{author}{\bibfnamefont{E.}~\bibnamefont{Ruiz~Arriola}},
  \bibinfo{journal}{Nucl. Phys.} \textbf{\bibinfo{volume}{A679}},
  \bibinfo{pages}{57} (\bibinfo{year}{2000}), \eprint{hep-ph/9907469}.

\bibitem[{\citenamefont{Nieves et~al.}(2002)\citenamefont{Nieves,
  Pavon~Valderrama, and Ruiz~Arriola}}]{Nieves:2001de}
\bibinfo{author}{\bibfnamefont{J.}~\bibnamefont{Nieves}},
  \bibinfo{author}{\bibfnamefont{M.}~\bibnamefont{Pavon~Valderrama}},
  \bibnamefont{and}
  \bibinfo{author}{\bibfnamefont{E.}~\bibnamefont{Ruiz~Arriola}},
  \bibinfo{journal}{Phys. Rev.} \textbf{\bibinfo{volume}{D65}},
  \bibinfo{pages}{036002} (\bibinfo{year}{2002}), \eprint{hep-ph/0109077}.

\bibitem[{\citenamefont{Partovi and Lomon}(1970)}]{Partovi:1969wd}
\bibinfo{author}{\bibfnamefont{M.~H.} \bibnamefont{Partovi}} \bibnamefont{and}
  \bibinfo{author}{\bibfnamefont{E.~L.} \bibnamefont{Lomon}},
  \bibinfo{journal}{Phys. Rev.} \textbf{\bibinfo{volume}{D2}},
  \bibinfo{pages}{1999} (\bibinfo{year}{1970}).

\bibitem[{\citenamefont{Lin and Serot}(1990)}]{Lin:1990cx}
\bibinfo{author}{\bibfnamefont{W.}~\bibnamefont{Lin}} \bibnamefont{and}
  \bibinfo{author}{\bibfnamefont{B.~D.} \bibnamefont{Serot}},
  \bibinfo{journal}{Nucl. Phys.} \textbf{\bibinfo{volume}{A512}},
  \bibinfo{pages}{637} (\bibinfo{year}{1990}).

\bibitem[{\citenamefont{Kim et~al.}(1994)\citenamefont{Kim, Durso, and
  Holinde}}]{Kim:1994ce}
\bibinfo{author}{\bibfnamefont{H.-C.} \bibnamefont{Kim}},
  \bibinfo{author}{\bibfnamefont{J.~W.} \bibnamefont{Durso}}, \bibnamefont{and}
  \bibinfo{author}{\bibfnamefont{K.}~\bibnamefont{Holinde}},
  \bibinfo{journal}{Phys. Rev.} \textbf{\bibinfo{volume}{C49}},
  \bibinfo{pages}{2355} (\bibinfo{year}{1994}).

\bibitem[{\citenamefont{Oset et~al.}(2000)\citenamefont{Oset, Toki, Mizobe, and
  Takahashi}}]{Oset:2000gn}
\bibinfo{author}{\bibfnamefont{E.}~\bibnamefont{Oset}},
  \bibinfo{author}{\bibfnamefont{H.}~\bibnamefont{Toki}},
  \bibinfo{author}{\bibfnamefont{M.}~\bibnamefont{Mizobe}}, \bibnamefont{and}
  \bibinfo{author}{\bibfnamefont{T.~T.} \bibnamefont{Takahashi}},
  \bibinfo{journal}{Prog. Theor. Phys.} \textbf{\bibinfo{volume}{103}},
  \bibinfo{pages}{351} (\bibinfo{year}{2000}), \eprint{nucl-th/0011008}.

\bibitem[{\citenamefont{Kaskulov and Clement}(2004)}]{Kaskulov:2004kr}
\bibinfo{author}{\bibfnamefont{M.~M.} \bibnamefont{Kaskulov}} \bibnamefont{and}
  \bibinfo{author}{\bibfnamefont{H.}~\bibnamefont{Clement}},
  \bibinfo{journal}{Phys. Rev.} \textbf{\bibinfo{volume}{C70}},
  \bibinfo{pages}{014002} (\bibinfo{year}{2004}), \eprint{nucl-th/0401061}.

\bibitem[{\citenamefont{Donoghue}(2006)}]{Donoghue:2006rg}
\bibinfo{author}{\bibfnamefont{J.~F.} \bibnamefont{Donoghue}},
  \bibinfo{journal}{Phys. Lett.} \textbf{\bibinfo{volume}{B643}},
  \bibinfo{pages}{165} (\bibinfo{year}{2006}), \eprint{nucl-th/0602074}.

\bibitem[{\citenamefont{Stoks and Rijken}(1997)}]{Stoks:1996yj}
\bibinfo{author}{\bibfnamefont{V.~G.~J.} \bibnamefont{Stoks}} \bibnamefont{and}
  \bibinfo{author}{\bibfnamefont{T.~A.} \bibnamefont{Rijken}},
  \bibinfo{journal}{Nucl. Phys.} \textbf{\bibinfo{volume}{A613}},
  \bibinfo{pages}{311} (\bibinfo{year}{1997}), \eprint{nucl-th/9611002}.

\bibitem[{\citenamefont{Furnstahl et~al.}(1997)\citenamefont{Furnstahl, Serot,
  and Tang}}]{Furnstahl:1996wv}
\bibinfo{author}{\bibfnamefont{R.~J.} \bibnamefont{Furnstahl}},
  \bibinfo{author}{\bibfnamefont{B.~D.} \bibnamefont{Serot}}, \bibnamefont{and}
  \bibinfo{author}{\bibfnamefont{H.-B.} \bibnamefont{Tang}},
  \bibinfo{journal}{Nucl. Phys.} \textbf{\bibinfo{volume}{A615}},
  \bibinfo{pages}{441} (\bibinfo{year}{1997}), \eprint{nucl-th/9608035}.

\bibitem[{\citenamefont{Papazoglou et~al.}(1999)}]{Papazoglou:1998vr}
\bibinfo{author}{\bibfnamefont{P.}~\bibnamefont{Papazoglou}}
  \bibnamefont{et~al.}, \bibinfo{journal}{Phys. Rev.}
  \textbf{\bibinfo{volume}{C59}}, \bibinfo{pages}{411} (\bibinfo{year}{1999}),
  \eprint{nucl-th/9806087}.

\bibitem[{\citenamefont{Binstock and Bryan}(1971)}]{Binstock:1972gx}
\bibinfo{author}{\bibfnamefont{J.}~\bibnamefont{Binstock}} \bibnamefont{and}
  \bibinfo{author}{\bibfnamefont{R.}~\bibnamefont{Bryan}},
  \bibinfo{journal}{Phys. Rev.} \textbf{\bibinfo{volume}{D4}},
  \bibinfo{pages}{1341} (\bibinfo{year}{1971}).

\bibitem[{\citenamefont{Ericson and Weise}(1988)}]{Ericson:1988gk}
\bibinfo{author}{\bibfnamefont{T.~E.~O.} \bibnamefont{Ericson}}
  \bibnamefont{and} \bibinfo{author}{\bibfnamefont{W.}~\bibnamefont{Weise}},
  \emph{\bibinfo{title}{Pions and Nuclei}} (\bibinfo{publisher}{Oxford, UK:
  Clarendon (1988)}, \bibinfo{year}{1988}).

\bibitem[{\citenamefont{Johnson and Teller}(1955)}]{PhysRev.98.783}
\bibinfo{author}{\bibfnamefont{M.~H.} \bibnamefont{Johnson}} \bibnamefont{and}
  \bibinfo{author}{\bibfnamefont{E.}~\bibnamefont{Teller}},
  \bibinfo{journal}{Phys. Rev.} \textbf{\bibinfo{volume}{98}},
  \bibinfo{pages}{783} (\bibinfo{year}{1955}).

\bibitem[{\citenamefont{Tornqvist and Roos}(1996)}]{Tornqvist:1995ay}
\bibinfo{author}{\bibfnamefont{N.~A.} \bibnamefont{Tornqvist}}
  \bibnamefont{and} \bibinfo{author}{\bibfnamefont{M.}~\bibnamefont{Roos}},
  \bibinfo{journal}{Phys. Rev. Lett.} \textbf{\bibinfo{volume}{76}},
  \bibinfo{pages}{1575} (\bibinfo{year}{1996}), \eprint{hep-ph/9511210}.

\bibitem[{\citenamefont{Yao et~al.}(2006)}]{Yao:2006px}
\bibinfo{author}{\bibfnamefont{W.~M.} \bibnamefont{Yao}} \bibnamefont{et~al.}
  (\bibinfo{collaboration}{Particle Data Group}), \bibinfo{journal}{J. Phys.}
  \textbf{\bibinfo{volume}{G33}}, \bibinfo{pages}{1} (\bibinfo{year}{2006}).

\bibitem[{\citenamefont{van Beveren et~al.}(2002)\citenamefont{van Beveren,
  Kleefeld, Rupp, and Scadron}}]{vanBeveren:2002mc}
\bibinfo{author}{\bibfnamefont{E.}~\bibnamefont{van Beveren}},
  \bibinfo{author}{\bibfnamefont{F.}~\bibnamefont{Kleefeld}},
  \bibinfo{author}{\bibfnamefont{G.}~\bibnamefont{Rupp}}, \bibnamefont{and}
  \bibinfo{author}{\bibfnamefont{M.~D.} \bibnamefont{Scadron}},
  \bibinfo{journal}{Mod. Phys. Lett.} \textbf{\bibinfo{volume}{A17}},
  \bibinfo{pages}{1673} (\bibinfo{year}{2002}), \eprint{hep-ph/0204139}.

\bibitem[{\citenamefont{Calle~Cordon and
  Ruiz~Arriola}(2008{\natexlab{b}})}]{CalleCordon:2008eu}
\bibinfo{author}{\bibfnamefont{A.}~\bibnamefont{Calle~Cordon}}
  \bibnamefont{and}
  \bibinfo{author}{\bibfnamefont{E.}~\bibnamefont{Ruiz~Arriola}},
  \bibinfo{journal}{AIP Conf. Proc.} \textbf{\bibinfo{volume}{1030}},
  \bibinfo{pages}{334} (\bibinfo{year}{2008}{\natexlab{b}}),
  \eprint{0804.2350}.

\bibitem[{\citenamefont{Pelaez}(2004)}]{Pelaez:2003dy}
\bibinfo{author}{\bibfnamefont{J.~R.} \bibnamefont{Pelaez}},
  \bibinfo{journal}{Phys. Rev. Lett.} \textbf{\bibinfo{volume}{92}},
  \bibinfo{pages}{102001} (\bibinfo{year}{2004}), \eprint{hep-ph/0309292}.

\bibitem[{\citenamefont{Pelaez and Rios}(2006)}]{Pelaez:2006nj}
\bibinfo{author}{\bibfnamefont{J.~R.} \bibnamefont{Pelaez}} \bibnamefont{and}
  \bibinfo{author}{\bibfnamefont{G.}~\bibnamefont{Rios}},
  \bibinfo{journal}{Phys. Rev. Lett.} \textbf{\bibinfo{volume}{97}},
  \bibinfo{pages}{242002} (\bibinfo{year}{2006}), \eprint{hep-ph/0610397}.

\bibitem[{\citenamefont{Nieves and Ruiz~Arriola}(2009)}]{JuanEnrique}
\bibinfo{author}{\bibfnamefont{J.}~\bibnamefont{Nieves}} \bibnamefont{and}
  \bibinfo{author}{\bibfnamefont{E.}~\bibnamefont{Ruiz~Arriola}}
  (\bibinfo{year}{2009}), \eprint{In preparation}.

\bibitem[{\citenamefont{Flambaum and Shuryak}(2007)}]{Flambaum:2007xj}
\bibinfo{author}{\bibfnamefont{V.~V.} \bibnamefont{Flambaum}} \bibnamefont{and}
  \bibinfo{author}{\bibfnamefont{E.~V.} \bibnamefont{Shuryak}},
  \bibinfo{journal}{Phys. Rev.} \textbf{\bibinfo{volume}{C76}},
  \bibinfo{pages}{065206} (\bibinfo{year}{2007}), \eprint{nucl-th/0702038}.

\bibitem[{\citenamefont{Yndurain et~al.}(2007)\citenamefont{Yndurain,
  Garcia-Martin, and Pelaez}}]{Yndurain:2007qm}
\bibinfo{author}{\bibfnamefont{F.~J.} \bibnamefont{Yndurain}},
  \bibinfo{author}{\bibfnamefont{R.}~\bibnamefont{Garcia-Martin}},
  \bibnamefont{and} \bibinfo{author}{\bibfnamefont{J.~R.}
  \bibnamefont{Pelaez}}, \bibinfo{journal}{Phys. Rev.}
  \textbf{\bibinfo{volume}{D76}}, \bibinfo{pages}{074034}
  (\bibinfo{year}{2007}), \eprint{hep-ph/0701025}.

\bibitem[{\citenamefont{Caprini}(2008)}]{Caprini:2008fc}
\bibinfo{author}{\bibfnamefont{I.}~\bibnamefont{Caprini}},
  \bibinfo{journal}{Phys. Rev.} \textbf{\bibinfo{volume}{D77}},
  \bibinfo{pages}{114019} (\bibinfo{year}{2008}), \eprint{0804.3504}.

\bibitem[{\citenamefont{Weinberg}(1969)}]{Weinberg:1969hw}
\bibinfo{author}{\bibfnamefont{S.}~\bibnamefont{Weinberg}},
  \bibinfo{journal}{Phys. Rev.} \textbf{\bibinfo{volume}{177}},
  \bibinfo{pages}{2604} (\bibinfo{year}{1969}).

\bibitem[{\citenamefont{Weinberg}(1994)}]{Weinberg:1994tu}
\bibinfo{author}{\bibfnamefont{S.}~\bibnamefont{Weinberg}}
  (\bibinfo{year}{1994}), \eprint{hep-ph/9412326}.

\bibitem[{\citenamefont{Protopopescu et~al.}(1973)}]{Protopopescu:1973sh}
\bibinfo{author}{\bibfnamefont{S.~D.} \bibnamefont{Protopopescu}}
  \bibnamefont{et~al.}, \bibinfo{journal}{Phys. Rev.}
  \textbf{\bibinfo{volume}{D7}}, \bibinfo{pages}{1279} (\bibinfo{year}{1973}).

\bibitem[{\citenamefont{Hyams et~al.}(1973)}]{Hyams:1973zf}
\bibinfo{author}{\bibfnamefont{B.}~\bibnamefont{Hyams}} \bibnamefont{et~al.},
  \bibinfo{journal}{Nucl. Phys.} \textbf{\bibinfo{volume}{B64}},
  \bibinfo{pages}{134} (\bibinfo{year}{1973}).

\bibitem[{\citenamefont{Estabrooks and Martin}(1974)}]{Estabrooks:1974vu}
\bibinfo{author}{\bibfnamefont{P.}~\bibnamefont{Estabrooks}} \bibnamefont{and}
  \bibinfo{author}{\bibfnamefont{A.~D.} \bibnamefont{Martin}},
  \bibinfo{journal}{Nucl. Phys.} \textbf{\bibinfo{volume}{B79}},
  \bibinfo{pages}{301} (\bibinfo{year}{1974}).

\bibitem[{\citenamefont{Srinivasan et~al.}(1975)}]{Srinivasan:1975tj}
\bibinfo{author}{\bibfnamefont{V.}~\bibnamefont{Srinivasan}}
  \bibnamefont{et~al.}, \bibinfo{journal}{Phys. Rev.}
  \textbf{\bibinfo{volume}{D12}}, \bibinfo{pages}{681} (\bibinfo{year}{1975}).

\bibitem[{\citenamefont{Kaminski et~al.}(1997)\citenamefont{Kaminski, Lesniak,
  and Rybicki}}]{Kaminski:1996da}
\bibinfo{author}{\bibfnamefont{R.}~\bibnamefont{Kaminski}},
  \bibinfo{author}{\bibfnamefont{L.}~\bibnamefont{Lesniak}}, \bibnamefont{and}
  \bibinfo{author}{\bibfnamefont{K.}~\bibnamefont{Rybicki}},
  \bibinfo{journal}{Z. Phys.} \textbf{\bibinfo{volume}{C74}},
  \bibinfo{pages}{79} (\bibinfo{year}{1997}), \eprint{hep-ph/9606362}.

\bibitem[{\citenamefont{Hoogland et~al.}(1977)}]{Hoogland:1977kt}
\bibinfo{author}{\bibfnamefont{W.}~\bibnamefont{Hoogland}}
  \bibnamefont{et~al.}, \bibinfo{journal}{Nucl. Phys.}
  \textbf{\bibinfo{volume}{B126}}, \bibinfo{pages}{109} (\bibinfo{year}{1977}).

\bibitem[{\citenamefont{Froggatt and Petersen}(1977)}]{Froggatt:1977hu}
\bibinfo{author}{\bibfnamefont{C.~D.} \bibnamefont{Froggatt}} \bibnamefont{and}
  \bibinfo{author}{\bibfnamefont{J.~L.} \bibnamefont{Petersen}},
  \bibinfo{journal}{Nucl. Phys.} \textbf{\bibinfo{volume}{B129}},
  \bibinfo{pages}{89} (\bibinfo{year}{1977}).

\bibitem[{\citenamefont{Calle~Cord\'on and Ruiz~Arriola}(2009)}]{Alvaro2009}
\bibinfo{author}{\bibfnamefont{M.}~\bibnamefont{Calle~Cord\'on}}
  \bibnamefont{and}
  \bibinfo{author}{\bibfnamefont{E.}~\bibnamefont{Ruiz~Arriola}}
  (\bibinfo{year}{2009}), \eprint{In preparation}.

\bibitem[{\citenamefont{Teper}(2008)}]{Teper:2008yi}
\bibinfo{author}{\bibfnamefont{M.}~\bibnamefont{Teper}} (\bibinfo{year}{2008}),
  \eprint{0812.0085}.

\bibitem[{\citenamefont{Bali and Bursa}(2008)}]{Bali:2008an}
\bibinfo{author}{\bibfnamefont{G.~S.} \bibnamefont{Bali}} \bibnamefont{and}
  \bibinfo{author}{\bibfnamefont{F.}~\bibnamefont{Bursa}},
  \bibinfo{journal}{JHEP} \textbf{\bibinfo{volume}{09}}, \bibinfo{pages}{110}
  (\bibinfo{year}{2008}), \eprint{0806.2278}.

\bibitem[{\citenamefont{Del~Debbio et~al.}(2008)\citenamefont{Del~Debbio,
  Lucini, Patella, and Pica}}]{DelDebbio:2007wk}
\bibinfo{author}{\bibfnamefont{L.}~\bibnamefont{Del~Debbio}},
  \bibinfo{author}{\bibfnamefont{B.}~\bibnamefont{Lucini}},
  \bibinfo{author}{\bibfnamefont{A.}~\bibnamefont{Patella}}, \bibnamefont{and}
  \bibinfo{author}{\bibfnamefont{C.}~\bibnamefont{Pica}},
  \bibinfo{journal}{JHEP} \textbf{\bibinfo{volume}{03}}, \bibinfo{pages}{062}
  (\bibinfo{year}{2008}), \eprint{0712.3036}.

\bibitem[{\citenamefont{Leutwyler}(2008)}]{Leutwyler:2008xd}
\bibinfo{author}{\bibfnamefont{H.}~\bibnamefont{Leutwyler}},
  \bibinfo{journal}{AIP Conf. Proc.} \textbf{\bibinfo{volume}{1030}},
  \bibinfo{pages}{46} (\bibinfo{year}{2008}), \eprint{0804.3182}.

\end{thebibliography}

\end{document}